\documentclass[12pt]{article}
\pdfoutput=1
\usepackage{jheppub}
\usepackage{amsmath,bbm,array,amsfonts,graphicx,wrapfig,lscape,float,mathtools,multirow,longtable,rotating,subfigure,ulem,cancel}
\usepackage[section]{placeins}

\newcommand{\be}{\begin{equation}}
\newcommand{\ee}{\end{equation}}
\newcommand{\beq}{\begin{equation}}
\newcommand{\beql}[1]{\begin{equation}\label{#1}}
\newcommand{\eeq}{\end{equation}}
\newcommand{\ba}{\begin{array}}
\newcommand{\ea}{\end{array}}
\newcommand{\bea}{\begin{eqnarray}}
\newcommand{\beal}[1]{\begin{eqnarray}\label{#1}}
\newcommand{\eea}{\end{eqnarray}}
\newcommand{\ben}{\begin{enumerate}}
\newcommand{\een}{\end{enumerate}}
\newcommand{\bean}{\begin{eqnarray*}}
\newcommand{\eean}{\end{eqnarray*}}

\newcommand{\btab}[1]{\begin{tabular}{#1}}
\newcommand{\etab}{\end{tabular}}

\newcommand{\comment}[1]{}

\let\n=\nu

  \let\D=\Delta

\newcommand{\qed}{\nobreak \ifvmode \relax \else
      \ifdim\lastskip<1.5em \hskip-\lastskip
      \hskip1.5em plus0em minus0.5em \fi \nobreak
      \vrule height0.75em width0.5em depth0.25em\fi}

\numberwithin{equation}{section}

\title{Quiver Theories for Moduli Spaces of Classical Group Nilpotent Orbits}

\author{Amihay Hanany, }
\author{Rudolph Kalveks}

\affiliation{
Theoretical Physics Group, The Blackett Laboratory,
Imperial College London, \\
Prince Consort Road, London SW7 2AZ, United Kingdom
}

\emailAdd{a.hanany@imperial.ac.uk, rudolph.kalveks09@imperial.ac.uk}

\preprint{Imperial/TP/16/AH/02}

\abstract{ 
We approach the topic of Classical group nilpotent orbits from the perspective of their moduli spaces, described in terms of Hilbert series and generating functions. We review the established Higgs and Coulomb branch quiver theory constructions for $A$ series nilpotent orbits. We present systematic constructions for $BCD$ series nilpotent orbits on the Higgs branches of quiver theories defined by canonical partitions; this paper collects earlier work into a systematic framework, filling in gaps and providing a complete treatment. We find new Coulomb branch constructions for above minimal nilpotent orbits, including some based upon twisted affine Dynkin diagrams. We also discuss aspects of $3d$ mirror symmetry between these Higgs and Coulomb branch constructions and explore dualities and other relationships, such as HyperK\"ahler quotients, between quivers. We analyse all Classical group nilpotent orbit moduli spaces up to rank 4 by giving their unrefined Hilbert series and the Highest Weight Generating functions for their decompositions into characters of irreducible representations and/or Hall Littlewood polynomials.
\\

~\today}

\begin{document}

\maketitle

\listoftables

\listoffigures

\clearpage
\section{Introduction}
\label{sec:intro}

An intriguing avenue for research into the relationships between supersymmetric (``SUSY") quiver gauge theories and the nilpotent orbits of Lie groups has been opened up by a number of recent papers \cite{Gaiotto:2008ak, Chacaltana:2012zy, Cremonesi:2014uva}. The theory of nilpotent orbits  \cite{Collingwood:1993fk} provides a language for classifying and describing the moduli spaces associated with the nilpotent generators of a Classical or Exceptional group.\footnote{Recall that the nilpotent matrices of a group are nilpotent linear combinations of its raising and lowering operators, relative to some chosen basis of Cartan operators, and correspond to linear combinations of its roots.}  Nilpotent orbits are increasingly being recognised as being relevant to many topics, ranging from supergravity (``SUGRA") theories involving $G/H$ coset spaces, whose field content can be characterised by nilpotent orbits of $G$ \cite{Kim:2010bf}, to counting massive vacua in ${\cal N}=1$ Super Yang-Mills (``SYM") theory \cite{Bourget:2015lua}, where the number of vacua is derived from the structure of the nilpotent orbits of the gauge group. In \cite{Benini:2010uu}, nilpotent orbits are used as building blocks in the construction of 3d Sicilian theories and their mirrors.

We choose to approach the topic of nilpotent orbits from the perspective of their moduli spaces and Hilbert series, which we analyse using the tools of the Plethystics Program \cite{Feng:2007ur,Hanany:2014dia}. Each such Hilbert series counts holomorphic functions on the closure of a nilpotent orbit \cite{Namikawa:2016xe}.

Remarkably, it appears that all the nilpotent orbits of any Classical group $G$ correspond to the moduli spaces of particular SUSY quiver gauge theories that can be constructed on the root lattice of $G$ and which are determined by the canonical parameters associated with the orbits.

There are two extremal non-trivial nilpotent orbits, the {\it minimal} nilpotent orbit and the {\it maximal} nilpotent orbit  \cite{Collingwood:1993fk}. In the case of classical groups, the dimensions of the minimal and maximal nilpotent orbits match the dimensions of the Hilbert series of the Higgs branches of particular SUSY quiver gauge theories. These correspond to the Hilbert series of reduced single instanton moduli spaces (``RSIMS") and to the Hilbert series of $T(G)$ quiver theories, respectively. 

Instanton moduli spaces have been studied extensively, following early work in \cite{Atiyah:1978tv, Atiyah:1978uz}. The connection between instanton moduli spaces and nilpotent orbits was made in  \cite{Kronheimer:1990ay}. The minimal nilpotent orbit of a group $G$ corresponds to a reduced single $G$ instanton moduli space and, for Classical groups, the Higgs branch quiver theory constructions of such moduli spaces are well known \cite{Douglas:1995bn, Witten:1995gx}. The Hilbert series of these moduli spaces have been calculated \cite{Benvenuti:2010pq}. Indeed, many different constructions for the Hilbert series of RSIMS are known, including Coulomb branch and other types \cite{Cremonesi:2013lqa, Cremonesi:2014xha, Hanany:2015hxa}.

The Hilbert series of maximal nilpotent orbits are also known, as are their constructions for Classical groups from $T(G)$ quiver theories with maximal partitions. The Higgs branches of maximal $T(G)$ quiver theories can be calculated from linear chains comprised of gauge fields and bifundamental hypermultiplets transforming in basic representations of unitary or alternating orthogonal and symplectic groups, building on structures outlined in \cite{Gaiotto:2008ak, kobak1996classical, Kraft:1982fk}. These $T(G)$ quiver theories can be self dual under $3d$ mirror symmetry, and their Hilbert series correspond to modified Hall Littlewood polynomials transforming in the singlet representation of $G$ \cite{Cremonesi:2014kwa}.

While the minimal and maximal nilpotent orbits coincide for $SU(2)$, a group generally has a characteristic set of distinct nilpotent orbits, which is bounded by these extremal cases, and which increases in number with rank. This opens up a rich landscape for study. These nilpotent orbits can be described canonically, in terms of partition data, Dynkin labels, their dimensions, or a partial ordering using Hasse diagrams \cite{Dynkin:1957um, Kraft:1982fk, Collingwood:1993fk}. Methods have also been proposed for mapping Classical group nilpotent orbits to particular quiver theories and their Hilbert series \cite{Cremonesi:2014uva}. One aim of this paper is to recapitulate the established method for mapping $A$ series nilpotent orbits and to develop a comparable method for a complete and consistent mapping of $BCD$ series nilpotent orbits to quivers and their Hilbert series.

We focus mainly on Higgs branch constructions involving $4d$ ${\cal N}=2$ SUSY, however, by principles of $3d$ mirror symmetry \cite{Intriligator:1996ex, Hanany:1996ie, Borokhov:2002cg, Gaiotto:2008ak, Hanany:2011db}, these can have counterparts in the form of Coulomb branch constructions on dual quiver theories involving ${\cal N}=4$ SUSY in three dimensions. Consequently, our investigations also shed light on aspects of $3d$ mirror symmetry. 

We introduce a number of systematic improvements to the analysis of the moduli spaces of $BCD$ nilpotent orbits and develop and implement methods for the systematic decomposition of the Hilbert series (``HS") of Classical group nilpotent orbits into their representation content, which we describe in terms of highest weight generating functions (``HWGs"), either for irreducible representations (``irreps") or modified Hall Littlewood polynomials (``mHL") of $G$.

In section \ref{sec:orbits}, we summarise relevant aspects of the theory of nilpotent orbits presented in the mathematical literature \cite{Dynkin:1957um, Collingwood:1993fk} and give simple algorithms for identifying the nilpotent orbits of any Classical group $G$ and calculating their dimensions, by finding homomorphisms from $SU(2)$ to $G$ using character maps and selection rules. The labelling of nilpotent orbits that we adopt is consistent with that in the mathematical literature. Each nilpotent orbit of $G$ is associated with a moduli space of representations of $G$ and our objective is to identify and describe these spaces in terms of their Hilbert series and their decompositions into representations of $G$. Presented in refined form, a Hilbert series faithfully encodes the representation content of a nilpotent orbit, up to isomorphisms, and, for brevity, we may identify a nilpotent orbit by either its refined Hilbert series or quiver.

In section \ref{sec:Aseries}, we carry out a complete Higgs branch analysis of $A$ series quiver chains with unitary gauge nodes, corresponding to $A$ series nilpotent orbits, up to and including rank 4. We describe the moduli spaces of these chains in terms of Hilbert series and their decompositions into characters and/or modified Hall Littlewood polynomials of $A_n$. Appendix \ref{apxHLP} contains some basic information about our use of modified Hall Littlewood polynomials and their generating functions in decompositions. The reader is also referred to \cite{Hanany:2015hxa} for a fuller exposition of our method of working with modified Hall Littlewood polynomials. We confirm identities between the Higgs branches of these quivers and the Coulomb branches of their $3d$ mirror duals and examine the relationship between the Coulomb branch quivers and the corresponding canonical nilpotent orbit descriptions.

In section \ref{sec:BCDseries}, we carry out a complete Higgs branch analysis of quiver chains with alternating $O/USp$ gauge nodes, corresponding to $B,C$ and $D$ series nilpotent orbits, up to and including rank 4. We describe the moduli spaces of these chains in terms of Hilbert series and their decompositions into characters of $G$ and/or into the modified Hall Littlewood polynomials of $G$. We also find new Coulomb branch constructions for the moduli spaces of supra-minimal (i.e. next to minimal in the Hasse diagram) and other close to minimal $BCD$ series nilpotent orbits, using methods based upon twisted affine Dynkin diagrams, amongst others.

In the concluding section \ref{sec:conclusions}, we summarise key findings and the many dualities that can be identified and also discuss the implications for aspects of $3d$ mirror symmetry.

\paragraph{Notation and Terminology}
We freely use the terminology and concepts of the Plethystics Program, including the Plethystic Exponential (``PE"), its inverse, the Plethystic Logarithm (``PL"), the Fermionic Plethystic Exponential (``PEF") and, its inverse, the Fermionic Plethystic Logarithm(``PFL"). For our purposes:
\begin{equation} 
\label{eq:intro1}
\begin{aligned}
PE\left[ {\sum\limits_{i = 1}^d {{A_i}} ,t} \right] & \equiv \prod\limits_{i = 1}^d {\frac{1}{{\left( {1 - {A_i}t} \right)}}},\\
PE\left[ { - \sum\limits_{i = 1}^d {{A_i}} ,t} \right] & \equiv \prod\limits_{i = 1}^d {\left( {1 - {A_i}t} \right)},\\
PE\left[ {\sum\limits_{i = 1}^d {{A_i}} , - t} \right] & \equiv \prod\limits_{i = 1}^d {\frac{1}{{\left( {1 + {A_i}t} \right)}}} ,\\
 PE\left[ { - \sum\limits_{i = 1}^d {{A_i}} , - t} \right] & \equiv PEF\left[ {\sum\limits_{i = 1}^d {{A_i}} ,t} \right] & \equiv \prod\limits_{i = 1}^d {\left( {1 + {A_i}t} \right)},\\
 \end{aligned}
\end{equation}
where $A_i$ are monomials in weight or root coordinates or fugacities. The reader is referred to \cite{Benvenuti:2006qr} or \cite{Hanany:2014dia} for further detail.

We present the characters of a group $G$ either in the generic form ${\cal X}_{G}(x_i)$, or as $[irrep]_{G}$, or using Dynkin labels as ${[ {{n_1}, \ldots , {n_r}}]_{G}}$, where $r$ is the rank of $G$. We may refer to \textit{series}, such as $1 + f + {f^2} + \ldots $, by their \textit{generating functions} $1/\left( {1 - f} \right)$. We use distinct coordinates/variables to help distinguish the different types of generating function, as indicated in table \ref{tab1}.
\begin{table}[htp]
\caption{Types of Generating Function}
\begin{center}
\begin{tabular}{|c|c|c|}
\hline
$\text{Generating~Function}$&$ \text{Notation}$&$ \text{Definition} $\\
\hline
$\text{Refined~HS~(Weight~coordinates)}$&$ {{g^{G}_{HS}}( {{t},{x}} )}$&$\sum\limits_{n = 0}^\infty {{a_n}({x})}{t^n} $\\
$\text{Refined~HS~(Simple root~coordinates)}$&$ {{g^{G}_{HS}}( {{t},{z}} )}$&$\sum\limits_{n = 0}^\infty {{a_n}({z})}{t^n} $\\
$\text{Unrefined~HS}$&$ {{g^{G}_{HS}}\left( t \right)}$&$ \sum\limits_{n = 0}^\infty {{a_n}} {t^n} \equiv \sum\limits_{n = 0}^\infty {{a_n}({1})}{t^n} $\\
\hline
$\text{HWG (Character) for HS}$&$g_{HS}^{G}( {{t},{m}} ) $&$ \sum\limits_{{n_1}, \ldots ,{n_r} = 0}^\infty {{a_{{n_1}, \ldots ,{n_r}}}( {{t}} )} ~m_1^{{n_1}} \ldots m_r^{{n_r}}$\\
$\text{HWG (mHL) for HS}$&$g_{HS}^{G}( {{t},{h}} ) $&$ \sum\limits_{{n_1}, \ldots ,{n_r} = 0}^\infty {{a_{{n_1}, \ldots ,{n_r}}}( {{t}} )} ~h_1^{{n_1}} \ldots h_r^{{n_r}}$\\
\hline
$\text{Character}$&$ {{g^{G}_{\cal X}}( {{m},{x}} )}$&$\sum\limits_{{n_1}, \ldots ,{n_r} = 0}^\infty {[ {{n_1}, \ldots ,{n_r}}]_G(x)} ~m_1^{{n_1}} \ldots m_r^{{n_r}} $\\
\hline
$\text{modified~Hall~Littlewood}$&$ {{g^{G}_{mHL}}( {{h},{x,t}} )}$&$\sum\limits_{{n_1}, \ldots ,{n_r} = 0}^\infty {mHL^G_{[ {{n_1}, \ldots ,{n_r}}]}} \left( {{x},t} \right)~h_1^{{n_1}} \ldots h_r^{{n_r}}$\\
\hline
\end{tabular}
\end{center}
\label{tab1}
\end{table}

These different types of generating function are related and can be considered as a hierarchy in which the refined HS, HWG, character and mHL generating functions fully encode the group theoretic information about a moduli space. We typically label unimodular Cartan subalgebra (``CSA") coordinates for weights within characters by $x \equiv (x_1 \ldots x_r)$ and simple root coordinates by $z \equiv (z_1 \ldots z_r)$, dropping subscripts if no ambiguities arise. We use the Cartan matrix $A_{ij}$ to define the canonical relationships between simple root and CSA coordinates as ${z_i} = \prod\limits_j {x_j^{{A_{ij}}}}$ and ${x_i} = \prod\limits_j {z_j^{{A^{ - 1}}_{ij}}}$. We generally label field counting variables with $t$, adding subscripts if necessary.

Finally, we deploy highest weight notation \cite{Hanany:2014dia}, which uses fugacities to track highest weight Dynkin labels, and describes the structure of a Hilbert series in terms of the highest weights of its constituent irreps. We typically denote such Dynkin label counting variables by $m \equiv (m_1 \ldots m_r)$ for representations based on characters $[n]_G \equiv {[ {{n_1}, \ldots ,{n_r}}]_G}$ and by $h \equiv (h_1 \ldots h_r)$ for representations based on Hall-Littlewood polynomials ${mHL^G_{[n]}}$, although we may also use other letters, where this is helpful. We define these counting variables to have a complex modulus of less than unity and follow established practice in referring to them as ``fugacities", along with the monomials formed from the products of CSA or root coordinates.


\section{Nilpotent Orbits}
\label{sec:orbits}

We will limit ourselves to a brief summary that is pertinent to the enumeration of nilpotent orbits for Classical groups; the reader is referred to \cite{Collingwood:1993fk} for a full exposition. We start from the Jacobson Morozov Theorem, which states that the nilpotent orbits of a group $G$ are in one to one correspondence, up to conjugation, with the homomorphisms $\rho$ from $SU(2)$ to $G$.

\subsection{Homomorphisms as Character maps}
Such homomorphisms lead to character maps from $G$ to $SU(2)$, under which every representation of $G$ decomposes into an exact {\it sum} of representations of $SU(2)$:

\begin{equation} 
\label{eq:orbits1}
\begin{aligned}
\rho :\left( {{x_1}, \ldots {x_r}} \right) & \to \left( {{x^{{n_1}}}, \ldots ,{x^{{n_r}}}} \right),\\
\rho \left( {{{{\cal X}}_G}\left( {{x_1}, \ldots {x_r}} \right)} \right) & \to \sum\nolimits_{}^ \oplus  {mult_{\left[ n \right]}{{\left[ n \right]}_{{A_1}}}} \left( x \right),
 \end{aligned}
\end{equation}
where we have taken the CSA coordinates of $G$ and $SU(2)$ as $\{x_1,\ldots,x_{rank(G)}\}$ and $\{x\}$, respectively. The enumeration of nilpotent orbits therefore reduces to the problem of identifying all such valid character maps.

We can refine the problem as follows. The exponents of $x$ that appear in a valid map $\rho(R)$ of some representation $R$ of $G$ are weight space Dynkin labels of SU(2) and must therefore be integers. Moreover, the highest exponent of $x$ that can appear must be an integer below $Dim[R]$, otherwise the monomials within $\rho(R)$ could not form a complete representation. Furthermore, once we establish that a map $\rho$ is valid for all the basic representations of $G$ (those with highest weight Dynkin labels of the form $[0,\ldots,1,\ldots,0]$), it follows that the map must be valid for all representations of $G$ \cite{Fuchs:1997bb}. This limits the number of possible maps at most to the product of the dimensions of the basic representations of $G$.

Indeed, the number of possible maps can be limited further by a theorem \cite{Dynkin:1957um}, which states that the map $\rho$, when expressed in terms of the simple roots $\{ z_1,\dots,z_r \}$ of $G$ and $\{z\}$ of $SU(2)$, must be conjugate under the action of the Weyl group of $G$ to a map of the form:
\begin{equation} 
\label{eq:orbits2}
\begin{aligned}
\rho :\left( {{z_1}, \ldots {z_r}} \right) \to \left( {{z^{\frac{q_1}{2}}}, \ldots ,{z^{\frac{q_r}{2}}}} \right),
 \end{aligned}
\end{equation}
where ${q_i} \in \left\{ {0,1,2} \right\}$. The labels $[q_1,\ldots,q_r]$ are termed the Dynkin labels of the nilpotent orbit\footnote{As distinct from the highest weight Dynkin labels of a representation}. Thus, there are at most $3^{rank[G]}$ possible character maps that need to be tested for a given group $G$, which is a straightforward computational procedure.

These homomorphisms can be labelled by the decomposition in $SU(2)$ of $\rho(R)$, where $R$ is some representation of $G$. For $B$ and $D$ series groups, $R$ is usually chosen to be the vector representation, or, for $A$ and $C$ series groups, the fundamental representation. These decompositions are conventionally expressed using condensed partition notation, under which each $SU(2)$ irrep in $\rho(R)$ is assigned an element in the partition equal to its dimension, and the partition elements for any irreps with non-zero multiplicities are assigned exponents equal to their multiplicities \footnote{The labelling of partitions $\rho$ can be refined to assign the multiplicities $\{a_0, \ldots,a_{max}\}$ of $SU(2)$ representations to representations of the group $H$ that is generated by the subalgebra of $G$ that commutes with the $SU(2)$ subalgebra, termed the commutant of $\rho$ in $G$.}:

\begin{equation} 
\label{eq:orbits3}
\begin{aligned}
\rho(R) & = \sum\limits_{n = 0}^{max } {{a_n}\left[ n \right]},\\ 
\rho(R) & \Leftrightarrow ( {Dim{{\left[ {max } \right]}^{{a_{max }}}}, \ldots ,Dim{{\left[ n \right]}^{{a_n}}}, \ldots ,{1^{{a_0}}}}).
 \end{aligned}
\end{equation}

Additional selection rules are required to ensure that the partitions $\rho(R)$ assigned to each irrep of $G$ by the homomorphism $\rho$ are consistent with its bilinear invariants. Recall that an irrep can be classified as (i) real, (ii) pseudo real or (iii) complex, depending, respectively, on whether it has (i) a symmetric bilinear invariant with itself, (ii) an antisymmetric bilinear invariant with itself, or (iii) a bilinear invariant with its complex conjugate. As shown in \cite{Collingwood:1993fk}, when $R$ has bilinear symmetric or antisymmetric invariants, this leads to selection rules that exclude homomorphisms $\rho$ containing partitions $\rho(R)$ that do not meet specified criteria which depend on the bilinears of $R$:

\begin{enumerate}
\item Real $R$. If an even element appears, it must appear an even number of times. These are often referred to as $B$ partitions or $D$ partitions.
\item Pseudo real $R$. If an odd element appears, it must appear an even number of times. These are often referred to as $C$ partitions.
\end{enumerate}

It is important to appreciate that the selection rules depend crucially on the representation $R$ of the parent group upon which $\rho$ acts, since several groups contain both real and pseudo real representations. We set out in appendix \ref{apxHom} a full set of these homomorphisms, up to conjugation, along with their action on the key basic irreps and the adjoint irrep of Classical groups up to rank 5. While partial tables are regularly presented in the literature \cite{Collingwood:1993fk, Chacaltana:2012zy}, we believe that a fuller presentation, including spinors and the adjoint representation in particular, is helpful to an understanding of nilpotent orbits.

Thus, taking $A_3$ as an example, there are five nilpotent orbits and these can be referred to uniquely, either by the partition data assigned to one of the basic representations, or by the Dynkin labels of the root coordinate map, or by the CSA coordinate map under the homomorphism $\rho$. Taking the fundamental character of $A_3$ as $[1,0,0]=x_1+x_2/x_1+x_3/x_2+1/x_3$ and its simple roots as $\{ z_1={x^2}_1/x_2, z_2={x^2}_2/x_1/x_3, z_3={x_3}^2/x_2 \}$, all obtained from the Cartan matrix for $A_3$, we can express the homomorphism $\rho \equiv (4)$ in any one of the following equivalent ways:
\begin{equation} 
\label{eq:orbits4}
\begin{aligned}
\rho :\left( {{x_1},{x_2},{x_3}} \right) &\to \left( {{x^3},{x^4},{x^3}} \right),\\
\rho :\left( {{z_1},{z_2},{z_3}} \right) &\to \left( {{z},{z},{z}} \right),\\
\rho :\left( {{x_1} + {x_2}/{x_1} + {x_3}/{x_2} + 1/{x_3}} \right) &\to \left( {{x^3} + x + 1/x + 1/{x^3}} \right),\\
\rho :\left[ {1,0,0} \right] &\to [3].\\
 \end{aligned}
\end{equation}

Since there is a bijective correspondence between partitions and homomorphisms \cite{Collingwood:1993fk} , the possible partitions can also be found from generating functions that encapsulate the selection rules. We introduce fugacities $(n_1,\ldots, n_{N})$, where $N$ is the fundamental/vector dimension of the flavour group, to identify the dimensions of the $SU(2)$ irreps appearing in the homomorphism $\rho$, such that the exponents of the fugacities correspond to the multiplicities of each irrep. For example, $\rho \equiv (4)$ maps to the monomial  $n_4$ and $\rho \equiv (1^2,2)$ maps to the monomial  $n_1^2 n_2$. We use an overall counting fugacity $t$. A short calculation then leads to the generating functions for partitions set out in table \ref{tab:orbits0}.

\begin{table}[htp]
\caption{Generating Functions for Partitions of Classical Group Nilpotent Orbits}
\begin{center}
\begin{tabular}{|c|c|c|}
\hline
 $\text{Flavour Group}$&$\text{Partition Series}$& $\text{Generating Function}$\\
\hline
$ {SU(N)}$&$
{\sum\limits_{i = 1}^\infty  {{P_{SU}}\left( {{n _1}, \ldots ,{n _\infty}} \right){t^i}} }
$&$
{\prod\limits_{i = 1}^\infty  {\frac{1}{{1 - {n _i}{t^i}}} - 1} }
$\\
\hline
$ {USp(N)}$ &${\sum\limits_{i = 1}^\infty  {{P_{USp}}\left( {{n _1}, \ldots ,{n _\infty}} \right){t^i}} }$ &$ {\prod\limits_{i = 1}^\infty  {\frac{1}{{1 - {n _i}{t^i}}}} \prod\limits_{j = 0}^\infty  {\frac{1}{{1 + {n _{2j + 1}}{t^{2j + 1}}}} - 1} }  $\\
\hline
$ {SO(N)}$ &${\sum\limits_{i = 1}^\infty  {{P_{SO}}\left( {{n _1}, \ldots ,{n _\infty}} \right){t^i}} }$ &${\prod\limits_{i = 1}^\infty  {\frac{1}{{1 - {n _i}{t^i}}}} \prod\limits_{j = 1}^\infty  {\frac{1}{{1 + {n _{2j}}{t^{2j}}}} + \prod\limits_{j = 1}^\infty  {\frac{1}{{1 + n _{2j}^2{t^{4j}}}} - 2} } }$\\
\hline
\end{tabular}
\end{center}
\label{tab:orbits0}
\end{table}
Thus, to obtain the partitions for the fundamental of $SU(4)$, we find the coefficient of $t^4$ in the Taylor expansion of the generating function for ${\sum\limits_{i = 1}^\infty  {{P_{SU}}\left( {{n _1}, \ldots ,{n _\infty}} \right){t^i}} }$. This is $n_1^4+n_1^2 n_2+n_1 n_3+n_2^2+n_4$, corresponding to the set of five partitions $\{(1^4), (1^2, 2),(1, 3),( 2^2),(4)\}$.
\subsection{Dimensions of Nilpotent Orbits}
Each nilpotent orbit $\cal O_{\rho}$ has a characteristic dimension $|\cal O_{\rho}|$, which can be calculated from the partition data, as set out in \cite{Collingwood:1993fk}. Consider an ordered partition (in standard notation) $\rho = ( {{\rho _1}, \ldots ,{\rho _n}} )$, with $max$ being the greatest element appearing in $\rho$. The transpose partition $\sigma \equiv\rho^T$, where $\sigma = ( {{\sigma _1}, \ldots ,{\sigma_{max}}} )$, can be obtained using Young's diagrams. It is convenient, for our purposes, to restate (6.1.4) of \cite{Collingwood:1993fk} more simply in terms of rank $n$ and the transposed partition $\sigma$, to obtain the dimension formulae shown in table \ref{tab:orbits1}. These dimensions are based on a lattice over a complex space and are always even.

\begin{table}[htp]
\caption{Dimension Formulae for Nilpotent Orbits of Classical Groups}
\begin{center}
\begin{tabular}{|c|c|}
\hline
 $\text{Group}$& $\text{Dimension of Nilpotent Orbit}~|\cal O_{\rho}|$\\
\hline
$ {A_n}$ & ${\left( {n + 1} \right)^2} - \sum\limits_{i = 1}^{max } {\sigma_i^2} $\\
\hline
$ {B_n}$ &$n\left( {2n + 1} \right) - \frac{1}{2}\sum\limits_{i~odd} {{\sigma_i}\left( {{\sigma_i} - 1} \right)} - \frac{1}{2}\sum\limits_{i~even} {{\sigma_i}\left( {{\sigma_i} + 1} \right)} $\\
\hline
$ {C_n}$ &$n\left( {2n + 1} \right) - \frac{1}{2}\sum\limits_{i~odd} {{\sigma_i}\left( {{\sigma_i} + 1} \right)} - \frac{1}{2}\sum\limits_{i~even} {{\sigma_i}\left( {{\sigma_i} - 1} \right)} $\\
\hline
$ {D_n}$ &$n\left( {2n - 1} \right) - \frac{1}{2}\sum\limits_{i~odd} {{\sigma_i}\left( {{\sigma_i} - 1} \right)} - \frac{1}{2}\sum\limits_{i~even} {{\sigma_i}\left( {{\sigma_i} + 1} \right)} $\\
\hline
\end{tabular}
\end{center}
\label{tab:orbits1}
\end{table}
We can identify within the expressions for $|\cal O_{\rho}|$, the dimension of the flavour group of rank $n$, reduced by a sequence of dimensions of square matrices defined by the elements $\sigma_i$ from the partition data. For the $A$ series, this sequence is associated with unitary matrices, while for $BCD$ series, this sequence is associated with alternating symmetric and antisymmetric real matrices.

We note that the dimensions of any nilpotent orbit can be found more directly by subtracting from $Dim[G]$ the number of $SU(2)$ representations into which the adjoint representation of $G$ is split by the homomorphism $\rho$ :
\begin{equation} 
\label{eq:orbits5}
\begin{aligned}
\left| {{{{\cal O}}_\rho }} \right| = Dim \left[G \right] - Length [\rho \left( {adj\left( G \right)} \right)],
 \end{aligned}
\end{equation}
as can be checked by inspection of appendix \ref{apxHom}. Importantly, identical dimensions can also be obtained by assigning a quiver theory to any partition that satisfies the B/D and C-partition selection rules, as will be shown below.

The dimensions of nilpotent orbits have a partial ordering, which is often expressed using Hasse diagrams. There are a number of characteristic orbits within this partial ordering:
\begin{enumerate}
\item The trivial orbit. This is associated with the partition $(1^{Dim[irrep]})$ and always has zero dimension.
\item The minimal orbit. This is the first orbit with non-zero dimension and is always unique. Its complex dimension is equal to twice the sum of the dual Coxeter labels of the Dynkin diagram for $G$. This equals the dimension of the reduced single instanton moduli space of $G$.
\item The sub-regular orbit. This is the orbit with next to highest dimension, which is always unique, having a complex dimension equal to the number of the roots, less $2$.
\item The maximal orbit. This is the orbit with highest dimension and is always unique. Its complex dimension is equal to the number of roots of the group. 

\end{enumerate}

The moduli space of the maximal orbit is equal to the modified Hall-Littlewood polynomial of $G$ transforming in the singlet representation, $mHL^G_{[0,\ldots,0]}$. This obeys the important identity \cite{Cremonesi:2014kwa} involving the Casimirs of $G$ \footnote{The Casimirs of a group are given by the degrees of the symmetric invariant tensors of the adjoint representation, being $A_n:\{2,\ldots,n,n+1\}$, $B/C_n:\{2, 4,\ldots,2\}$, $D_n:\{2,4,\ldots,2n-2,n\}$, $G_2: \{2,6\}$, $F_4: \{2,6,8,12\}$, $E_6:\{2,5,6,8,9,12\}$, $E_7:\{2,6,8,10,12,14,18\}$ and $E_8:\{2,8,12,14,18,20,24,30\}$} \footnote{We use mHL polynomials with a $t^2$ fugacity in order to match the Higgs branch constructions.}:
\begin{equation} 
\label{eq:Aseries3}
\begin{aligned}
mHL^{G}_{[0,\ldots,0]}(t^2) & =\left( \prod\limits_{Casimirs} {\left( {1 - {t^{2~degree(Casimir)}}} \right)} \right) PE[adjoint,{t^2}],
 \end{aligned}
\end{equation}
The above orbits are not distinct for low rank groups. For example, in the case of $A_1$, the minimal and maximal orbits coincide, as do the trivial and sub-regular. It is also significant that a description of nilpotent orbits, by partitions of the vector representation alone, does not give a unique labelling for $D$ series groups of even rank. Recalling that the spinor is a more fundamental representation than a vector, we can see in appendix  \ref{apxD}, for example, that the $(2^4)$ and $(4^2)$ vector partitions of $D_4$ both correspond to pairs of nilpotent orbits that are distinguished by the partition data for the spinors.

\subsection{Quiver Theories for Nilpotent Orbits as Moduli Spaces}
\label{subsec:quivers}
\begin{figure}[htbp]
\begin{center}
\includegraphics[scale=0.5]{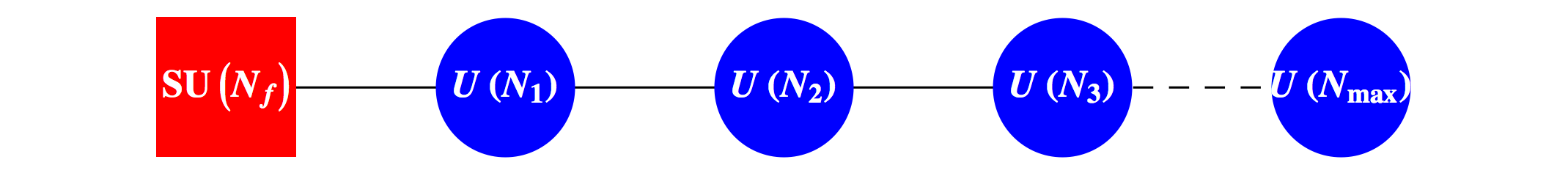}\\
\caption[Unitary Linear Quiver]{Unitary Linear Quiver. Square (red) nodes denote flavour nodes. Round (blue) nodes denote gauge nodes. The links represent pairs of bifundamental fields transforming in the fundamental or antifundamental representations. The quiver is ordered such that $N_f > N_1 > N_i > \ldots > N_{max}$.}
\label{fig:Aquiver}
\end{center}
\end{figure}

SUSY quiver gauge theories whose Higgs branches correspond to nilpotent orbits are all described by an $SU(N_f)$ flavour node linked to certain linear chains of unitary gauge nodes $U(N_i)$ \cite{kobak1996classical}. Such quivers are a subset of the set of quivers with a descending sequence of unitary gauge nodes, as shown in figure \ref{fig:Aquiver}. We shall often use the notation $[N_f]-(N_1)-\ldots (N_{max})$ to describe such quivers.

If we consider the Higgs branch of such a quiver: each link represents a bifundamental hypermultiplet containing a conjugate pair of scalar fields $X_{ij}$ and $X_{ji}$ transforming under the flavour and/or gauge groups associated with its nodes; each gauge node is associated with a scalar field $\Phi_{ii}$ transforming in the adjoint representation of the gauge group. A superpotential can be formed by contracting the bifundamental and adjoint fields. The F-terms obtained by application of vacuum minima conditions to the superpotential lead to the association to each node of a HyperK\"ahler quotient (``HKQ"). The ring of gauge invariant operators that is formed by symmetrising the bifundamental fields, modulo the HKQ, can be enumerated in a Hilbert series.

The Higgs branch formula for this Hilbert series, expressed in terms of characters ${\cal X}(x)$ and a counting fugacity $t$, is:
\begin{equation} 
\label{eq:orbits5a}
\begin{aligned}
{g_{Higgs}^A}\left( {\cal X}(x),t \right) = \oint\limits_{gauge} {d\mu } \prod\limits_{i > j} {\frac{{PE\left[ {{{\cal X}}\left( {{X_{ij}} + {X_{ji}}} \right),t} \right]}}{{{g_{HK}}({{{\cal X}}_{gauge\left( i \right)}},t)}}}.\\
 \end{aligned}
\end{equation}
One delicate aspect of this calculation is that of the HyperK\"ahler quotient $g_{HK}$. This has the effect of ensuring, for each Weyl integration, that the flavour group Hilbert series excludes any flavour group singlets, (which might otherwise result under the PE from invariants of the gauge group). As noted above, the usual method of calculation involves applying vacuum conditions to the superpotential terms that can be constructed from the bifundamental fields and adjoint gauge fields. A more direct route, which we adopt here, is to find the HKQ from the refined Hilbert series of the gauge fields that correspond to the flavour group singlets that we wish to exclude:

\begin{equation} 
\label{eq:orbits6}
\begin{aligned}
{g_{HK}}\left( {{{{\cal X}}_{U\left( {{N_i}} \right)}},t} \right) &= \oint\limits_{U\left( {{N_j}} \right)} {d\mu }~ PE\left[ {{{\cal X}}\left( {{X_{ij}}} \right) + {{\cal X}}\left( {{X_{ji}}} \right),t} \right].\\
 \end{aligned}
\end{equation}
For a linear $A$ series quiver, this HyperK\"ahler quotient is usually equal to the PE of the adjoint of the gauge group: ${g_{HK}}({\cal X}_{U(N_i)},t ) = PE[{\cal X}( \Phi_{ii}),t^2]$. The Hilbert series for the Higgs branch is then given by:
\begin{equation} 
\label{eq:orbits7}
\begin{aligned}
{g_{Higgs}^A}\left( {\cal X}(x),t \right) = \oint\limits_{gauge} {d\mu } \left( {\prod\limits_{i>j} {\frac{{PE\left[ {{{\cal X}}\left( {{X_{ij}} + {X_{ji}}} \right),t} \right]}}{{PE\left[ {{{\cal X}}\left( {{\Phi _{ii}}} \right),{t^2}} \right]}}} } \right)
 \end{aligned}
\end{equation}
The dimension of this Hilbert series, when unrefined by setting all the flavour group CSA coordinates $x$ to unity, is given by the formula:
\begin{equation} 
\label{eq:orbits8}
\begin{aligned}
Dim \left[ {{g_{Higgs}^A}\left({\cal X}(1), t \right)} \right] = \sum\limits_{ij}^{} {Dim \left[ {{{\cal X}}\left( {{X_{ij}}} \right)} \right]} - \sum\limits_i^{} {Dim \left[ {{{\cal X}}\left( {{\Phi _{ii}}} \right)} \right] - \sum\limits_{gauge}^{} {Dim [ {adjoint} ]} }
 \end{aligned}
\end{equation}
The last two terms on the RHS follow from the HyperK\"ahler quotient and the Weyl integration over each gauge group, respectively, and have identical dimensions. Since we have assumed that the sequence of node dimensions $\{N_f, N_1,\ldots,N_{max}\}$ is non-increasing, we can assign unordered partition data to the quiver using the rule:
\begin{equation} 
\label{eq:orbits9}
\begin{aligned}
\sigma = \left\{ {{\sigma _i}:{\sigma _1} = {N_f} - {N_1};{\sigma _{i}} = {N_{i - 1}} - {N_i};{\sigma _{max }} = {N_{max }}} \right\}.
 \end{aligned}
\end{equation}
Note that the $\sigma _{i}$ from this construction are non-negative, but are not necessarily ordered. We now use the identity, ${N_f} = \sum\limits_{i = 1}^{max } {{\sigma _i}}$, to rearrange the dimension formula \ref{eq:orbits8} as:
\begin{equation} 
\label{eq:orbits10}
\begin{aligned}
Dim \left[ {{g^{A}_{Higgs}}\left(1,t \right)} \right] &= \sum\limits_{n = 1}^{max - 1} {\underbrace {2\left( {{N_f} - \sum\limits_{i = 1}^{n - 1} {{\sigma _{i}}} } \right)\left( {{N_f} - \sum\limits_{i = 1}^n {{\sigma _{i}}} } \right)}_{hypers} - } 2\underbrace {{{\left( {{N_f} - \sum\limits_{i = 1}^n {{\sigma _{i}}} } \right)}^2}}_{vectors}\\
& = {N_f}^2 - \sum\limits_{i = 1}^{max} {{\sigma ^2}_i}.\\
 \end{aligned}
\end{equation}
Thus, we have recovered the dimension of the $A$ series nilpotent orbit in table \ref{tab:orbits1} from the unitary quiver defined by the sequence $\{N_f, N_1,\ldots,N_{max}\}$. So, we can use the partition data associated with an $A$ series nilpotent orbit to identify a unitary linear quiver, whose moduli space has a Hilbert series of the same dimension as the nilpotent orbit.

The process of matching partition data from the nilpotent orbits of $BCD$ series groups to quiver theories is similar, albeit less straightforward. The dimension formulae in table \ref{tab:orbits1} for $BCD$ groups invite association with alternating $O/USp$ groups. As a natural development from diagrams outlined in \cite{kobak1996classical}, it was proposed in \cite{Gaiotto:2008ak} that linear quivers for $BCD$ groups could take the form of alternating chains of $O/USp$ groups. It is therefore natural to examine the mapping of partition data from nilpotent orbits to the vector/fundamental dimensions of an alternating chain of $O/USp$ groups. 

One issue that arises is that some partitions could require $USp$ groups with odd fundamental dimension; however, homomorphisms $\rho$ with such partitions are precisely those excluded by the B/D and C-partition selection rules. So the B/D and C-partition selection rules in effect correspond to the restriction of nilpotent orbits for $BCD$ groups to homomorphisms $\rho$ that can meaningfully be described by an alternating $O/USp$ chain.

\begin{figure}[htbp]
\begin{center}
\includegraphics[scale=0.5]{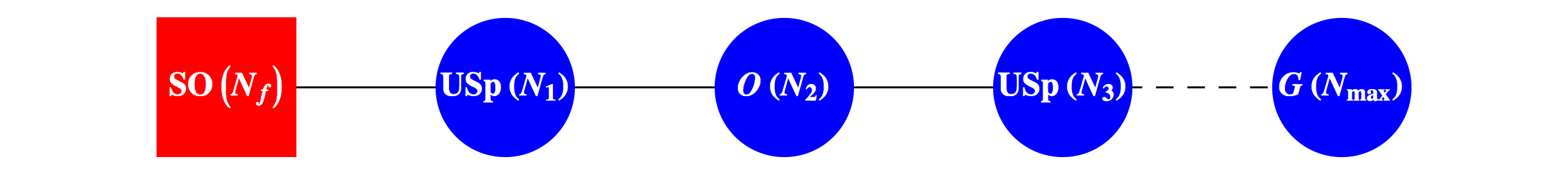}\\
\caption[Orthogonal Linear Quiver]{Orthogonal Linear Quiver. Square (red) nodes denote flavour nodes. Round (blue) nodes denote gauge nodes. The links represent bifundamental fields transforming in the vector/fundamental representations. The quiver is ordered such that $N_f > N_1 > N_i > \ldots > N_{max}$.}
\label{fig:BDquiver}
\end{center}
\end{figure}

\begin{figure}[htbp]
\begin{center}
\includegraphics[scale=0.5]{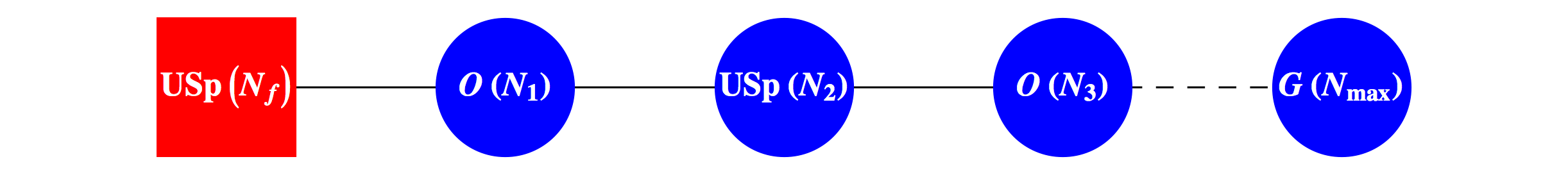}\\
\caption[Symplectic Linear Quiver]{Symplectic Linear Quiver. Square (red) nodes denote flavour nodes. Round (blue) nodes denote gauge nodes. The links represent bifundamental fields transforming in the vector/fundamental representations. The quiver is ordered such that $N_f > N_1 > N_i > \ldots > N_{max}$.}
\label{fig:Cquiver}
\end{center}
\end{figure}

The linear $BCD$ quivers that we investigate all take the form of chains of alternating $O/USp$ nodes, with the first node being a flavour node and the remaining nodes being gauge nodes, ordered with non-increasing vector/fundamental dimension, as in figures \ref{fig:BDquiver} and \ref{fig:Cquiver}.\\


We can calculate the Hilbert series for the Higgs branches of such $BCD$ series quivers and find their dimensions using prescriptions similar to \ref{eq:orbits7} and \ref{eq:orbits8}, with certain modifications. The fields $X_{jk}$ are now half-hypermultiplets, so that there is just one field $X_{jk}$ between nodes $\{j,k \}$. There are complications relating to the structure of the HyperK\"ahler quotient and the use of orthogonal rather than SO groups; these do not, however, affect the dimensions of a Hilbert series, so we defer a discussion of these topics to section \ref{sec:BCDseries}. The application of the dimensional formula \ref{eq:orbits10} necessarily reflects both the series of the flavour group and the position of a node, with the gauge group series matching (or complementing) the flavour group on even (or odd) indexed $N_i$ nodes. Otherwise the Higgs branch dimension formula for $BCD$ quivers follows in a similar manner to that for $A$ series quivers:
\begin{equation} 
\label{eq:orbits11}
\begin{aligned}
Dim \left[ {{g^{B/D}_{Higgs}}\left(1,t \right)} \right] &=\sum\limits_n {\left( {{N_f} - \sum\limits_{k = 1}^{n - 1} {{\sigma _k}} } \right)\left( {{N_f} - \sum\limits_{k = 1}^n {{\sigma _k}} } \right) }\\
&~~~~~~~~~~ - \sum\limits_{n~odd} {\left( {{N_f} + 1 - \sum\limits_{k = 1}^n {{\sigma _k}} } \right)\left( {{N_f} - \sum\limits_{k = 1}^n {{\sigma _k}} } \right)}\\
&~~~~~~~~~~ - \sum\limits_{n~even} {\left( {{N_f} - 1 - \sum\limits_{k = 1}^n {{\sigma _k}} } \right)\left( {{N_f} - \sum\limits_{k = 1}^n {{\sigma _k}} } \right)}\\
& = \frac{1}{2}{N_f}\left( {{N_f} - 1} \right) - \frac{1}{2}\sum\limits_{i~odd} {{\sigma_i}\left( {{\sigma_i} - 1} \right)} - \frac{1}{2}\sum\limits_{i~even} {{\sigma_i}\left( {{\sigma_i} + 1} \right)}
 \end{aligned}
\end{equation}
\begin{equation} 
\label{eq:orbits12}
\begin{aligned}
Dim \left[ {{g^{C}_{Higgs}}\left(1,t \right)} \right] &=\sum\limits_n {\left( {{N_f} - \sum\limits_{k = 1}^{n - 1} {{\sigma _k}} } \right)\left( {{N_f} - \sum\limits_{k = 1}^n {{\sigma _k}} } \right) }\\
&~~~~~~~~~~ - \sum\limits_{n~odd} {\left( {{N_f} - 1 - \sum\limits_{k = 1}^n {{\sigma _k}} } \right)\left( {{N_f} - \sum\limits_{k = 1}^n {{\sigma _k}} } \right)}\\
&~~~~~~~~~~ - \sum\limits_{n~even} {\left( {{N_f} + 1 - \sum\limits_{k = 1}^n {{\sigma _k}} } \right)\left( {{N_f} - \sum\limits_{k = 1}^n {{\sigma _k}} } \right)}\\
& = \frac{1}{2}{N_f}\left( {{N_f} + 1} \right) - \frac{1}{2}\sum\limits_{i~odd} {{\sigma_i}\left( {{\sigma_i} + 1} \right)} - \frac{1}{2}\sum\limits_{i~even} {{\sigma_i}\left( {{\sigma_i} - 1} \right)}
 \end{aligned}
\end{equation}
Thus, we can recover the dimensions of the $BCD$ series nilpotent orbits in table \ref{tab:orbits1} from quivers with alternating $O/USp$ nodes, in a similar manner to the $A$ series. We can, therefore, use the canonical partition data from a $BCD$ series nilpotent orbit to identify a linear $BCD$ quiver, whose moduli space should have a Hilbert series with the same dimension as the nilpotent orbit.

We construct these moduli spaces in the following sections and examine their structures in terms of their Hilbert series and the representations of $G$ which they contain. We analyse representations both in terms of characters of $G$ and also in terms of the modified Hall-Littlewood polynomials of $G$, which provide a useful basis for their finite decomposition.

Clearly the set of well-ordered partitions does not exhaust the set of all possible quivers and so it is also interesting to ask whether there are dualities, such that different $A$ or $BCD$ quivers share the same moduli space. The dimension formulae in table \ref{tab:orbits1} do not depend upon the strict ordering of the partition data, so dualities can indeed arise.

Consider the general case of a link $USp-O-USp$ between two symplectic nodes in a quiver described by the partition data $(\ldots,\sigma_i, \sigma_{i+1},\ldots)$. It follows directly from the dimension formulae \ref{eq:orbits11} and \ref{eq:orbits12} that the mapping $(\ldots,\sigma_i, \sigma_{i+1},\ldots) \to (\ldots,\sigma_{i+1}-1, \sigma_{i}+1,\ldots)$ preserves the dimension of the Hilbert series as calculated from the partition data, while switching between $USp-O(even)-USp$ and $USp-O(odd)-USp$ or vice versa. The issue is that the resulting partition data is not well ordered and so detailed calculations are necessary to verify whether or not the Hilbert series of the Higgs branches of the two related quivers are the same.

\section{Quivers for $A$ Series Nilpotent Orbits }
\label{sec:Aseries}
\subsection{Minimal and Maximal Higgs Branch: $A$ Series}
For the reduced single $A_n$ instanton moduli space, or minimal nilpotent orbit of $A_n$, the Higgs branch construction is given by a bifundamental hypermultiplet containing a pair of chiral multiplets, with one transforming in the fundamental of an $SU(N_f) \otimes U(1)$ product group and the other transforming in the corresponding antifundamental. Applying the formula \ref{eq:orbits7}, we obtain, upon evaluation of the contour integral:
\begin{equation} 
\label{eq:Aseries1}
\begin{aligned}
g^{[N_f]-(1)}_{Higgs}\left( {{{\cal X}},t} \right)& = \oint \limits_{U(1)} {dq/q }~PE\left[ {\left[ {1,0, \ldots ,0} \right]q + \left[ {0,0, \ldots ,1} \right]/q,t} \right]/PE\left[ {1,{t^2}} \right]\\
& = g^{SU{(N_f)}}_{instanton}\left( {{{\cal X}},t} \right),\\
 \end{aligned}
\end{equation}
where we have taken $q$ as our $U(1)$ CSA coordinate and $t$ as our fugacity, corresponding to highest weights under $SU(2)_R$. The construction follows from the formation of the adjoint representation of $A_n$ as the contraction of its fundamental and anti-fundamental representations, giving a $U(1)$ singlet under the product group. The Weyl integral selects such U(1) singlets from amongst the combinations of the bifundamental fields that have been symmetrised by the PE, and the HyperK\"ahler quotient eliminates unwanted $A_n$ singlets. In all cases, we obtain the RSIMS for $A_n$, whose Hilbert series dimension is $2n$, i.e. twice the sum of the dual Coxeter numbers of the nodes of $A_n$. \cite{Benvenuti:2010pq} \footnote{The dual Coxeter number of a group is equal to one plus the sum of the dual Coxeter numbers of the nodes in its Dynkin diagram}

Carrying out this calculation for $[2]-(1)$, we obtain the RSIMS of $A_1$. Since the minimal and maximal nilpotent orbits coincide, this also equals the modified Hall Littlewood polynomial \ref{eq:Aseries3} transforming in the singlet representation $mHL^{A}_{[0]}$.

\begin{figure}[htbp]
\begin{center}
\includegraphics[scale=0.57]{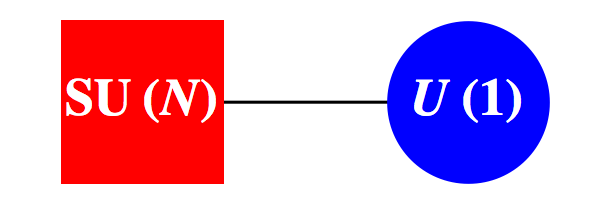}\\
\text{}\\
\includegraphics[scale=0.52]{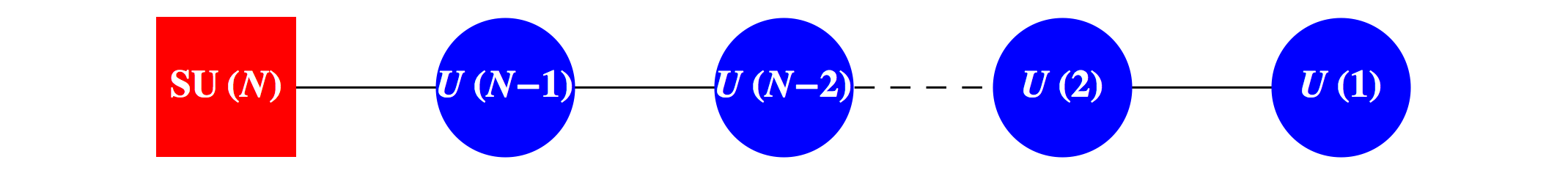}\\
\caption[Quivers for $A$ Series Minimal and Maximal Nilpotent Orbits]{Quivers for $A$ Series Minimal and Maximal Nilpotent Orbits. Square (red) nodes denote flavour nodes. Round (blue) nodes denote gauge nodes. The links represent pairs of bifundamental chiral scalars transforming in the fundamental and antifundamental representations. The minimal nilpotent orbit, with two nodes, corresponds to the reduced single instanton moduli space of $SU(N)$. The maximal nilpotent orbit, with $N$ nodes, corresponds to the modified Hall Littlewood polynomial $mHL^{A}_{[0,\ldots, 0]}$ of $SU(N)$.}
\label{fig:Aminmax}
\end{center}
\end{figure}

If we repeat a similar Higgs branch calculation for $A_2$, decomposing its $U(2)$ subgroup as $A_1 \otimes U(1)$, while also including the $mHL^{A}_{[0]}$ polynomial in the integrand, we obtain the linear quiver $[3]-(2)-(1)$. This moduli space evaluates as the maximal nilpotent orbit of $A_2$:
\begin{equation} 
\label{eq:Aseries2}
\begin{aligned}
g^{[3]-(2)-(1)}_{Higgs}( {{{\cal X}},t} ) &= \oint \limits_{ A_{1} \otimes U(1)} {d\mu } ~PE[ {[ {1,0} ]q + [ {0,1} ]/q,t} ]/PE[ {[ {1,1} ] + 1,{t^2}} ]~mHL^{A}_{[0]}(t^2)\\
&= mHL^{A}_{[0,0]}(t^2), \\
 \end{aligned}
\end{equation}
This is the first in a chain of recursive relations that can be used to generate $mHL^{A}_{[0,\ldots, 0]}$ for any $A$ series group from a linear quiver consisting of a chain of unitary nodes. Thus, we can construct the moduli spaces of the minimal and maximal nilpotent orbits of any $A$ series group from the quivers shown in figure \ref{fig:Aminmax}.
Consistent with section \ref{sec:orbits}, these have dimensions corresponding to those of the minimal and maximal nilpotent orbits of $A_n$, described by the partitions $(2,1^{n-1})$ and $(n+1)$, respectively.

\subsection{General Higgs Branch: $A$ Series}
\label{sec:AseriesGeneral}
The quivers in figure \ref{fig:Aminmax} lie at the extremal points of the set of quivers that can be defined by an increasing linear sequence of unitary node dimensions, as in figure \ref{fig:Aquiver}. Similarly, the minimal and maximal nilpotent orbits define extremal points on the set of non-trivial nilpotent orbits. As discussed in the previous section, the partition data associated with a nilpotent orbit defines a sequence of dimensions, whose separation is non-increasing, such that $N_i - N_{i+1} \ge N_{i+1} - N_{i+2}$. However, these quivers represent only a subset of those within the more general schema in figure \ref{fig:Aquiver}, so we include the full set in our analysis in order to examine dualities between quiver theories.

We can analyse the moduli space defined by the Higgs branch of a quiver theory in a number of different ways. Once we have calculated a generating function $g^{G}_{Higgs}\left( {{{\cal X}(x)},t} \right)$ for its refined Hilbert series, we can restate this in a number of different ways:

\begin{enumerate}

\item As an unrefined Hilbert series ${g_{Higgs}^G}\left({\cal X}(1),t \right)$ in terms of the fugacity $t$, by setting all the CSA coordinates in the refined Hilbert series to unity. An unrefined Hilbert series permits the counting of dimensions, generators and relations. When the generating function for an unrefined Hilbert series has a palindromic numerator, this indicates a correspondence with a Calabi-Yau surface \cite{Hanany:2008sb}.

\item As a character expansion in irreps of $G$. These infinite series can be described by an HWG ${g_{Higgs}^G} (m,t)$ for the coefficients of each irrep, identified by its Dynkin labels.

\item As an expansion in terms of modified Hall Littlewood polynomials of $G$. These series can be described by an HWG ${g_{Higgs}^G} (h,t)$ for the coefficients of each modified Hall Littlewood polynomial, identified by its Dynkin labels.

\end{enumerate}

Both characters and modified Hall Littlewood polynomials provide complete basis sets of orthogonal functions that can be used to decompose the class functions represented by refined Hilbert series. We find that, for low rank groups, the moduli space defined by a quiver often has a simple description in terms of one, but not always both, of these bases, which thus provide complementary modes of analysis.

Applying the general prescription in \ref{eq:orbits7}, we obtain the formula for the refined Hilbert series of an $A$ type quiver:
\begin{equation} 
\label{eq:Aseries4}
\begin{aligned}
& g_{Higgs}^{[N_f] - (N_1) - \ldots (N_{max})}\left( {{\cal X}_{SU(N_f)},t} \right)\\
&= \oint\limits_{U\left( {{N_1}} \right) \otimes \ldots U\left( {{N_{max }}} \right)} {d\mu }~\prod\limits_{n = 1}^{max } {\frac{{PE\left[ {{{\left[ {fund} \right]}_{U\left( {{N_{n - 1}}} \right)}} \otimes {{\left[ {anti} \right]}_{U\left( {{N_n}} \right)}} + {{\left[ {anti} \right]}_{U\left( {{N_{n - 1}}} \right)}} \otimes {{\left[ {fund} \right]}_{U\left( {{N_n}} \right)}},t} \right]}}{{PE\left[ {{{\left[ {adjoint} \right]}_{U\left( {{N_n}} \right)}},{t^2}} \right]}}},
\end{aligned}
\end{equation}
where we have defined $U(N_0) \equiv U(N_f)$, and note that gauge invariance entails that the resulting quiver is one for $SU(N_f) \equiv A_{N_f-1}$.

The Hilbert series ${g^G_{Higgs}}\left({\cal X},t \right)$ for such a linear quiver can be decomposed in the form:
\begin{equation} 
\label{eq:Aseries5}
\begin{aligned}
g^{G}_{Higgs}\left( {{{\cal X}},t} \right) \equiv P^{G}_{Higgs}\left( {{{\cal X}},t} \right) ~ PE\left[ {{{\left[ {adjoint} \right]}_G} - rank(G),t^2} \right],
\end{aligned}
\end{equation}
where $P^{G}_{Higgs}\left({\cal X},t \right)$ is the character expansion of some finite polynomial class function. We often find it helpful to express the Hilbert series ${g^G_{Higgs}}\left({\cal X}(x_i),t \right)$ in unrefined form as ${g^G_{Higgs}}\left(Dim({\cal X}),t \right)$ by mapping the CSA coordinates $x_i$ to unity. This facilitates the calculation of the dimension of the moduli space and the identification of its structural features, such as palindromy.

We can also analyse the representation structure of the moduli space in terms of characters or in terms of Hall Littlewood polynomials. The HWG $g^G_{Higgs}\left({m},t \right)$ for the full character expansion of the Hilbert series is obtained from the projection of $g^G_{Higgs}({\cal X} (x),t)$ onto a generating function $g^G_{\cal X} \left( {m},x* \right)$ for characters of $G$, parameterised by Dynkin label fugacities $m \equiv (m_1,\ldots, m_{rank(G)})$, as described in \cite{Hanany:2014dia}:
\begin{equation} 
\label{eq:Aseries6}
\begin{aligned}
g^G_{Higgs}\left({m},t \right) = \oint\limits_G {d\mu}~ g^G_{\cal X}\left( {{m},x*} \right) g^G_{Higgs}\left({\cal X}\left(x \right),t \right).
\end{aligned}
\end{equation}
The HWG $g^G_{Higgs}\left({h},t \right)$ for the expansion of the Hilbert series in terms of modified Hall Littlewood polynomials is obtained in a comparable manner from the projection of $g^G_{Higgs}({\cal X} (x),t)$ onto a generating function $\overline{g^G_{ {mHL} }} \left(h,x,t^2 \right)$ for the orthonormal modified Hall Littlewood polynomials of $G$, parameterised by Dynkin label fugacities $h \equiv (h_1,\ldots, h_{rank(G)})$, as described in appendix \ref{apxHLP}:
\begin{equation} 
\label{eq:Aseries7}
\begin{aligned}
g^G_{Higgs}\left({h},t \right) = \oint\limits_G {d{\mu _{HL}}}~ \overline{g^G_{mHL} }\left(h,x,t^2 \right)~ g^G_{Higgs}\left({\cal X}\left( {x} \right),t \right).
\end{aligned}
\end{equation}
We set out in tables \ref{tab:Aseries1} and \ref{tab:Aseries2}, the results of calculations of \ref{eq:Aseries4} through \ref{eq:Aseries7} for $A_{1}$, $A_{2}$, $A_{3}$ and $A_{4}$, for all the possible quivers associated with descending sequences of unitary gauge nodes as per figure \ref{fig:Aquiver}. 
\begin{sidewaystable}[htp]
\caption{Quivers for Nilpotent Orbits of $A_1$, $A_2$ and $A_3$}
\begin{center}
\footnotesize
\begin{tabular}{|c|c|c|c|c|c|}
\hline
$\text{Orbit} $&$\text{Quiver}$&$\text{Dim.}$&$\text{Hilbert Series}$&$ \text{Character HWG}$&$\text{mHL HWG}$\\
\hline
$ (1^2)$&$[2]$&$0$&$1$&$1$&$1 - {h^2}{t^2}$\\
\hline
$ (2) $&$\text{[2]-(1)} $&$ 2 $&$ {\frac{{1 - {t^4}}}{{{{\left( {1 - {t^2}} \right)}^3}}}}$&${\frac{1}{{1 - {m^2}{t^2}}}}$&$1$\\
\hline
\hline
$(1^3)$
&$ [3]  $
&$ 0	$
&$ 1 $
&$ 1 $
&$
\begin{array}{c}
 1 - {h_1}{h_2}{t^2} + {h_1}^3{t^4}\\
 - {h_1}{h_2}{t^4} + {h_2}^3{t^4} - {h_1}^2{h_2}^2{t^6}
 \end{array}$\\
\hline
$ ( 2,1)  $
&$
\begin{array}{c}
\text{[3]-(1)}\\
\text{[3]-(2)}
\end{array}
 $
&$ 4 	$
&${\frac{{1+ 4{t^2} +{t^4} }}{{{{\left( {1 - {t^2}} \right)}^4}}}}	$
&${\frac{1}{{1 - {{{m}}_{{1}}}{{m}}_{{2}}{t^2}}}}$
&${1 - {{{h}}_{{1}}}{{{h}}_{{2}}}{t^4}}$\\
\hline
$ (3)    $&$\text{[3]-(2)-(1)} 	$&$ 6 	$
&$ {\frac{{(1 - t^4)\left( {1 - {t^6}} \right)}}{{{{\left( {1 - {t^2}} \right)}^8}}}}	$
&$
 {\frac{{1 - {{{m}}_{{1}}}^3{{{m}}_{{2}}}^3{t^{12}}}}{{\left( {1 - {{{m}}_{{1}}}{{{m}}_{{2}}}{t^2}} \right)\left( {1 - {{{m}}_{{1}}}{{{m}}_{{2}}}{t^4}} \right)\left( {1 - {{{m}}_{{1}}}^3{t^6}} \right)\left( {1 - {{{m}}_{{2}}}^3{t^6}} \right)}}}$
&$1$\\
\hline
\hline
$ (1^4)    $
&$ [4]$
&$ 0	$
&$ 1	$
&$1 $
&$
\scriptsize{
\begin{array}{c}
1-h_1 h_3 t^2-h_1 h_3 t^4+h_1^2 h_2 t^4+h_2 h_3^2 t^4\\
-h_1 h_3 t^6-h_1 h_2^2 h_3 t^6+h_1^2 h_2 t^6+h_2 h_3^2 t^6\\
-h_1^2 h_3^2 t^6+h_2^2 t^6-h_3^4 t^6-h_1^4 t^6+h_1^3 h_2 h_3 t^8\\
-h_1 h_2^2 h_3 t^8-h_1^2 h_3^2 t^8+h_1 h_2 h_3^3 t^8+h_2^4 t^8\\
+h_1 h_2 h_3^3 t^{10}-h_2^3 h_3^2 t^{10}+h_1^3 h_2 h_3 t^{10}\\
-h_1^3 h_3^3 t^{10}-h_1^2 h_2^3 t^{10}+h_1^2 h_2^2 h_3^2 t^{12}\\
\end{array}
}
$\\
\hline
$(2,1^2)    $
&$ \text{[4]-(1)}$
&$ 6	$
&$ {\frac{{\left( {1 + {t^2}} \right)\left( {1 + 8{t^2}+{t^4} } \right)}}{{{{\left( {1 - {t^2}} \right)}^6}}}}	$
&${\frac{1}{{1 - {{{m}}_{{1}}}{{{m}}_{{3}}}{t^2}}}} $
&$
\begin{array}{c}
 1 - {h_1}{h_3} {t^4} - {h_2}^2 {t^4}\\
 - {h_1} {h_3} {t^6}+ {h_1}^2 {h_2} {t^6}+ {h_2}{h_3}^2 {t^6}\\
 + {h_2}^2 {t^6}- {h_1}^2 {h_3}^2 {t^8}
\end{array}$\\
\hline
$(2^2)    $
&$\text{[4]-(2)}$
&$ 8	$
&$ \frac{{{{\left( {1 + {t^2}} \right)}^2}\left( {1 + 5{t^2}+{t^4} } \right)}}{{{{\left( {1 - {t^2}} \right)}^{8}}}}$
&$ {\frac{1}{{\left( {1 - {{{m}}_{{1}}}{{{m}}_{{3}}}{t^2}} \right)\left( {1 - {{{m}}_{{2}}}^2{t^4}} \right)}}} $
&${1 - {{{h}}_{{1}}}{{{h}}_{{3}}}{t^4} - {{{h}}_{{1}}}{{{h}}_{{3}}}{t^6} + {{{h}}_{{2}}}^2{t^6}}$\\
\hline
$(3,1)    $
&$
\begin{array}{c}
\text{[4]-(2)-(1)}\\
\text{[4]-(3)-(1)}\\
\text{[4]-(3)-(2)}
\end{array}
$
&$ 10	$
&${\frac{{\left( {1 + {t^2}} \right)\left( {1 + 4{t^2} + 10{t^4} + 4{t^6} + {t^8}} \right)}}{{{{\left( {1 - {t^2}} \right)}^{10}}}}}$
&
$
\scriptsize{
\frac{{1 - {m_1}^2{m_2}^2{m_3}^2{t^{12}}}}{{\left( \begin{array}{c}
\left( {1 - {m_1}{m_3}{t^2}} \right)\left( {1 - {m_2}^2{t^4}} \right)\left( {1 - {m_1}{m_3}{t^4}} \right)\\
 \times \left( {1 - {m_1}^2{m_2}{t^6}} \right)\left( {1 - {m_2}{m_3}^2{t^6}} \right)
\end{array} \right)}}
}
$
&${1 - {{{h}}_{{1}}}{{{h}}_{{3}}}{t^6}}$\\
\hline
$(4)    $
&$
\text{[4]-(3)-(2)-(1)}
$
&$ 12	$
&$ \frac {(1 - {t^4})(1 - {t^6})(1 - {t^8})}{(1 - t^2)^{15}}	$
&$
\scriptsize{
\frac{{1 + \ldots 56~terms \ldots+ m_1^6m_2^4m_3^6{t^{42}}}}{{\left( \begin{array}{c}
\left( {1 - {m_1}{m_3}{t^2}} \right)\left( {1 - m_2^2{t^4}} \right)\left( {1 - {m_1}{m_3}{t^4}} \right)\\
 \times \left( {1 - m_1^2{m_2}{t^6}} \right)\left( {1 - {m_2}m_3^2{t^6}} \right)\\
 \times \left( {1 - m_2^2{t^8}} \right)\left( {1 - m_1^4{t^{12}}} \right)\left( {1 - m_3^4{t^{12}}} \right)
\end{array} \right)}}
}
$
&$1$\\
\hline
\end{tabular}
\end{center}
\text{[N] denotes an SU(N) flavour node. (N) denotes a U(N) gauge node.}\\
\text{An mHL HWG of 1 denotes $mHL_{[0, \ldots, 0]}(t^2)$.}\\
\text{Where multiple quivers have the same Hilbert series, the canonical partition is given first.}\\
\text{The numerator of the character HWG for the [4]-(3)-(2)-(1) quiver has been truncated.}
\label{tab:Aseries1}
\end{sidewaystable}
\begin{sidewaystable}[htp]
\caption{Quivers for Nilpotent Orbits of $A_4$}
\begin{center}
\footnotesize
\begin{tabular}{|c|c|c|c|c|c|}
\hline
$\text{Orbit} $&$\text{Quiver}$&$\text{Dim}$&$\text{Hilbert Series}$&$ \text{Character HWG}$&$\text{mHL HWG}$\\
\hline 
$(1^5) $&$[5]$&$ 0 $&$ 1$&$ 1 $&$ \ldots $\\
\hline
$(2,1^3) $&$\text{ [5]-(1)} $&$ 8 $&$
\frac{1+16 t^2+36 t^4+16 t^6+t^8}{\left(1-t^2\right)^8} 
$&$
\frac{1}{1-{m_1} {m_4} t^2}
$&$ \ldots
$ \\
\hline
$(2^2,1) $&$\begin{array}{c}\text{[5]-(2)}\\\text{[5]-(3)}\end{array} $&$ 12 $&$
 \frac{1+12 t^2+53 t^4+88 t^6+53 t^8+12 t^{10}+t^{12}}{\left(1-t^2\right)^{12}} 
 $&$
 \frac{1}{\left(1-{m_1} {m_4} t^2\right) \left(1-{m_2} {m_3} t^4\right)} 
 $&$ \scriptsize{
 \begin{array}{c}
 1-h_1 h_4 t^4-h_1 h_4 t^6 -h_1 h_4 t^8\\
+h_2 h_3 t^8+h_1^2 h_3 t^8 +h_2 h_4^2 t^8\\
 +h_2 h_3 t^{10}-h_3^2 h_4 t^{10}\\
-h_1^2 h_4^2 t^{10}-h_1 h_2^2 t^{10}+h_1 h_2 h_3 h_4 t^{14}
 \end{array} }$ \\
\hline
$(3,1^2)$&$\text{[5]-(2)-(1)}$&$ 14 $&$
\frac{1+9 t^2+45 t^4+65 t^6+45 t^8+9 t^{10}+t^{12}}{\left(1-t^2\right)^{15} \left(1-t^4\right)^{-1}}
$&$
\scriptsize{
 \frac{1-{m^2_1} {m_2} {m_3} {m^2_4} t^{12}}{\left( \begin{array}{c} (1-{m_1}{m_4}t^2) (1-{m_1}{m_4}t^4)(1-{m_2}{m_3}t^4)\\ \times (1-{m_1}^2{m_3}t^6)(1-{m_2}{m_4}^2 t^6)\end{array} \right)}
 }
 $&$\scriptsize{
 \begin{array}{c}
1-h_1 h_4 t^6 -h_2 h_3 t^6\\
-h_1 h_4 t^8+h_1^2 h_3 t^8+h_2 h_4^2 t^8\\
+h_2 h_3 t^{10} -h_1^2 h_4^2 t^{10}\\
 \end{array} }$\\
\hline
$(3, 2)$&$\begin{array}{c}\text{[5]-(3)-(1)}\\\text{[5]-(3)-(2)}\\\text{[5]-(4)-(2)}\end{array}$&$ 16 $&$
 \frac{1+6 t^2+22 t^4+37 t^6+22 t^8+6 t^{10}+t^{12}}{\left(1-t^2\right)^{18} \left(1-t^4\right)^{-2}}
 $&$ \ldots $&$\scriptsize{
 \begin{array}{c}
 1-{h_1} {h_4} t^6\\
 -{h_1} {h_4} t^8+{h_2} {h_3} t^{10}
 \end{array} }$\\
\hline
$(4, 1) $&$\begin{array}{c}\text{[5]-(3)-(2)-(1)}\\\text{[5]-(4)-(2)-(1)}\\\text{[5]-(4)-(3)-(1)}\\\text {[5]-(4)-(3)-(2)}\end{array}$&$ 18 $&$
\frac{1+4 t^2+10 t^4+20 t^6+10 t^8+4 t^{10}+ t^{12}}{\left(1-t^2\right)^{20} \left(1-t^4\right)^{-1} \left(1-t^6\right)^{-1}}
$&$ \dots $&$\scriptsize{1-{h_1} {h_4} t^8}$\\
\hline
$(5) $&$\text{[5]-(4)-(3)-(2)-(1)}$&$ 20 $&$
\frac{\left(1-t^4\right) \left(1-t^6\right) \left(1-t^8\right) \left(1-t^{10}\right)}{\left(1-t^2\right)^{24}}
$&$ \ldots $&$ 1 $\\
\hline
\end{tabular}
\end{center}
\text{[N] denotes an SU(N) flavour node. (N) denotes a U(N) gauge node.}\\
\text{An mHL HWG of 1 denotes $mHL_{[0, \ldots, 0]}(t^2)$.}\\
\text{Where multiple quivers have the same Hilbert series, the canonical partition is given first.}\\
\text{The character or mHL HWGs for some quivers have been omitted.}
\label{tab:Aseries2}
\end{sidewaystable}
%
%
There are many observations that can be made:

\begin{enumerate}

\item All moduli spaces of Higgs branch quiver theories constructed using nilpotent orbit partition data have dimensions equal to those of the nilpotent orbits.

\item The Hilbert series of these moduli spaces are all palindromic, indicating Calabi Yau surfaces, and consistent with the property of being HyperK\"ahler. The maximal nilpotent orbits all have Hilbert series that are complete intersections \cite{Hanany:2011db}.

\item The moduli space decompositions into characters identify their generators, such as the $A_1$ generator $m^2 t^2$ or the $A_2$ generator $m_{1} m_{2} t^2$.
Each generator (or monomial) is a root lattice object, having {\it N-ality} zero, i.e. the property that the sum of the indices of the $m_i$ fugacities raised to their exponents, modulo the fundamental dimension $N$ of the flavour group, is zero. 

\item All the moduli spaces can also be decomposed into finite sums of modified Hall Littlewood polynomials. Again, each monomial is a root lattice object having {\it N-ality} zero with respect to the $h_i$ fugacities.

\item The character and mHL descriptions are complementary; orbits close to the minimal nilpotent orbit have character HWGs that are freely generated or complete intersections; orbits close to the maximal nilpotent orbit decompose to a small number of mHL functions.

\item The characteristic $A$ series nilpotent orbits have distinct signatures in terms of Hilbert series, character HWGs and/or mHL HWGs and we summarise these in table \ref{tab:Aseries3} for future reference.

\item There are some interesting dualities, where multiple quivers correspond to the same nilpotent orbit. The circumstances under which these arise are discussed further below.

\item There are inclusion relations between the moduli spaces that can be read off most easily from the character HWGs. These follow exactly the canonical partial orderings of the nilpotent orbits according to their Hasse diagrams, as given in \cite{Collingwood:1993fk}, for example.
\end{enumerate}

\begin{sidewaystable}[htp]
\caption{Generalised $A$ Series Nilpotent Orbit Moduli Spaces}
\begin{center}
\begin{tabular}{|c|c|c|c|c|c|}
\hline
$\text{Orbit}$&$\text{Dimension}$&$\text{Quiver}$&$\text{Hilbert Series}$&$\text{PL[Character HWG]}$&$\text{mHL HWG}$\\
\hline
$\text{Trivial} $&$ 0 $&$ [n+1] $&$ 1$&$1$&$ \ldots $ \\
\hline
$\text{Minimal}$&${2n} $&$ [n+1]-(1) $&
$\frac{{\prod\limits_{i = 0}^n   {\left( \begin{array}{l}n\\ i \end{array} \right)^2}{t^{2i}}   }}{{{{\left( {1 - {t^2}} \right)}^{2n}}}} $&${ {{m_1}{m_n}{t^2}}}$
 &$ \ldots $\\
\hline
$
\begin{array}{c}
\text{Supra Minimal}\\
n \ge 3
\end{array}
$&${4n - 4}$&$[n+1]-(2)$&$ \ldots $&$ { {{m_1}{m_n}{t^2} + {m_2}{m_{n - 1}}{t^4}}}$&$ \ldots $\\
\hline
$
\begin{array}{c}
\text{Supra Supra}\\
\text{Minimal}\\
n \ge 3
\end{array}
$&${4n - 2}$&$[n+1]-(2)-(1) $&$ \ldots $&$
\footnotesize{
\begin{array}{c}
{m_1}{m_n}{t^2}\\
 + {m_1}{m_n}{t^4} + {m_2}{m_{n - 1}}{t^4} \\
+ m_1^2{m_{n - 1}}{t^6} + {m_2}m_n^2{t^6} \\
- m_1^2{m_2}{m_{n - 1}}m_n^2{t^{12}} 
\end{array}
}
$&$ \ldots $\\
\hline
\hline
$\begin{array}{c}
\text{2-Node Quiver}\\
n+1\ge 2k
\end{array}
$&$2k(n+1-k)$&$[n+1]-(k)$&$ \ldots $&$ \sum\limits_{i = 1}^k {{m_i}{m_{n + 1 - i}}{t^{2i}}} $&$\ldots $\\
\hline
\hline
$
\begin{array}{c}
\text{Sub Sub}\\
\text{Regular}\\
n \ge 3
\end{array}
$&${n\left( {n + 1} \right) - 4}$&$
\begin{array}{c}
[n+1]-(n-1)-\\
(n-3)-\ldots(1)
\end{array}
$&$ \ldots $&$ \ldots $&$
\footnotesize{
\begin{array}{c}
1 + {h_2}{h_{n - 1}}{t^{4n - 6}}\\
- {h_1}{h_n}{t^{2n - 2}} - {h_1}{h_n}{t^{2n}}
 \end{array}
 }
$\\
\hline
$\text{Sub Regular}$&${n\left( {n + 1} \right) - 2}$&$
\begin{array}{c}
[n+1]-(n-1)-\\
(n-2)-\ldots(1)
\end{array}
$&$ \ldots $&$ \ldots $&${1 - {h_1}{h_n}{t^{2n}}}$\\
\hline
$\text{Maximal}$&${n\left( {n + 1} \right)}$&$[n+1]-\ldots(1) $&$
{\frac{{\prod\limits_{i = 1}^n {\left( {1 - {t^{2(i + 1)}}} \right)} }}{{\left( {1 - {t^2}} \right)}^{n\left( {n + 2} \right)}}}
$&$ \ldots $&$1$\\
\hline
\end{tabular}
\end{center}
\text{Within a quiver, $\ldots$ denotes a maximal sub-chain}
\label{tab:Aseries3}
\end{sidewaystable}%

It is significant that there are a number of quivers, such as $[3]-(2)$, $[4]-(3)-(1)$ and $[4]-(3)-(2)$, that cannot be described by partition data, since their decrements are non-decreasing. These may nonetheless have the same Hilbert series as the {\it canonical} quivers calculated directly from the nilpotent orbit partition data. Any linear quiver that has decreasing dimensions can be dualised to one of the canonical quivers by reordering its dimensional increments. Noting that the Hilbert series dimensions set out in table \ref{tab:orbits1} are insensitive to the order of the increments $\sigma_i$, this dualisation often leaves moduli space dimensions invariant. Furthermore, we find, by full calculation, that in many cases the refined Hilbert series of these non-partition quivers match those of their canonical duals, including the aforementioned examples.

\FloatBarrier

There are, however, limits to the extent to which the $\sigma_i$ can be reordered to obtain a dual quiver with the same Hilbert series. For example, a calculation of the Hilbert series of the quiver $[4]-(3)$, using the procedure given, yields an incorrect non-palindromic result that does not match $[4]-(1)$. It is noteworthy that the related concepts of quiver {\it balance} and {\it conformal dimension} developed in \cite{Gaiotto:2008ak} can be used to identify when the moduli spaces of such dual quivers match those of the canonical quivers.

The {\it balance} of a $U(N)$ gauge node $i$ in a simply laced ADE Series quiver is defined as:
\begin{equation}
\label{eq:Aseries8}
Balance_{ADE}(i)= -2{N_i}+ \sum\limits_{j \in \left\{ {\scriptstyle adjacent \atop \scriptstyle nodes} \right\}}^{} {{N_j}}.
\end{equation}
If all gauge nodes in a quiver have a balance of zero, the quiver is termed {\it balanced}. If one or more gauge nodes have a positive balance and no gauge nodes have a negative balance, the quiver is described as {\it positively balanced}. If one or more gauge nodes have a shortage of at most one link, i.e. a balance of $-1$, the quiver is described as {\it minimally unbalanced}. If one or more gauge nodes has a shortage of two or more links, i.e. a balance of $-2$ or less, the quiver is described as {\it unbalanced}.

When calculating the Coulomb branch of an ADE Series quiver, the shift in conformal dimension associated with the first non-zero monopole charge on a gauge node is given by:
\begin{equation}
\label{eq:Aseries9}
\delta \left(Conformal~Dimension \right)(i) = \frac{1}{2}\sum\limits_{j \in \left\{ {\scriptstyle adjacent \atop \scriptstyle nodes} \right\}} {{N_j}} - \left( {{N_i} - 1} \right) = \frac{1}{2} Balance_{ADE}(i) + 1.
\end{equation}
If a quiver is balanced, a single monopole gauge charge has a conformal dimension of 1, and corresponds to an integer shift around the root lattice. When a quiver is minimally unbalanced, a single monopole gauge charge has a conformal dimension of $1/2$. When a quiver is unbalanced, the conformal dimension is zero or negative (which is meaningless). In \cite{Gaiotto:2008ak}, balanced or positively balanced quivers are termed ``good", minimally unbalanced quivers are termed ``ugly" and unbalanced quivers are termed ``bad".

We observe from inspection of tables \ref{tab:Aseries1} and \ref{tab:Aseries2} that:
\begin{enumerate}

\item The quivers specified by the partition data from a nilpotent orbit are either balanced or positively balanced.

\item The quivers that do not correspond to canonically ordered partitions are either minimally unbalanced or unbalanced.

\item Minimally unbalanced quivers have Hilbert series that match those of the nilpotent orbits given by a canonical reordering of their partition data.

\item Unbalanced quivers, if evaluated using \ref{eq:Aseries4}, have Hilbert series that are non-palindromic and do not match those of the nilpotent orbits given by a normal ordering of the quiver partition data. \footnote{As discussed in \cite{Cremonesi:2014uva} for the case of $[N_f]-(N_c)$ quivers, where $N_f = 2N_c-k$, the extra dimensions of the moduli space result from incomplete breaking of the gauge group, for values of $k>1$, when the theory becomes {\it unbalanced}. These extra dimensions and non-palindromic features of the moduli space can be eliminated by the introduction of FI terms.}
\end{enumerate}
This pattern of Higgs branch dualities between $A$ series quivers is consistent with findings in \cite{Yaakov:2013fza}.


\FloatBarrier

\subsection{Coulomb Branch and Mirror Symmetry: $A$ Series}

\begin{figure}[htbp]
\begin{center}
\includegraphics[scale=0.50]{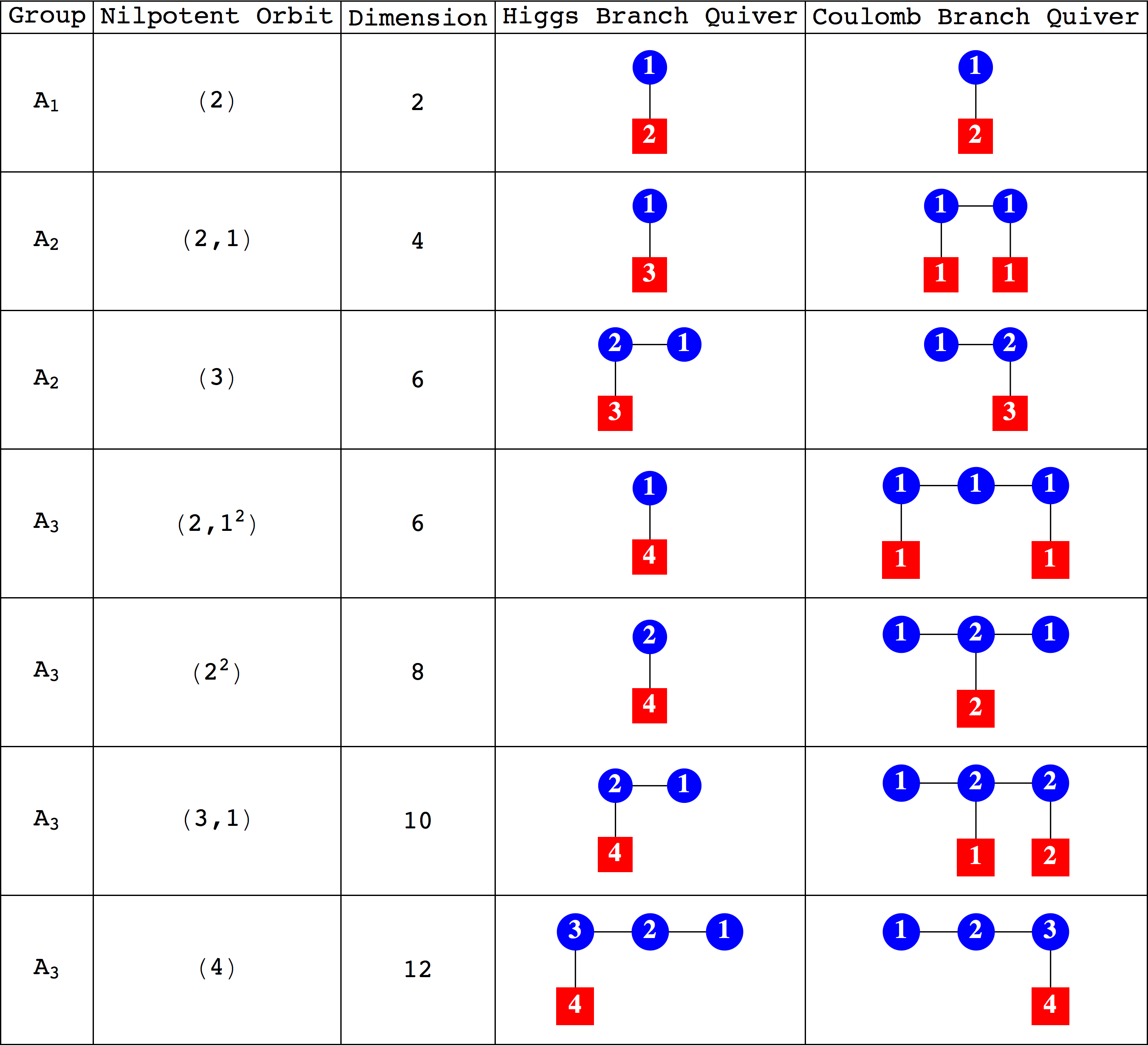}\\
\caption[Mirror Dual Quivers for $A$ Series Nilpotent Orbits]{Mirror Dual Quivers for $A$ Series Nilpotent Orbits. Round (blue) nodes denote unitary gauge nodes of the indicated rank. Square (red) nodes denote numbers of uncharged flavour nodes.}
\label{fig:Amirrors}
\end{center}
\end{figure}

It is well known that the Higgs branch constructions on the quivers described above have moduli spaces that are identical to Coulomb branch constructions on unitary quivers that are dual under $3d$ mirror symmetry and that the mirror symmetric dual quivers can be calculated by working with their brane constructions \cite{Hanany:1996ie, Hanany:2011db, Cremonesi:2014uva}. We display in figure \ref{fig:Amirrors} those quivers that yield the nilpotent orbits for $A_1$ through $A_3$ on their Higgs branches along with their mirror duals that define the same moduli spaces on on their Coulomb branches.

Some useful observations can be made about the structure of these Coulomb branch quivers for $A$ series nilpotent orbits.
\begin{enumerate}

\item The Coulomb branch quivers have a number of gauge nodes equal to the rank of the group.

\item The dimension of each nilpotent orbit is equal to twice the sum of the ranks of the gauge nodes of its Coulomb branch quiver.

\item The Coulomb branch quivers are all {\it balanced}, as defined earlier. Consequently, the labels of the flavour and gauge nodes are mapped into each other by the $A_n$ Cartan matrix $A^{ij}$. So, taking the flavour node dimensions as the n-vector $(f_1,\ldots,f_n)$ and the gauge node ranks as $(g_1,\ldots, g_n)$, we have ${f^i} = {A^{ij}}{g_j}$.

\item The Coulomb branch quivers of the minimal nilpotent orbits (or RSIMS) match the affine Dynkin diagrams of their groups, once their flavour nodes are identified.

\item The Coulomb branch quivers of the maximal nilpotent orbits are all self-mirror and match those of $T(SU(N))$ quiver theories.

\item Interestingly, the flavour and gauge node vectors for near-minimal nilpotent orbits match, respectively, the root and weight maps given in appendix  \ref{apxHom}.

\end{enumerate}
We defer further discussion of the monopole formula until section \ref{subsec:$BCD$ Mirrors}. 

\FloatBarrier
\section{Quivers for $BCD$ Series Nilpotent Orbits}
\label{sec:BCDseries}
We now turn to the more intricate matter of carrying through a comparable analysis for $BCD$ series groups. Orthogonal and symplectic matrices are complementary in terms of constructing the matrix generators of nilpotent orbits, which are all members of $GL(N,R)$, and the interplay between the two series is necessary to construct moduli spaces that match the dimensions of $B$, $C$ and $D$ series nilpotent orbits. As observed in \cite{Gaiotto:2008ak}, unitary and orthosymplectic quivers are related by a ${\cal{Z}_2}$ orbifold action and the orthogonal and symplectic groups in a quiver must alternate so that this action can be defined. This does not, however, entail that a quiver should not contain both $B$ and $D$ series groups and, accordingly, we work with quivers that can be of mixed $BCD$ type.

We also encounter a number of complications relating to the necessity, in several cases, of using character representations of $O(N)$ gauge groups \cite{kobak1996classical}, rather than $SO(N)$, to obtain palindromic moduli spaces, and also to the calculation of HyperK\"ahler quotients for $O(N)$ groups. These complications are least severe for minimal and maximal nilpotent orbits and so these are a good place to start.

\subsection{Minimal and Maximal Higgs Branch: $BCD$ Series}
\label{subsec:BCDminmax}

\begin{figure}[htbp]
\begin{center}
\includegraphics[scale=0.50]{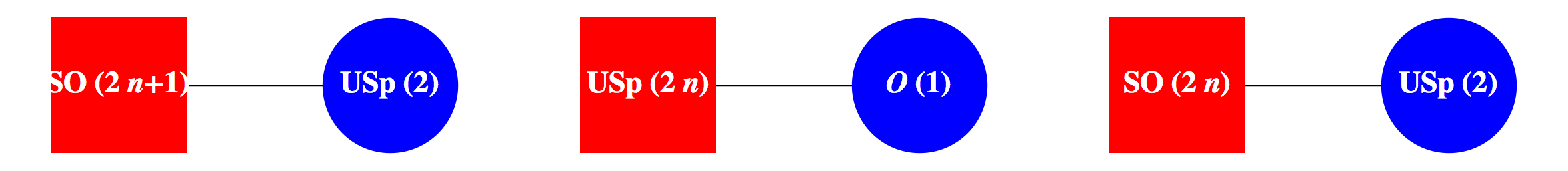}\\
\caption[Quivers for $BCD$ Series Minimal Nilpotent Orbits]{Quivers for $BCD$ Series Minimal Nilpotent Orbits. Square (red) nodes denote flavour nodes. Round (blue) nodes denote gauge nodes. The links represent bifundamental half-hypermultiplets with scalar fields transforming in the vector/fundamental representations.}
\label{fig:minBCDgrid}
\end{center}
\end{figure}

As noted earlier, minimal nilpotent orbits for $BCD$ series are well known and correspond to RSIMS and their Higgs branch constructions. For $BCD$ series groups, these quivers consist of a bifundamental half-hypermultiplet containing a scalar transforming in the vector $\otimes$ fundamental of an $O \otimes USp$ product group. In all cases, the vector/fundamental of the $O/USp$ flavour group is coupled with a fundamental/vector of a minimal rank $USp/O$ gauge group. The quivers are shown in figure \ref{fig:minBCDgrid}.

The precise evaluations are provided by adapting formula \ref{eq:orbits7}:
\begin{equation} 
\label{eq:BCDseries1}
\begin{aligned}
g_{Higgs~min}^{{B_n}}\left( {{{{\cal X}}_B},t} \right) & = \oint\limits_{{C_1}} {d\mu}~PE\left[ {{{\left[ {1,0, \ldots ,0} \right]}_B} \otimes {{\left[ 1 \right]}_C},t} \right]/PE\left[ {{{\left[ 2 \right]}_C},{t^2}} \right],\\
g_{Higgs~min}^{{D_n}}\left( {{{{\cal X}}_D},t} \right) & = \oint\limits_{{C_1}} {d\mu }~PE\left[ {{{\left[ {1,0, \ldots ,0} \right]}_D} \otimes {{\left[ 1 \right]}_C},t} \right]/PE\left[ {{{\left[ 2 \right]}_C},{t^2}} \right],\\
g_{Higgs~min}^{{C_n}}\left( {{{{\cal X}}_C},t} \right) &= \frac{1}{2}\left( {PE\left[ {{{\left[ {1,0, \ldots ,0} \right]}_C},t} \right] + PE\left[ {{{\left[ {1,0, \ldots ,0} \right]}_C}, - t} \right]} \right).\\
 \end{aligned}
\end{equation}
Minimal nilpotent orbits for $SO$ flavour groups are based on Weyl integrations over a $C_1$ gauge group, whereas those for symplectic flavour groups are based on Molien sums over a $B_0 \cong O(1)$ gauge group. $B_0$ is a finite group with two elements that can be represented by the characters $\{ 1,-1 \}$, so the group average is provided by a Molien sum, rather than by Weyl integration \cite{Gray:2008yu}. The HyperK\"ahler quotient in the integrations is given by the adjoint of the gauge group, with counting fugacity $t^2$.

\begin{figure}[htbp]
\begin{center}
\includegraphics[scale=0.50]{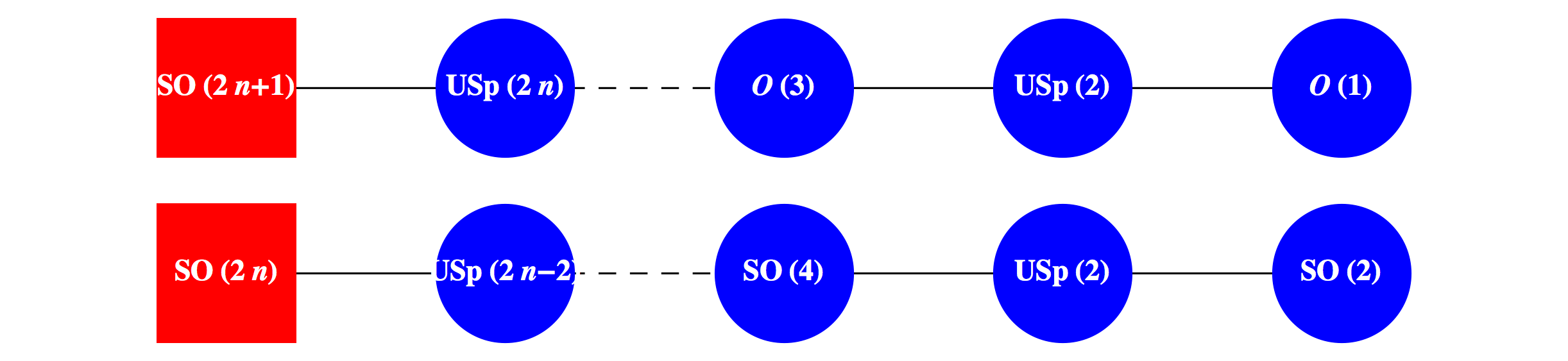}\\
\caption[Quivers for $BCD$ Series Maximal Nilpotent Orbits]{Quivers for $BCD$ Series Maximal Nilpotent Orbits. Square (red) nodes denote flavour nodes. Round (blue) nodes denote gauge nodes. The links represent bifundamental fields transforming in the vector/fundamental representations. A maximal chain for a symplectic group can be obtained by truncating either the $BC$ or $DC$ chain and taking the highest rank symplectic group as the new flavour group.}
\label{fig:maxBCDgrid}
\end{center}
\end{figure}

As for $A_1$, a minimal nilpotent orbit for $B_1$ or $C_1$ is also maximal. Otherwise, the maximal nilpotent orbits for $BCD$ groups are provided by chains of $O/USp$ groups with adjacent dimensions, as shown in figure \ref{fig:maxBCDgrid}.
In the case of the $BC$ chain, the fundamental dimension decreases by one between adjacent nodes, whereas in a $DC$ chain the fundamental dimension decreases by alternating steps of zero or two; it is important to note the ordering, with the $C$ series, which has higher group dimension, taking precedence. These two types of maximal chain: $BC$ and $DC$, represent special cases, since we can substitute between $C_n D_n C_{n-1}$ and $C_n B_{n-1} C_{n-1}$ links in a maximal chain without affecting the moduli space.

Once again, the precise evaluations are provided by adapting formula \ref{eq:orbits7}:

\begin{equation} 
\label{eq:BCDseries2}
\begin{aligned}
g_{Higgs~max}^{{B_n}}\left( {{{{\cal X}}_{B_n}},t} \right) &= \oint\limits_{{C_n} \otimes {B_{n - 1}} \ldots \otimes {C_1}} {d\mu }~\prod\limits_{i = 1}^n {\frac{{PE\left[ {{{\left[ {vec} \right]}_{{B_i}}} \otimes {{\left[ {fund} \right]}_{{C_i}}},t} \right]}}{{PE\left[ {{{\left[ {adjoint} \right]}_{{C_i}}},{t^2}} \right]}}} \\
 &~~~~~~~~~~~~~~~\times \prod\limits_{i = 1}^n \frac{1}{2}\sum\limits_{t = \left\{ {t, - t} \right\}}^{} {\frac{{PE\left[ {{{\left[ {fund} \right]}_{{C_i}}} \otimes {{\left[ {vec} \right]}_{{B_{i - 1}}}},t} \right]}}{{PE\left[ {{{\left[ {adjoint} \right]}_{{B_{i - 1}}}},{t^2}} \right]}}} \\
 \end{aligned}
\end{equation}

\begin{equation} 
\label{eq:BCDseries3}
\begin{aligned}
g_{Higgs~max}^{{C_n}}\left( {{{\cal X}_{{C_n}}},t} \right){\rm{ }} &=\oint\limits_{{B_{n - 1}} \otimes {C_{n - 1}} \ldots \otimes {C_1}} {d\mu } ~\prod\limits_{i = 1}^n {\frac{1}{2}\sum\limits_{t = \left\{ {t, - t} \right\}}^{} {\frac{{PE\left[ {{{\left[ {fund} \right]}_{{C_i}}} \otimes {{\left[ {vec} \right]}_{{B_{i - 1}}}},t} \right]}}{{PE\left[ {{{\left[ {adjoint} \right]}_{{B_{i - 1}}}},{t^2}} \right]}}} }\\
&~~~~~~~~~~~~~~~ \times \prod\limits_{i = 1}^{n - 1} \frac{{PE\left[ {{{\left[ {vec} \right]}_{{B_i}}} \otimes {{\left[ {fund} \right]}_{{C_i}}},t} \right]}}{{PE\left[ {{{\left[ {adjoint} \right]}_{{C_i}}},{t^2}} \right]}}
\end{aligned}
\end{equation}

\begin{equation} 
\label{eq:BCDseries4}
\begin{aligned}
g_{Higgs~max}^{{C_n}}\left( {{{{\cal X}}_{{C_n}}},t} \right) & = \oint\limits_{{D_n} \otimes {C_{n-1}} \ldots \otimes {D_1}} {d\mu }~ \prod\limits_{i = 1}^n \frac{{PE\left[ {{{\left[ {fund} \right]}_{{C_i}}} \otimes {{\left[ {vec} \right]}_{{D_i}}},t} \right]}}{{PE\left[ {{{\left[ {adjoint} \right]}_{{D_i}}},{t^2}} \right]}}\\
&~~~~~~~~~~~~~~~ \times \prod\limits_{i = 1}^{n - 1} \frac{{PE\left[ {{{\left[ {vec} \right]}_{{D_{i + 1}}}} \otimes {{\left[ {fund} \right]}_{{C_i}}},t} \right]}}{{PE\left[ {{{\left[ {adjoint} \right]}_{{C_i}}},{t^2}} \right]}}\\
 \end{aligned}
\end{equation}

\begin{equation} 
\label{eq:BCDseries5}
\begin{aligned}
g_{Higgs~max}^{{D_n}}\left( {{{{\cal X}}_{{D_n}}},t} \right)& = \oint\limits_{ {C_{n - 1}} \otimes {D_{n - 1}} \ldots \otimes {D_1}} {d\mu }~\prod\limits_{i = 1}^{n - 1} \frac{{PE\left[ {{{\left[ {vec} \right]}_{{D_{i + 1}}}} \otimes {{\left[ {fund} \right]}_{{C_i}}},t} \right]}}{{PE\left[ {{{\left[ {adjoint} \right]}_{{C_i}}},{t^2}} \right]}}\\
 &~~~~~~~~~~~~~~~\times \prod\limits_{i = 1}^{n - 1} \frac{{PE\left[ {{{\left[ {fund} \right]}_{{C_i}}} \otimes {{\left[ {vec} \right]}_{{D_i}}},t} \right]}}{{PE\left[ {{{\left[ {adjoint} \right]}_{{D_i}}},{t^2}} \right]}}\\
 \end{aligned}
\end{equation}
The maximal nilpotent orbit for a symplectic group can be constructed from a $BC$ chain or a $DC$ chain, or a combination.

In order for the moduli spaces to be HyperK\"ahler, the gauge groups must be connected \cite{kobak1996classical}, which in turn entails that a quiver should contain orthogonal $O$, rather than $SO$, gauge groups. This requirement is met by means of a Molien sum accompanying each Weyl integration. The Molien sum performs a group average over the ${\mathbb Z}_2$ factor corresponding to the sign of the determinant of the orthogonal group representation matrix. For $B$ series gauge groups the ${\mathbb Z}_2$ factor is $-1$ times the identity matrix, which commutes with the representation matrices and has no effect on the structure of the characters. Algebraically, the ${\mathbb Z}_2$ factor for the $B$ series is introduced by changing the sign of the fugacity $t$ within the PE function. For $D$ series gauge groups, {\it in the case of maximal nilpotent orbits}, the introduction of a ${\mathbb Z}_2$ factor has no effect on the Molien-Weyl integrals, and so we defer further discussion of this topic, which is pertinent to the calculation of general $BCD$ nilpotent orbits, to the next section.

Calculation shows that the $BCD$ maximal nilpotent orbits correspond to the modified Hall Littlewood polynomials ${mHL^{G}_{[singlet]}}(t^2)$, which encode both the Casimirs and the adjoint of the flavour group $G$, as in \ref{eq:Aseries3}. So, since $mHL^G_{[singlet]}(t^2)$ and the HyperK\"ahler quotient associated with each gauge group contain offsetting factors of $PE[adj, t^2]$, it follows, by recursion, that this correspondence holds for all $BCD$ maximal nilpotent orbits, similar to the case for $A$ series maximal nilpotent orbits \cite{Hanany:2011db}.

\FloatBarrier
\subsection{$O(2n)$ Gauge Groups}
\label{subsec:BCDO2n}

In most respects, the evaluation of a general $BCD$ quiver follows the methodology for minimal or maximal quivers, set out above. To obtain a HyperK\"ahler moduli space from the partition data associated with a general nilpotent orbit, it is, however, necessary to work with all the components of a gauge group, which entails using $O$ rather than $SO$ gauge groups throughout.\footnote{This issue does not arise for nodes with $Sp$ gauge groups, since these are simply connected.} Whilst the Molien sum, introduced in the previous section, deals with $O(2n+1)$ gauge groups, $O(2n)$ gauge groups need a more sophisticated treatment.

\subsubsection{Characters of $O(2n)$}

Recall that an orthogonal representation matrix $O$ obeys the defining identity $O.O^T =I$ and so $Det[O]= \pm 1$. A complication arises when constructing the character of an $O(2n)$ representation matrix, since the ${\mathbb Z}_2$ factor which acts to change the sign of its determinant is not a multiple of the identity matrix and therefore does not commute with it. As a consequence, the character (i.e. sum of the eigenvalues) of an $O(2n)$ matrix with negative determinant, denoted $O(2n)^-$, does not have the same structure as the character of an $SO(2n)$ matrix. Indeed, it is necessary to calculate the character of an $O(2n)^-$ matrix from first principles. While the calculation for $O(2)^-$ is relatively straightforward, the general result for $O(2n)^-$ is surprising, since it involves both a reduction in rank and a partly symplectic character.

An illuminating method of calculating the character of a representation matrix is to find its eigenvalues, or at least their structure, as encoded in the characteristic polynomial. Consider the $D_1 \cong SO(2) \cong U(1)$ matrix, $O = \left( {\begin{array}{*{20}{c}}{\cos \theta }&{\sin \theta }\\{ - \sin \theta }&{\cos \theta }\end{array}} \right)$. The characteristic polynomial $Det[O - \lambda I]=0$ evaluates as $1-(e^{-i \theta } -e^{i \theta }) \lambda +\lambda ^2=0$ and the eigenvalues of $O$ follow as $\lambda = e^{\pm i \theta }$, corresponding to the $D_1$ character $x+1/x$. If we now apply the ${\mathbb Z}_2$ factor $\left(\begin{array}{cc} 0 & 1 \\1 & 0 \\\end{array}\right)$, the characteristic polynomial becomes $1- \lambda ^2=0$, with the eigenvalues $\lambda =\pm 1$. Thus the character for $O(2)^-$ has zero rank and is just $1 + (-1)$. An equivalent treatment is given in \cite{Hanany:2012dm}.

Now consider O(4) and O(6) matrices acting on the vector representation. The structures of their eigenvalues differ between $SO$ and $O^-$ matrices, since the characteristic polynomials of $SO(2n)$ matrices are palindromic, while those of $O(2n)^-$ are anti-palindromic. Their eigenvalues rearrange to the forms in table \ref{tab:BCDseries1}, where we use unimodular coordinates to indicate the groups from which characters are taken: $\{x,y,\ldots \}$ for $D_{n}$ and $\{a,b,\ldots \}$ for $C_n$.
\begin{table}[htp]
\caption{Characteristic Polynomials and Eigenvalues of O(2n)}
\begin{center}
\begin{tabular}{|c|c|c|}
\hline
$\text{Matrix}$&$\text{Characteristic~Polynomial}$&$\text{Eigenvalues} {\left( \lambda \right)}$\\
\hline
${SO(2)}$&${1 - {a_1}\lambda + {\lambda ^2} = 0}$&${\left\{ {x,1/x} \right\}}$\\
${O{{(2)}^ - }}$&${1 - {\lambda ^2} = 0}$&${\left\{ {1, - 1} \right\}}$\\
\hline
${SO(4)}$&${1 - {a_1}\lambda + {a_2}{\lambda ^2} - {a_1}{\lambda ^3} + {\lambda ^4}= 0}$&${\left\{ {xy,1/xy,x/y,y/x} \right\}}$\\
${O{{(4)}^ - }}$&${1 - {a_1}\lambda + {a_1}{\lambda ^3} - {\lambda ^4} = 0}$&${\left\{ {1, - 1,a,1/a} \right\}}$\\
\hline
${SO(6)}$&${1 - {a_1}\lambda + {a_2}{\lambda ^2} - {a_3}{\lambda ^3} + {a_2}{\lambda ^4} - {a_1}{\lambda ^5} + {\lambda ^6} = 0}$&${\left\{ {\frac{x}{{yz}},\frac{{yz}}{x},x,\frac{1}{x},\frac{z}{y},\frac{y}{z}} \right\}}$\\
${O{{(6)}^ - }}$&${1 - {a_1}\lambda + {a_2}{\lambda ^2} - {a_2}{\lambda ^4} + {a_1}{\lambda ^5} - {\lambda ^6} = 0}$&${\left\{ {1, - 1,a,\frac{1}{a},\frac{a}{b},\frac{b}{a}} \right\}}$\\
\hline
\end{tabular}
\end{center}
\label{tab:BCDseries1}
\end{table}
Importantly, this decomposition of the character of an $O(2n)^-$ matrix in the vector representation generalises to higher rank $O(2n)$ groups, as $[vec]_{O(2n)^-} \cong [vec]_{O(2)^-} \oplus [fund]_{C_{n-1}}$.

Before proceeding, it is useful to verify that the use of the characters $[vec]_{O(2n)^-}$ and $[vec]_{SO(2n)}$ for the two types of $O(2n)$ vector representation leads to the required invariants. We obtain the Hilbert series for symmetric and antisymmetric invariants by applying the PE or PEF, respectively, to a character, in both cases followed by Weyl integration. The Weyl integration is carried out using the Haar measures for the corresponding $D$ or $C$ groups and we obtain the results in table \ref{tab:BCDseries2}. The exponents of the fugacity $t$ give the degrees of the invariants \cite{Hanany:2014dia, Hanany:2008kn} and show that both types of $O(2n)$ vector representation matrices are associated with symmetric and antisymmetric invariants in the form of delta and epsilon tensors, but with a change of sign in the antisymmetric invariants (i.e. determinants). Thus, when we take a group average over $O(2n)$, the antisymmetric invariants encoded in a Hilbert series cancel out.
%
\begin{table}[htp]
\caption{Invariants of O(2n) Matrices}
\label{tab:BCDseries2}
\begin{center}
\begin{tabular}{|c|c|c|c|c|}
\hline
$\text{Matrix}$&$\text{Determinant}$&${\cal X} \equiv {\left[ {vec} \right]_{O}}$&${\oint\limits_{} {d \mu}~PE\left[ {{{\cal X}},t} \right]}$&${\oint\limits_{} {d \mu}~PEF\left[ {{{\cal X}},t} \right]}$\\
\hline
$SO(2)$&${ + 1}$&${q + 1/q}$&${\frac{1}{{1 - {t^2}}}}$&${1 + {t^2}}$\\
$O(2)^-$&${ - 1}$&${1 + \left( { - 1} \right)}$&${\frac{1}{{1 - {t^2}}}}$&${1 - {t^2}}$\\
\hline
$SO(4)$&${ + 1}$&${x y + \frac{1}{x y}+\frac{x}{y}+\frac{y}{x}}$&${\frac{1}{{1 - {t^2}}}}$&${1 + {t^4}}$\\
$O(4)^-$&${ - 1}$&${1 + \left( { - 1} \right)+ x +\frac{1}{x}}$&${\frac{1}{{1 - {t^2}}}}$&${1 - {t^4}}$\\
\hline
$SO(6)$&${ + 1}$&${\frac{x}{y z}+\frac{y z}{x}+x+\frac{1}{x}+\frac{z}{y}+\frac{y}{z}}$&${\frac{1}{{1 - {t^2}}}}$&${1 + {t^6}}$\\
$O(6)^-$&${ - 1}$&${1 + \left( { - 1} \right)+x+\frac{1}{x}+ \frac{x}{y}+\frac{y}{x}}$&${\frac{1}{{1 - {t^2}}}}$&${1 - {t^6}}$\\
\hline
$SO(2n)$&${ + 1}$&${\left[ {vec} \right]_{{D_n}}} $&${\frac{1}{{1 - {t^2}}}}$&${1 + {t^{2n}}}$\\
$O(2n)^-$&${ - 1}$&$1 + \left(- 1\right)+ {\left[ {fund} \right]_{{C_{n - 1}}}}$&${\frac{1}{{1 - {t^2}}}}$&${1 - {t^{2n}}}$\\
\hline
\end{tabular}
\end{center}
\end{table}
\FloatBarrier

\subsubsection{HyperK\"ahler Quotients for $C_k - O(2n)$}

The peculiar form of character for $[vec]_{O(2n)^-}$ leads to a HyperK\"ahler quotient for a quiver with $C_k$ flavour group and  $O(2n)^-$ gauge group that varies from the usual $PE[[adjoint]_{SO(2n)},t^2]$. We can find this HKQ by integrating over the $C_k$ flavour group, where $k \ge n$ for the quivers under study: 

\begin{equation} 
\label{eq:BCDseries6}
\begin{aligned}
{g_{HK}}\left( {{{{\cal X}}_{O{{\left( {2n} \right)}^ - }}},t} \right) = \oint\limits_{{C_k}} {d\mu}~PE\left[ {{{\left[ {fund} \right]}_{{C_k}}} \otimes {{\left[ {vec} \right]}_{O{{\left( {2n} \right)}^ - }}},t} \right]\\
 \end{aligned}
\end{equation}

Carrying out the calculation for $O(2n)^-$ characters up to $n=5$ gives the results in table \ref{tab:BCDseries3}. Based on these, we conjecture that the HKQ for higher rank $O(2n)^-$ characters is as shown.
\begin{table}[htp]
\caption{HyperK\"ahler Quotients for $O(2n)^-$}
\begin{center}
\begin{tabular}{|c|c|c|c|c|}
\hline
$\text{Bifundamental}$&${g_{HK}^{O{{\left( {2n} \right)}^ - }}\left( {{{{\cal X}}_{{C_{n - 1}}}},t} \right)}$\\
\hline
${{C_{k \ge 1}} \otimes O{{(2)}^ - }}$&${1/(1 + {t^2})}$\\
\hline
${{C_{k \ge 2}} \otimes O{{(4)}^ - }}$&${PE[{{[2]}_C},{t^4}]}$\\
\hline
${{C_{k \ge 3}} \otimes O{{(6)}^ - }}$&${PE[{{[0,1]}_C},{t^2}]~PE[{{[2,0]}_C} - {{[0,1]}_C},{t^4}]}$\\
\hline
${{C_{k \ge 4}} \otimes O{{(8)}^ - }}$&${PE[{{[0,1,0]}_C},{t^2}]~PE[{{[2,0,0]}_C} - {{[0,1,0]}_C},{t^4}]}$\\
\hline
${{C_{k \ge 5}} \otimes O{{(10)}^ - }}$&${PE[{{[0,1,0,0]}_C},{t^2}]~PE[{{[2,0,0,0]}_C} - {{[0,1,0,0]}_C},{t^4}]}$\\
\hline
$C_{k \ge n} \otimes O(2n)^-$
&
\footnotesize{
$ PE\left[ {{{[0,1,0, \ldots  \ldots ,0]}_{{C_{n - 1}}}},{t^2}} \right]PE\left[ {{{[2,0,0, \ldots ,0]}_{{C_{n - 1}}}} - {{[0,1,0 \ldots ,0]}_{{C_{n - 1}}}},{t^4}} \right]$
}
\\
\hline
\end{tabular}
\end{center}
\label{tab:BCDseries3}
\end{table}

The structure of the HKQ terms follows from the invariants of the $C_k$ flavour group fundamental, which are generated by antisymmetric tensors of degree $2,4, \ldots, 2k$. Under the PE of the bifundamental of the $C_{k} \otimes O(2n)^-$ product group, these $C_k$ invariants map the character $[vec]_{O(2n)^-} \cong [vec]_{O(2)^-} \oplus [fund]_{C_{n-1}}$ to a series of characters of $C_{n-1}$ irreps. The PEs in table \ref{tab:BCDseries3} that generate this series contain terms at $t^4$, in addition to the usual term at $t^2$ from the anti-symmetrisation of an orthogonal group vector representation.
\FloatBarrier

Based on the foregoing, we can express the group averaged Weyl integration over a quiver containing a bifundamental field with $C_k$ flavour group and $O(2)$ gauge group, as:
\begin{equation} 
\label{eq:BCDseries7a}
\begin{aligned}
g_{Higgs}^{C_k-O(2)}\left({\cal X}_{C_k},t \right) &= \frac{1}{2} \left(g_{Higgs}^{C_k-SO(2)}\left( {\cal X}_{C_k},t \right)+ g_{Higgs}^{C_k-O(2)^-}\left({\cal X}_{C_k},t \right)\right),\\ 
 \end{aligned}
\end{equation}
where
\begin{equation} 
\label{eq:BCDseries7b}
\begin{aligned}
g_{Higgs}^{C_k-SO(2)}\left( {\cal X}_{C_k},t \right) &= {\oint\limits_{{SO(2)}} {d\mu } \frac{{PE\left[ {{{\left[ {fund} \right]}_{{C_k}}} \otimes {{\left[ {vec} \right]}_{{SO(2)}}},t} \right]}}{{PE\left[ {1,{t^2}} \right]}}}\\
 \end{aligned}
\end{equation}
and
\begin{equation} 
\label{eq:BCDseries7c}
\begin{aligned} 
g_{Higgs}^{C_k-O(2)^-}\left({\cal X}_{C_k},t \right)&= \frac{{PE\left[ {{{\left[ {fund} \right]}_{{C_k}}},t} \right]PE\left[ {{{\left[ {fund} \right]}_{{C_k}}}, - t} \right]}}{{1/\left( {1 + {t^2}} \right)}}.
 \end{aligned}
\end{equation}
We represent the vector character of $D_1 \cong SO(2)$ as $x+1/x$ and use the unitary Haar measure $1/x$, when calculating \ref{eq:BCDseries7b}. The action of the ${\mathbb Z}_2$ factor encoded in \ref{eq:BCDseries7c} is trivial for the maximal chain $C_1-D_1$, but has an impact on the Hilbert series for quivers containing non-maximal chains, from $C_2-D_1$ upwards.

The corresponding Weyl integral for a $C_k$ flavour group and $O(2n)$ gauge group where $k \ge n >1$ is:
\begin{equation} 
\label{eq:BCDseries8}
\begin{aligned}
& g_{Higgs}^{{C_k-O(2n)}}\left( {{{{\cal X}}_{{C_k}}},t} \right){\rm{ }} = \frac{1}{2}\oint\limits_{{D_n}} {d\mu } \frac{{PE\left[ {{{\left[ {fund} \right]}_{{C_k}}} \otimes {{\left[ {vec} \right]}_{{SO(2n)}}},t} \right]}}{{PE\left[ {{{\left[ {adjoint} \right]}_{{SO(2n)}}},{t^2}} \right]}}\\
& + \frac{1}{2}\oint\limits_{{C_{n - 1}}} {d\mu } \frac{{PE\left[ {{{\left[ {fund} \right]}_{{C_k}}} \otimes {{\left[ {fund} \right]}_{{C_{n - 1}}}},t} \right]PE\left[ {{{\left[ {fund} \right]}_{{C_k}}},t} \right]PE\left[ {{{\left[ {fund} \right]}_{{C_k}}}, - t} \right]}}{{g_{HK}^{O{{(2n)}^ - }}\left( {{{{\cal X}}_{{C_{n - 1}}}},t} \right)}},
 \end{aligned}
\end{equation}
where $g_{HK}^{O(2n)^-}$ is as in table \ref{tab:BCDseries3}. We incorporate these group averaging procedures, which do not affect the dimensions of a moduli space, but may affect its structure, within the results for general $BCD$ quivers in the following.

\subsection{General Higgs Branch: $BCD$ Series}
\label{subsec:BCDgen}

We have now assembled the analytic procedures necessary for the calculation of quivers associated with general $BCD$ nilpotent orbits. We have shown in section \ref{sec:orbits} how the partition data from a $BCD$ nilpotent orbit can be used to define a quiver with alternating O/USp nodes (figures \ref{fig:BDquiver} or \ref{fig:Cquiver}), that has a moduli space of the required dimension (\ref{eq:orbits11} and \ref{eq:orbits12}). Using the averaging procedures over orthogonal groups set out in sections \ref{subsec:BCDminmax} and \ref{subsec:BCDO2n}, we can calculate the Hilbert series of a $BCD$ quiver from the formula, adapted from \ref{eq:orbits7}:
\begin{equation} 
\label{eq:BCDseries9}
\begin{aligned}
&g_{Higgs}^{SO/USp\left( {{N_0}} \right)}\left( {{{\cal X}_{SO/USp\left( {{N_0}} \right)}},t} \right)\\
& = \frac{1}{{{2^{\# O}}}} \sum\limits_{O \pm } \oint\limits_{USp/O\left( {{N_1}} \right) \otimes \hfill \atop O/USp({N_2}) \otimes \ldots \hfill}{d\mu } \prod\limits_{G\left( i \right) = USp}^{} \frac{{PE\left[ {{{\left[ {vec} \right]}_{O\left( {{N_{i - 1}}} \right)}} \otimes {{\left[ {fund} \right]}_{USp\left( {{N_i}} \right)}},t} \right]}}{{PE\left[ {{{\left[ {adj} \right]}_{USp\left( {{N_i}} \right)}},{t^2}} \right]}} \\
&~~~~~~~~~~~~~~~~~~~~ ~~~~~~~~~~~~~~~\times \prod\limits_{G\left( i \right) = O}^{} \frac{{PE\left[ {{{\left[ {fund} \right]}_{USp\left( {{N_{i - 1}}} \right)}} \otimes {{\left[ {vec} \right]}_{O\left( {{N_i}} \right)}},t} \right]}}{{g_{HK}^{O\left( {{N_i}} \right)}\left( {{{{\cal X}}_{O\left( {{N_i}} \right)}},t} \right)}},
 \end{aligned}
\end{equation}
where ${\text \#}O$ equals the number of orthogonal gauge groups and the summation indicates that all possible combinations of $SO/O^-$ characters should be evaluated. Once the Hilbert series for a $BCD$ quiver has been calculated, it can be restated in a number of forms, as per \ref{eq:Aseries6} and \ref{eq:Aseries7}. We shall not digress further into the practical details of the calculations, but simply set out the results for $BCD$ groups of rank up to 4 in tables \ref{tab:Bseries1} to \ref{tab:Dseries2}.

\begin{sidewaystable}[htp]
\caption{Quivers for Nilpotent Orbits of $B_1$, $B_2$ and $B_3$}
\begin{center}
\footnotesize
\begin{tabular}{|c|c|c|c|c|c|}
 \hline
 $\text{Orbit} $&$ \text{Quiver} $&$ \text{Dim.} $&$ \text{Hilbert } \text{Series} $&$ \text{Character } \text{HWG} $&$ \text{mHL } \text{HWG} $\\
 \hline
 $(1^3)$&$B_1$&$ 0 $&$ 1 $&$ 1 $&$ 1-h^2 t^2 $\\
 \hline
 $(3)$&$B_1,C_1,B_0 $&$ 2 $&$ \frac{1+t^2}{(1-t^2)^2} $&$ \frac{1}{1-m^2 t^2} $&$ 1 $\\
\hline
\hline
$(1^5)$&$B_2$&$ 0 $&$ 1 $&$ 1 $&$
\scriptsize{
\begin{array}{c}
1 - {h_2}^2{t^2} - {h_1}{t^4}\\
 + {h_1}{h_2}^2{t^4} + {h_1}{h_2}^2{t^6}\\
 - {h_1}^3{t^6} - {h_2}^4{t^6} + {h_1}^2{h_2}^2{t^8}
\end{array}
 } 
 $\\
 \hline
$(2^2,1)$&$B_2,C_1$&$ 4 $&$ \frac{1+6 t^2+t^4}{(1-t^2)^4} $&$ \frac{1}{1-{m_2}^2 t^2} $&$1-{h_1} t^4-{h_1}^2 t^4+{h_1} {h_2}^2 t^6$\\
 \hline
$(3,1^2)$&$B_2,C_1,B_0$&$ 6 $&$ \frac{(1+t^2) (1+3 t^2+t^4)}{(1-t^2)^6} $&$ \frac{1}{(1-{m_2}^2 t^2) (1-{m_1}^2 t^4)} $&$ 1-{h_1} t^4 $\\
 \hline
$(5)$&$B_2,C_2,B_1,C_1,B_0$&$ 8 $&$ \frac{(1+t^2)^2 (1+t^4)}{(1-t^2)^8} $&$
\scriptsize{
 \frac{1+{m_1} {m_2}^2 t^8} {
 \left( \begin{array}{c}
(1-{m_2}^2 t^2) (1-{m_1} t^4)\\
\times (1-{m_1}^2 t^4) (1-{m_2}^2 t^6)
\end{array}\right)
 } }
 $&$ 1 $\\
\hline
\hline
$(1^7)$&$B_3$&$ 0 $&$ 1 $&$ 1 $&$ \ldots $\\
 \hline
$(2^2,1^3)$&$B_3,C_1$&$ 8 $&$ \frac{1 + 13{t^2} + 28{t^4} + 13{t^6} + {t^8}}{(1-t^2)^8} $&$ \frac{1}{1-{m_2} t^2} $&$ \ldots $\\
 \hline
$(3,1^4)$&$B_3,C_1,B_0$&$ 10 $&$ \frac{1+10 t^2+20 t^4+10 t^6+t^8}{(1-t^2)^{10} (1+t^2)^{-1}}$&$ \frac{1}{(1-{m_2} t^2) (1-{m_1}^2 t^4)} $&$
\scriptsize{
\begin{array}{c}
1 - {h_3}^2{t^4} - {h_2}{t^6} + {h_1}{h_2}{t^6}\\
 - {h_1}{t^6} + {h_1}{h_2}{t^8} + {h_3}^2{t^8}\\
 - {h_1}^3{t^8} - {h_2}^2{t^{10}} + {h_1}^2{h_2}{t^{12}}
\end{array}
%
 } 
 $\\
 \hline
$(3,2^2)$&$B_3,C_2,B_0$&$ 12 $&$ \frac{1 + 8{t^2} + 36{t^4} + 92{t^6} + {t^8} - 6{t^{10}}}{{(1-t^2)^{12}}{(1+t^2)}^{-1}} $&$ \frac{1+{m_1} {m_3}^2 t^6}{(1-{m_2} t^2) (1-{m_1}^2 t^4) (1-{m_3}^2 t^4)} $&$
\scriptsize{
\begin{array}{c}
1 - {h_1}{t^6} - {h_2}{t^6} - {h_2}^2{t^8} \\
+ {h_3}^2{t^8} + {h_1}^2{h_2}{t^{10}} \\
+ {h_2}{h_3}^2{t^{10}} - {h_1}^2{h_3}^2{t^{12}}
\end{array}
%
 } 
 $\\
 \hline
$(3^2,1)$&$B_3,C_2,D_1$&$ 14 $&$ \frac{1 + 6{t^2} + 21{t^4} + 28{t^6} + 21{t^8} + 6{t^{10}} + {t^{12}}}{{(1-t^2)^{14}}{(1+t^2)}^{-1}} $&$
\scriptsize{
 \frac{1+{m_1} {m_2} {m_3}^2 t^{10}}{
 \left( \begin{array}{c}
 (1-{m_2} t^2) (1-{m_1}^2 t^4) (1-{m_3}^2 t^4) \\
 \times (1-{m_1} {m_3}^2 t^6) (1-{m_2}^2 t^8)
 \end{array}\right)
 } }
$&$ 1 - {h_2}{t^6} - {h_1}{t^6} + {h_3}^2{t^8} $\\
 \hline
$(5,1^2)$&$B_3,C_2,B_1,C_1,B_0$&$ 16 $&$ \frac{1 + 3{t^2} + 6{t^4} + 3{t^6} + {t^8}}{{(1-t^2)^{16}}{(1+t^2)^{-2} (1+t^4)^{-1}}} $&$ \ldots $&$ 1-{h_1} t^6 $\\
 \hline
$(7)$&$B_3,C_3,B_2,C_2,B_1,C_1,B_0$&$ 18 $&$ \frac{(1-t^4) (1-t^8) (1-t^{12})}{(1-t^2)^{21}} $&$ \ldots $&$ 1 $\\
 \hline
\end{tabular}
\end{center}
\text{B/D gauge nodes in a quiver indicate averages over the corresponding O gauge groups}\\
\text{An mHL HWG of 1 denotes $mHL^B_{[0, \ldots, 0]}(t^2)$.}\\
\text{Some character and mHL HWGs have been omitted for brevity.}\\
\text{See text for a discussion of the non-palindromic $B_3-C_2-B_0$ quiver.}
\label{tab:Bseries1}
\end{sidewaystable}

\begin{sidewaystable}[htp]
\caption{Quivers for Nilpotent Orbits of $B_4$}
\begin{center}
\scriptsize
\begin{tabular}{|c|c|c|c|c|c|}
 \hline
 $\text{Orbit} $&$ \text{Quiver} $&$ \text{Dim.} $&$ \text{Hilbert } \text{Series} $&$ \text{Character } \text{HWG } $&$ \text{mHL } \text{HWG} $\\
\hline
$(1^9)$&$B_4$&$ 0 $&$ 1 $&$ 1 $&$ \ldots $\\
 \hline
 $(2^2,1^5)$&$B_4,C_1$&$ 12 $&$ \frac{1 + 24 t^2 + 129 t^4 + 220 t^6 + 129 t^8 + 24 t^{10} + t^{12}}{(1-t^2)^{12}} $&$ \frac{1}{1-{m_2} t^2} $&$ \ldots $\\
 \hline
 $(3,1^6)$&$B_4,C_1,B_0$&$ 14 $
&
$\frac{{1 + 21{t^2} + 105{t^4} + 175{t^6} + 105{t^8} + 21{t^{10}} + {t^{12}}}}{{{{( {1 - {t^2}} )}^{14}}{{( {1 + {t^2}} )}^{ - 1}}}}$
&
$ \frac{1}{(1-{m_1}^2 t^4) (1-{m_2} t^2) } $&$\ldots $\\
 \hline
 $(2^4,1)$&$B_4,C_2$&$ 16 $&\tiny{
 $\frac{
 \begin{array}{c}
 1 + 20 t^2 + 165 t^4 + 600 t^6 + 924 t^8 \\
 + 600 t^{10} + 165 t^{12} + 20 t^{14} + t^{16}
 \end{array}
 }{(1-t^2)^{16}} $
 }
 &$ \frac{1}{(1-{m_2} t^2) (1-{m_4}^2 t^4)} $&$\ldots $\\
 \hline
$(3,2^2,1^2)$&$B_4,C_2,B_0$&$ 20 $&\tiny{
$ \frac{
 \begin{array}{c}
1 + 14 t^2 + 106 t^4 + 454 t^6 + 788 t^8 \\
 +454 t^{10} + 106 t^{12} + 14 t^{14} + t^{16}
 \end{array}
}{(1-t^2)^{20} (1+t^2)^{-2}} $
}
&
\scriptsize{$
\frac{1+{m_1} {m_3} t^6}{(1-{m_1}^2 t^4) (1-{m_2} t^2) (1-{m_3}^2 t^8) (1-{m_4}^2 t^4)}
$ }

&$\ldots $\\
 \hline
 $(3^2,1^3)$&$B_4,C_2,D_1$&$ 22 $&\tiny{$
 \frac{
   \begin{array}{c}
 1+13 t^2+91 t^4+335 t^6+737 t^8+946 t^{10}\\
 +737 t^{12}+335 t^{14}+91 t^{16}+13 t^{18}+ t^{20}
 \end{array}
  } 
 {(1-t^2)^{22} (1+t^2)^{-1}
 }
 $}
 &
 \scriptsize{$
  \frac{1 + m_1 m_2 m_3 t^{10}}
  {\left( \begin{array}{c}
  (1 - m_2 t^2) (1 - m_1^2 t^4) (1 - m_4^2 t^4)\\
   (1 - m_1 m_3 t^6) (1 - m_2^2 t^8) (1 - m_3^2 t^8)
 \end{array}\right)}
%
%
 $}
 &$
  \ldots 
  $\\
 \hline
$(3^3)$&$B_4,C_3,B_1$&$ 24 $&\tiny{
$
\frac{
 \begin{array}{c}
1 + 10 t^2 + 56 t^4 + 194 t^6 + 438 t^8 + 578 t^{10} \\
 +438 t^{12} + 194 t^{14} + 56 t^{16} + 10 t^{18} + t^{20}
 \end{array}
}{(1-t^2)^{24} (1+t^2)^{-2}}
 $}
 &$ \ldots $&
 $
 \begin{array}{c}
 1 - h_2 t^6 - h_1 t^8 - h_2 t^{10}\\
  + h_3 t^{10} + h_1 h_3 t^{12} - h_2^2 t^{14}\\
  + h_3 t^{14} + h_1 h_3 t^{14} - h_1 h_4^2 t^{14} \\
 - h_1 h_4^2 t^{16} + h_2 h_3 t^{18} 
  \end{array}
 $\\
 \hline
 $(5,1^4)$&$B_4,C_2,B_1,C_1,B_0$&$ 24 $&\tiny{
 $
\frac{
 \begin{array}{c}
 1 + 10 t^2 + 55 t^4 + 136 t^6 + 190 t^8 \\
 + 136 t^{10} + 55 t^{12} + 10 t^{14} + t^{16}
 \end{array}
 }{(1-t^2)^{24} (1+t^2)^{-2} (1+t^4)^{-1}} 
 $}
&
$ \ldots $&$
 \begin{array}{c}
 1 - h_3 t^6 - h_1 t^8 + h_1 h_2 t^8 \\
 - h_1^3 t^{10} - h_2 t^{10} + h_1 h_2 t^{10} \\
 - h_2^2 t^{14} + h_3 t^{14} + h_1^2 h_2 t^{16} 
  \end{array}
 $\\
 \hline
 $(4^2,1)$&$B_4,C_3,D_2,C_1$&$ 26 $&\tiny{
 $
\frac{
 \begin{array}{c}
1 + 9 t^2 + 45 t^4 + 165 t^6 + 441 t^8 + 854 t^{10} + 1050 t^{12}\\
 + 854 t^{14} + 441 t^{16} + 165 t^{18} + 45 t^{20} + 9 t^{22} + t^{24}
 \end{array}
}{(1-t^2)^{26}(1+t^2)^{-1}}
$}
&
$ \ldots $&$
 \begin{array}{c}
 1 - h_1 t^8 - h_1^2 t^8 - h_2 t^{10} \\
 + h_1 h_2 t^{14} + h_3 t^{14} + h_1 h_3 t^{14} \\
 - h_1 h_4^2 t^{16} 
   \end{array}
 $\\
 \hline
 
 $(5,2^2)$&$B_4,C_3,B_1,C_1,B_0$&$ 26 $&\tiny{
 $
\frac{ 
 \begin{array}{c}
 1 + 7 t^2 + 30 t^4 + 98 t^6 + 259 t^8 + 554 t^{10} \\
 + 484 t^{12} + 71 t^{14} - 15 t^{16} - 2 t^{18} + t^{20}
 \end{array}
 }{(1-t^2)^{26} (1+t^2)^{-3}}
$}
&
$ \ldots $&$
 \begin{array}{c}
 1 - h_1 t^8 - h_2 t^{10} - h_2^2 t^{12} \\
 + h_1^2 h_2 t^{14} + h_3 t^{14} \\
 + h_2 h_3 t^{16} - h_1^2 h_3 t^{18} 
   \end{array}
 $\\
 \hline
 
$(5,3,1)$&$B_4,C_3,D_2,C_1,B_0$&$ 28 $&\tiny{
 $
\frac{
 \begin{array}{c}
1 + 6 t^2 + 22 t^4 + 62 t^6 + 138 t^8 + 227 t^{10} + 264 t^{12}\\ 
+227 t^{14} + 138 t^{16} + 62 t^{18} + 22 t^{20} + 6 t^{22} + t^{24}
 \end{array}
 }{(1-t^2)^{28} (1+t^2)^{-2} }
$}
&
$ \ldots $&$ 1 - h_1 t^8 - h_2 t^{10} + h_3 t^{14} $\\
 \hline
 
 $(7,1^2)$&$ \begin{array}{c}
 B_4,C_3,B_2,\\
 C_2,B_1,C_1,B_0
   \end{array}
 $&$ 30 $&
 $
\frac{ 1 + 3 t^2 + 6 t^4 + 10 t^6 + 6 t^8 + 3 t^{10} + t^{12}}{(1-t^2)^{33} (1-t^4)^{-1} (1-t^8)^{-1} (1-t^{12})^{-1}}$
&
$ \ldots $&$ 1 - h_1 t^8$\\
 \hline
 
 $(9)$&$
 \begin{array}{c}
 B_4,C_4,B_3,C_3,\\
 B_2,C_2,B_1,C_1,B_0
  \end{array}
 $&$ 32 $&
 $
\frac{(1-t^4) (1-t^8) (1-t^{12}) (1-t^{16})}{(1-t^2)^{36}}
$
&
$ \ldots $&$ 1 $\\
 \hline
\end{tabular}
\end{center}
\text{B/D gauge nodes in a quiver indicate averages over the corresponding O gauge groups}\\
\text{An mHL HWG of 1 denotes $mHL^B_{[0, \ldots, 0]}(t^2)$.}\\
\text{Some character and mHL HWGs have been omitted for brevity.}\\
\text{See text for a discussion of the non-palindromic $B_4-C_3-B_1-C_1-B_0$ quiver.}
\label{tab:Bseries2}
\end{sidewaystable}

\begin{sidewaystable}[htp]
\caption{Quivers for Nilpotent Orbits of $C_1$, $C_2$ and $C_3$}
\begin{center}
\footnotesize
\begin{tabular}{|c|c|c|c|c|c|}
 \hline
 $\text{Orbit} $&$ \text{Quiver} $&$ \text{Dim.} $&$ \text{Hilbert } \text{Series} $&$ \text{Character } \text{HWG} $&$ \text{mHL } \text{HWG} $\\
 \hline
 $(1^2) $&$C_1$&$ 0 $&$ 1 $&$ 1 $&$ 1-h^2 t^2 $\\
 \hline
 $(2)$&$C_1,B_0$&$ 2 $&$ \frac{1+t^2}{(1-t^2)^2} $&$ \frac{1}{1-m^2 t^2} $&$ 1 $\\
 \hline
\hline
 $(1^4)$&$C_2 $&$ 0 $&$ 1 $&$ 1 $&$
 \begin{array}{c}
1 - {h_1}^2{t^2} - {h_2}{t^4}\\
 + {h_1}^2{h_2}{t^4} + {h_1}^2{h_2}{t^6}\\
 - {h_2}^3{t^6} - {h_1}^4{t^6} + {h_1}^2{h_2}^2{t^8}
\end{array}
 $\\
 \hline
 $(2,1^2)$&$ C_2,B_0 $&$ 4 $&$ \frac{1+6 t^2+t^4}{(1-t^2)^4} $&$ \frac{1}{1-{m_1}^2 t^2} $&$1 - {h_2}{t^4} - {h_2}^2{t^4} + {h_1}^2{h_2}{t^6} $\\
 \hline
 $(2^2)$&$ C_2,D_1 $&$ 6 $&$ \frac{1+3 t^2+t^4}{{(1-t^2)^6}{(1+t^2)^{-1}}} $&$ \frac{1}{(1-{m_1}^2 t^2) (1-{m_2}^2 t^4)} $&$ 1-{h_2} t^4 $\\
 \hline
 $(4)$&$ C_2,B_1,C_1,B_0 $&$ 8 $&$ \frac{(1-t^4) (1-t^8)}{(1-t^2)^{10}} $&$\scriptsize{
 \frac{1+{m_1}^2 {m_2} t^8} { \left( \begin{array}{c}
 (1-{m_1}^2 t^2)(1-{m_2} t^4) \\ \times (1-{m_2}^2 t^4) (1-{m_1}^2 t^6)
  \end{array}\right) }} 
 $&$ 1 $\\
\hline
\hline
 $(1^6)$&$ C_3 $&$ 0 $&$ 1 $&$ 1 $&$ \ldots $\\
 \hline
 $(2,1^4)$&$ C_3,B_0 $&$ 6 $&$ \frac{1+14 t^2+t^4}{{(1-t^2)^6}{(1+t^2)}^{-1}} $&$ \frac{1}{1-{m_1}^2 t^2} $&$ \ldots $\\
 \hline
 $(2^2,1^2)$&$ C_3,D_1 $&$ 10 $&$ \frac{1 + 10{t^2} + 41{t^4} + 10{t^6} + {t^8}}{{(1-t^2)^{10}}{(1+t^2)^{-1}}} $&$ \frac{1}{(1-{m_1}^2 t^2) (1-{m_2}^2 t^4)} $&$
 \begin{array}{c}
1 - {h_2}{t^4} - {h_3}^2{t^6}\\
 + {h_1}{h_2}{h_3}{t^8} + {h_1}{h_3}{t^8} - {h_2}{t^8}\\
 + {h_1}{h_2}{h_3}{t^{10}} + {h_1}{h_3}{t^{10}} - {h_1}^3{h_3}{t^{10}}\\
 - {h_2}^2{t^{10}} - {h_2}^3{t^{10}} + {h_1}^2{h_2}^2{t^{12}}
\end{array}
 $\\
 \hline
 $(2^3)$&$ C_3,B_1 $&$ 12 $&$ \frac{1 + 7{t^2} + 15{t^4} + 7{t^6} + {t^8}}{{(1-t^2)^{12}}{(1+t^2)^{-2}}} $&$ \frac{1}{(1-{m_1}^2 t^2) (1-{m_2}^2 t^4) (1-{m_3}^2 t^6)} $&$
 \begin{array}{c}
1 - {h_2}{t^4} + {h_1}{h_3}{t^8}\\
 - {h_2}{t^8} + {h_1}{h_3}{t^{10}}\\
 - {h_2}^2{t^{10}}
\end{array}
 $\\
\hline
 $(3^2)$&$ C_3,D_2,C_1 $&$ 14 $&$ \frac{1 + 6{t^2} + 21{t^4} + 35{t^6} + 21{t^8} + 6{t^{10}} + {t^{12}}}{{(1-t^2)^{14}}{(1+t^2)^{-1}}} $&$\scriptsize{
 \frac{1+{m_1} {m_2} {m_3} t^8}{
  \left( \begin{array}{c}
 (1-{m_1}^2 t^2) (1-{m_2} t^4)(1-{m_2}^2 t^4)\\ \times (1-{m_1} {m_3} t^6) (1-{m_3}^2 t^6)
  \end{array}\right) } }
 $&$ 1-{h_1}^2 t^6-{h_2} t^8+{h_1} {h_3} t^{10} $\\
\hline
 $(4,1^2)$&$ C_3,B_1,C_1,B_0 $&$ 14 $&$ \frac{1 + 6{t^2} + 21{t^4} + 56{t^6} + 21{t^8} + 6{t^{10}} + {t^{12}}}{{(1-t^2)^{14}}{(1+t^2)^{-1}}} $&$ \ldots $&$ 1-{h_2} t^8-{h_2}^2 t^8 +{h_1}^2 {h_2} t^{10} $\\
 \hline
 $(4,2)$&$C_3,D_2,C_1,B_0$&$ 16 $&$ \frac{1 + 3{t^2} + 7{t^4} + 13{t^6} + 7{t^8} + 3{t^{10}} + {t^{12}}}{{(1-t^2)^{16}}{(1+t^2)^{-2}}} $&$ \ldots $&$ 1-{h_2} t^8 $\\
 \hline
 $(6)$&$C_3,B_2,C_2,B_1,C_1,B_0$&$ 18 $&$ \frac{(1-t^4) (1-t^8) (1-t^{12})}{(1-t^2)^{21}} $&$ \ldots $&$ 1 $\\
 \hline
\end{tabular}
\end{center}
\label{tab:Cseries1}
\text{B/D gauge nodes in a quiver indicate averages over the corresponding O gauge groups}\\
\text{An mHL HWG of 1 denotes $mHL^C_{[0, \ldots, 0]}(t^2)$.}\\
\text{Some character and mHL HWGs have been omitted for brevity.}
\end{sidewaystable}


\begin{sidewaystable}[htp]
\caption{Quivers for Nilpotent Orbits of $C_4$}
\begin{center}
\scriptsize
\begin{tabular}{|c|c|c|c|c|c|}
 \hline
 $\text{Orbit} $&$ \text{Quiver} $&$ \text{Dim.} $&$ \text{Hilbert } \text{Series} $&$ \text{Character } \text{HWG} $&$ \text{mHL } \text{HWG} $\\
\hline
$(1^8)$&$C_4$&$ 0 $&$ 1 $&$ 1 $&$ \ldots $\\
 \hline
 $(2,1^6)$&$C_4,B_0$&$8 $&$
\frac{1 + 28 t^2 + 70 t^4 + 28 t^6 + t^8}{(1-t^2)^8}
 $&$
\frac{1}{1-{m_1}^2 t^2}
  $&$ \ldots $\\
 \hline
 $(2^2,1^4)$&$C_4,D_1$&$ 14 $
&
$
\frac{
\tiny{ \begin{array}{c}
1 + 21 t^2 + 204 t^4 + 406 t^6 \\
+ 204 t^8 + 21 t^{10} + t^{12}
 \end{array}}
}{(1-t^2)^{14}(1+t^2)^{-1} }
$
&
$
\frac{1}{(1-{m_1}^2 t^2) (1-{m_2}^2 t^4)}
 $&$\ldots $\\
 \hline
 $(2^3,1^2)$&$C_4,B_1$&$ 18 $&
 $
\frac{ 
\tiny{ \begin{array}{c}
1 + 17 t^2 + 126 t^4 + 537 t^6 + 894 t^8 \\
+ 537 t^{10} + 126 t^{12} +17 t^{14} + t^{16} 
  \end{array}}
 }{(1-t^2)^{18} (1+t^2)^{-1}}
 $
 &$
 \frac{1}{(1-{m_1}^2 t^2) (1-{m_2}^2 t^4) (1-{m_3}^2 t^6)}
 $&$\ldots $\\
 \hline
$(2^4)$&$C_4,D_2$&$ 20 $&
$
\frac{ 
\tiny{ \begin{array}{c}
1 + 14 t^2 + 79 t^4 + 223 t^6 + 317 t^8 \\
+ 223 t^{10} + 79 t^{12} + 14 t^{14} + t^{16}
  \end{array}}
}{(1-t^2)^{20}(1+t^2)^{-2}}
$
&$
\frac{1}{
\tiny{ \left( \begin{array}{c}
(1-{m_1}^2 t^2) (1-{m_2}^2 t^4)\\
\times (1-{m_3}^2 t^6) (1-{m_4}^2 t^8)
  \end{array}\right)}
}$
&$\ldots $\\
 \hline
 $(4,1^4)$&$C_4,B_1,C_1,B_0$&$ 20 $&$
\frac{
\tiny{ \begin{array}{c}
1 + 14 t^2 + 106 t^4 + 574 t^6 + 722 t^8 \\
+ 574 t^{10} + 106 t^{12} + 14 t^{14} + t^{16}
  \end{array}}
}{(1-t^2)^{20}(t^2+1)^{-2}}
 $
 &
 $
 \frac{{m_1}^2 {m_2} t^8+{m_1} {m_2} {m_3} t^8+{m_1} {m_3} t^6+1}{
 \tiny{ \left( \begin{array}{c}
 (1-{m_1}^2 t^2) (1-{m_1}^2 t^6) (1-{m_2} t^4)\\
 \times (1-{m_2}^2 t^4) (1-{m_3}^2 t^6)
  \end{array}\right)}
 }
 $
 &$ \ldots $\\
 \hline
$(3^2,1^2)$&$C_4,D_2,C_1$&$ 22 $&
$
\frac{
\tiny{ \begin{array}{c}
1 + 13 t^2 + 91 t^4 + 419 t^6 + 1346 t^8 + 2365 t^{10} \\
+ 1841 t^{12} + 476 t^{14} - 56 t^{16} - 29 t^{18} + t^{20}
  \end{array}}
 }{(1-t^2)^{22} (1+t^2)^{-1}}
 $
 &$
 \frac{
 \tiny{ \begin{array}{c}
 1+{m_2} {m_4} t^8+{m_1} {m_2} {m_3} t^8+{m_1} {m_3} {m_4} t^{10}\\
 +{m_1} {m_2} {m_3} {m_4} t^{12}-{m_1} {m_2} {m_3} {m_4} t^{14}
   \end{array}}
 }{
 \tiny{ \left( \begin{array}{c}
 (1-{m_1}^2 t^2) (1-{m_2}^2 t^4) (1-{m_2} t^4)\\
 \times (1-{m_3}^2 t^6) (1-{m_1} {m_3} t^6) (1-{m_4}^2 t^8)
   \end{array}\right)}
 }
 $&
 $
 \ldots
 $\\
 \hline
 $(3^2,2)$&$C_4,B_2,C_1$&$ 24 $&
 $
\frac{
\tiny{ \begin{array}{c}
1 + 10 t^2 + 56 t^4 + 194 t^6 + 405 t^8 + 512 t^{10}\\
 + 405 t^{12} + 194 t^{14} + 56 t^{16} + 10 t^{18} + t^{20}
   \end{array}}
 }{(1-t^2)^{24} (1+t^2)^{-2}}
 $
&
$ \ldots $&$
 \begin{array}{c}
1 - h_1^2 t^6 - h_2 t^8 - h_4 t^8 + h_1 h_3 t^{10}\\
 - h_2 t^{12} + h_1 h_3 t^{12} + h_4 t^{12} - h_2^2 t^{14} \\
 + 2 h_1 h_3 t^{14} - h_2 h_4 t^{14} + h_3^2 t^{16} \\
 - h_2 h_4 t^{16}  \end{array}
 $\\
 \hline
 $(4,2,1^2)$&$C_4,D_2,C_1,B_0$&$ 24 $&
 $
\frac{ 
\tiny{ \begin{array}{c}
1 + 10 t^2 + 56 t^4 + 230 t^6 + 701 t^8 + 776 t^{10} \\
+ 701 t^{12} + 230 t^{14} + 56 t^{16} + 10 t^{18} + t^{20}
   \end{array}}
}{(1-t^2)^{24} (1+t^2)^{-2}}
$
&
$ \ldots $&$
 \begin{array}{c}
1 - h_2 t^8 - h_4 t^8 - h_3^2 t^{10} - h_2 t^{12} \\
+ h_1 h_3 t^{12} + h_1 h_2 h_3 t^{12} + h_4 t^{12} - h_2^2 t^{14}\\
 - h_2^3 t^{14} + h_1 h_3 t^{14} - h_1^3 h_3 t^{14}\\
 + h_1 h_2 h_3 t^{14} + h_1^2 h_2^2 t^{16}
 \end{array}
 $\\
 \hline 
 $(4,2^2)$&$C_4,B_2,C_1,B_0$&$ 26 $&
 $
\frac{
\tiny{ \begin{array}{c}
1 + 7 t^2 + 30 t^4 + 98 t^6 + 199 t^8 + 230 t^{10}\\
 + 199 t^{12} + 98 t^{14} + 30 t^{16} + 7 t^{18} + t^{20}
   \end{array} }
}{(1-t^2)^{29} (1-t^4)^{-3}}
$
&
$ \ldots $&$
 \begin{array}{c}
1 - h_2 t^8 - h_4 t^8 - h_2 t^{12} + h_1 h_3 t^{12}\\
 + h_4 t^{12} - h_2^2 t^{14} + h_1 h_3 t^{14}
 \end{array}
 $\\
 \hline
$(4^2)$&$C_4,D_3,C_2,D_1$&$ 28 $&
 $
\frac{
\tiny{ \begin{array}{c}
1 + 6 t^2 + 21 t^4 + 56 t^6 + 99 t^8 + 117 t^{10} \\
+ 99 t^{12} + 56 t^{14} + 21 t^{16} + 6 t^{18} + t^{20}
   \end{array} }
}{(1-t^2)^{30} (1-t^4)^{-1} (1-t^8)^{-1}}
$
&
$ \ldots $&$ 1 - h_2 t^8 - h_2 t^{12} + h_4 t^{12} $\\
 \hline
 $(6,1^2)$&$ 
  \begin{array}{c}
 C_4,B_2,C_2,\\
 B_1,C_1,B_0
   \end{array}
 $&$ 28 $&
 $
\frac{
\tiny{ \begin{array}{c}
1 + 6 t^2 + 21 t^4 + 56 t^6 + 126 t^8 + 252 t^{10}\\
 + 126 t^{12} + 56 t^{14} + 21 t^{16} + 6 t^{18} + t^{20}
   \end{array}}
}{(1-t^2)^{30}(1-t^4)^{-1} (1-t^8) ^{-1}}
$
&
$ \ldots $&$1 - h_2 t^{12} - h_2^2 t^{12} + h_1^2 h_2 t^{14}$\\
 \hline 
 $(6,2)$&$
 \begin{array}{c}
 C_4,D_3,C_2,\\
 B_1,C_1,B_0
  \end{array}
 $&$ 30 $&
 $
\frac{ 
\tiny{ \begin{array}{c}
1 + 3 t^2 + 7 t^4 + 13 t^6 + 22 t^8 + 34 t^{10}\\
 + 22 t^{12} + 13 t^{14} +7 t^{16} + 3 t^{18} + t^{20}
   \end{array}}
}{(1-t^2)^{33}(1-t^4)^{-2} (1-t^8)^{-1}}$
&
$ \ldots $&$ 1 - h_2 t^{12} $\\
 \hline
  $(8)$&$
 \begin{array}{c}
 C_4,B_3,C_3,B_2,\\
 C_2,B_1,C_1,B_0
  \end{array}
 $&$ 32 $&
 $
\frac{(1-t^4) (1-t^8) (1-t^{12}) (1-t^{16})}{(1-t^2)^{36}}
$
&
$ \ldots $&$ 1 $\\
 \hline
\end{tabular}
\end{center}
\label{tab:Cseries2}
\text{B/D gauge nodes in a quiver indicate averages over the corresponding O gauge groups}\\
\text{An mHL HWG of 1 denotes $mHL^C_{[0, \ldots, 0]}(t^2)$.}\\
\text{Some character and mHL HWGs have been omitted for brevity.}\\
\text{See text for a discussion of the non-palindromic $C_4-D_2-C_1$ quiver.}
\end{sidewaystable}

\begin{sidewaystable}[htp]
\caption{Quivers for Nilpotent Orbits of $D_2$ and $D_3$}
\begin{center}
\footnotesize
\begin{tabular}{|c|c|c|c|c|c|}
 \hline
 $\text{Orbit} $&$ \text{Quiver} $&$ \text{Dim.} $&$ \text{Hilbert Series} $&$ \text{Character HWG} $&$ \text{mHL HWG} $\\
 \hline
 $(1^4)$&$ D_2 $&$ 0 $&$ 1 $&$ 1 $&$1-{h_2}^2 t^2-{h_1}^2 t^2+ {h_1}^2 {h_2}^2 t^4 $\\
 \hline
 $(2^2)^{I/II}$&$ D_2,C_1 $&$ 2 $&$ \frac{1+4 t^2-t^4}{(1-t^2)^2} $&$ \frac{1-{m_1}^2 {m_2}^2 t^4}{(1-{m_1}^2 t^2) (1-{m_2}^2 t^2)} $&$ 1-{h_1}^2 {h_2}^2 t^4 $\\
 \hline
 $(3,1)$&$ D_2,C_1,B_0 $&$ 4 $&$ \frac{(1-t^4)^2}{(1-t^2)^6} $&$ \frac{1}{(1-{m_1}^2 t^2) (1-{m_2}^2 t^2)} $&$ 1 $\\
 \hline
 \hline
 $(1^6)$&$ D_3 $&$ 0 $&$ 1 $&$ 1 $&$
 \begin{array}{c}
1 - {h_2}{h_3}{t^2} - {h_2}{h_3}{t^4}\\
 + {h_1}{h_3}^2{t^4} + {h_1}{h_2}^2{t^4} - {h_2}{h_3}{t^6}\\
 - {h_1}^2{h_2}{h_3}{t^6} + {h_1}{h_3}^2{t^6} - {h_2}^2{h_3}^2{t^6}\\
 + {h_1}{h_2}^2{t^6} + {h_1}^2{t^6} - {h_3}^4{t^6}\\
 - {h_2}^4{t^6} - {h_1}^2{h_2}{h_3}{t^8} + {h_1}{h_2}^3{h_3}{t^8}\\
 - {h_2}^2{h_3}^2{t^8} + {h_1}{h_2}{h_3}^3{t^8} + {h_1}^4{t^8}\\
 + {h_1}{h_2}^3{h_3}{t^{10}} - {h_1}^3{h_3}^2{t^{10}} - {h_1}^3{h_2}^2{t^{10}}\\
 + {h_1}{h_2}{h_3}^3{t^{10}} - {h_2}^3{h_3}^3{t^{10}} + {h_1}^2{h_2}^2{h_3}^2{t^{12}}
\end{array}
 $\\
 \hline
 $(2^2,1^2)$&$ D_3,C_1 $&$ 6 $&$ \frac{1+8 t^2+t^4}{(1-t^2)^6 (1+t^2)^{-1}} $&$ \frac{1}{1-{m_2} {m_3} t^2} $&$
 \begin{array}{c}
1 - {h_2}{h_3}{t^4} - {h_1}^2{t^4}\\
 - {h_2}{h_3}{t^6} + {h_1}{h_3}^2{t^6}\\
 + {h_1}{h_2}^2{t^6} + {h_1}^2{t^6} - {h_2}^2{h_3}^2{t^8}
\end{array}
 $\\
 \hline
 $(3,1^3)$&$ D_3,C_1,B_0 $&$ 8 $&$ \frac{1+5 t^2+t^4}{(1-t^2)^8 (1+t^2)^{-2}} $&$ \frac{1}{(1-{m_2} {m_3} t^2) (1-{m_1}^2 t^4)} $&$ 1-{h_2} {h_3} t^4+{h_1}^2 t^6-{h_2} {h_3} t^6 $\\
 \hline
 $(3^2)$&$ D_3,C_2,D_1 $&$ 10 $&$ \frac{1+4 t^2+10 t^4+4 t^6+t^8}{(1-t^2)^{10}(1+t^2)^{-1}} $&$\scriptsize{ \frac{1-{m_1}^2 {m_2}^2 {m_3}^2 t^{12}}{
  \left( \begin{array}{c}
  (1-{m_2} {m_3} t^2) (1-{m_1}^2 t^4) (1-{m_2} {m_3} t^4)\\ \times (1-{m_1} {m_2}^2 t^6) (1-{m_1} {m_3}^2 t^6)
   \end{array}\right) } }
 $&$ 1-{h_2} {h_3} t^6 $\\
 \hline
$(5,1)$&$ D_3,C_2,B_1,C_1,B_0 $&$ 12 $&$ \frac{(1-t^4) (1-t^6) (1-t^8)}{(1-t^2)^{15}} $&$\ldots $&$ 1 $\\
 \hline
\end{tabular}
\end{center}
\label{tab:Dseries1}
\text{B/D gauge nodes in a quiver indicate averages over the corresponding O gauge groups}\\
\text{An mHL HWG of 1 denotes $mHL^D_{[0, \ldots, 0]}(t^2)$.}\\
\text{Some character and mHL HWGs have been omitted for brevity.}\\
\text{See text for a discussion of the non-palindromic $D_2-C_1$ spinor pair quiver.}
\end{sidewaystable}


\begin{sidewaystable}[htp]
\caption{Quivers for Nilpotent Orbits of $D_4$}
\begin{center}
\footnotesize
\begin{tabular}{|c|c|c|c|c|c|}
 \hline
 $\text{Orbit} $&$ \text{Quiver} $&$ \text{Dim.} $&$ \text{Hilbert } \text{Series} $&$ \text{Character } \text{HWG} $&$ \text{mHL } \text{HWG} $\\
\hline
$(1^8)$&$D_4$&$ 0 $&$ 1 $&$ 1 $&$ \ldots $\\
 \hline
 
 $(2^2,1^4)$&$D_4,C_1$&$10 $&$
\frac{1 + 17 t^2 + 48 t^4 + 17 t^6 + t^8}{(1-t^2)^{10}(1+t^2)^{-1} }
 $&$
\frac{1}{1-{m_2} t^2}
 $&$ \ldots $\\
 \hline
 
 $(2^4)^{I/II}$&$D_4,C_2$&$ 12 $&$
\frac{1 + 15 t^2 + 85 t^4 + 162 t^6 + 15 t^8 - 13 t^{10} - t^{12}}{(1-t^2)^{12}(1+t^2)^{-1}}
$&$
\frac{1-{m_3}^2 {m_4}^2 t^8}{(1-{m_2} t^2) (1-{m_3}^2 t^4) (1-{m_4}^2 t^4)}
$&$\ldots $\\
 \hline
 
$(3,1^5)$&$D_4,C_1,B_0$&$ 12 $&
 $
\frac{ 1 + 14 t^2 + 36 t^4 + 14 t^6 + t^8)}{(1-t^2)^{12} (1+t^2)^{-2}} $
 &$
 \frac{1}{(1-{m_1}^2 t^4) (1-{m_2} t^2)}
 $&$\ldots $\\
 \hline
 
$(3,2^2,1)$&$D_4,C_2,B_0$&$ 16 $&
$
\frac{
\scriptsize{ \begin{array}{c}
1 + 12 t^2 + 77 t^4 + 296 t^6 + 476 t^8 \\
+ 296 t^{10} + 77 t^{12} + 12 t^{14} + t^{16}
   \end{array}}
}{(1-t^2)^{16}}$
&$

\frac{1+{m_1} {m_3} {m_4} t^6}{
 \left( \scriptsize{ \begin{array}{c}

(1-{m_1}^2 t^4) (1-{m_2}t^2) \\
\times (1-{m_3}^2 t^4) (1-{m_4}^2 t^4)
 \end{array}} \right)}

$
&$
 \begin{array}{c}
 1 - 2 h_2 t^6 + h_1^2 t^8 - h_2^2 t^8 \\
 + h_3^2 t^8 + h_4^2 t^8 - h_2 t^{10} \\
 + h_1^2 h_2 t^{10} + h_2 h_3^2 t^{10} \\
+ h_2 h_4^2 t^{10} - h_1^2 h_3^2 t^{12}\\
 - h_1^2 h_4^2 t^{12} - h_3^2 h_4^2 t^{12} \\
 - h_1 h_3 h_4 t^{14} + h_1 h_2 h_3 h_4 t^{16}
 \end{array}$\\
 \hline
 
 $(3^2,1^2)$&$D_4,C_2,D_1$&$ 18 $&$
\frac{
\scriptsize{ \begin{array}{c}
1 + 9 t^2 + 45 t^4 + 109 t^6 + 152 t^8\\
 + 109 t^{10} + 45 t^{12} + 9 t^{14} + t^{16}
   \end{array}}
}{(1-t^2)^{18} (1+t^2)^{-1}} $
 &
 $
 \frac{1+{m_1} {m_2} {m_3} {m_4} t^{10}}{
 \scriptsize{\left( \begin{array}{c}
 (1-{m_1}^2 t^4) (1-{m_2} t^2)\\
 \times (1-{m_2}^2 t^8) (1-{m_3}^2 t^4) \\
 \times (1-{m_4}^2 t^4) (1-{m_1} {m_3} {m_4} t^6)
 \end{array}\right)}
 }
 $
 &$
  \begin{array}{c}
 1 - 2 h_2 t^6 + h_1^2 t^8 + h_3^2 t^8 \\
 + h_4^2 t^8 - h_2 t^{10} - h_1 h_3 h_4 t^{14} 
  \end{array}
 $\\
 \hline
 
$(4^2)^{I/II}$&$D_4,C_3,D_2,C_1$&$ 20 $&
$
\frac{
\scriptsize{\begin{array}{c}
1 + 7 t^2 + 28 t^4 + 84 t^6 + 173 t^8 + 238 t^{10} \\
+ 133 t^{12} + 28 t^{14} - 14 t^{16} - 5 t^{18} - t^{20}
   \end{array}}
}{(1-t^2)^{20}(1+t^2)^{-1} }
 $
 &$
\ldots
 $&
 $
1 - h_1^2 t^8 - h_2 t^{10} + h_1 h_3 h_4 t^{14}
 $\\
 \hline
 
 $(5,1^3)$&$D_4,C_2,B_1,C_1,B_0$&$ 20 $&
 $
\frac{
\scriptsize{ \begin{array}{c}
1 + 6 t^2 + 21 t^4 + 28 t^6 \\
+ 21 t^8 + 6 t^{10} + t^{12} 
  \end{array}}
}{(1-t^2)^{20}(1+t^2)^{-2} (1+t^4)^{-1}}
 $
&
$ \ldots $&$
1 - h_2 t^6 + h_1^2 t^8 - h_2 t^{10}
 $\\
 \hline
 
 $(5,3)$&$D_4,C_3,D_2,C_1,B_0$&$ 22 $&
 $
\frac{ 
\scriptsize{ \begin{array}{c}
1 + 3 t^2 + 8 t^4 + 16 t^6 + 28 t^8 \\
+ 16 t^{10} + 8 t^{12} + 3 t^{14} + t^{16} 
  \end{array}}
}{(1-t^2)^{22} (1+t^2)^{-3}}
$
&
$ \ldots $&$
1 - h_2 t^{10} $\\
 \hline
 
 $(7,1)$&$
 \begin{array}{c}
 D_4,C_3,B_2,\\
 C_2,B_1,C_1,B_0
 \end{array}
 $&$ 24 $&
 $
\frac{ (1-t^4) (1-t^8)^2 (1-t^12)}{(1-t^2)^{28}}
$
&
$ \ldots $&$
1 $\\
 \hline
 
\end{tabular}
\end{center}
\label{tab:Dseries2}
\text{B/D gauge nodes in a quiver indicate the corresponding O gauge groups}\\
\text{An mHL HWG of 1 denotes $mHL^D_{[0, \ldots, 0]}(t^2)$.}\\
\text{Some character and mHL HWGs have been omitted for brevity.}\\
\text{See text for a discussion of the non-palindromic $D_4-C_2$ and $D_4-C_3-D_2-C_1$  spinor pair quivers.}
\end{sidewaystable}

It is noteworthy that, for all the $BCD$ nilpotent orbit partitions, this construction yields moduli spaces that (i) have the correct dimensions, (ii) are unchanged under the usual group isomorphisms, (iii) have character expansions that are free of singlets (i.e. satisfy the vacuum conditions) and (iv) decompose into finite sums of modified Hall Littlewood polynomials. There are inclusion relations between the moduli spaces that can be read off either from the character HWGs or from the subgroup relations amongst the quivers. These confirm that all the lower dimensioned moduli spaces are contained in both the maximal and sub-regular nilpotent orbits. Almost all the moduli spaces have palindromic Hilbert series and we comment on those that do not below.

\FloatBarrier

In the case of $D$ groups of even rank, this construction does not yield palindromic moduli spaces for those nilpotent orbits associated with pairs of spinor partitions. Specifically, as can be seen from appendix  \ref{apxD}, the orbits with vector partitions $\{ (2^2), (2^4), (4^2) \}$ all correspond to pairs of orbits distinguished by their spinor partition data. While we can identify palindromic moduli spaces associated with each of the spinors, the union of these spaces is non-palindromic. Since the method of nilpotent orbit construction, which is based on bi-fundamental fields transforming in the vector representation, is symmetric with respect to the spinors, it naturally yields this union of two spinor moduli spaces. In the case of $D_2$, the palindromic 2 dimensional moduli spaces are provided by the 2 dimensional nilpotent orbits of the Weyl spinors, analysed in section \ref{sec:Aseries}. In the case of $D_4$, we can obtain 12 and 20 dimensional palindromic moduli spaces by applying triality to the palindromic moduli spaces from the nilpotent orbits with vector partitions $\{3,1^5\}$ and $\{5,1^3\}$. We describe these relations between moduli spaces in \ref{eq:BCDseries10a}, \ref{eq:BCDseries10b} and \ref{eq:BCDseries10c}. The algebraic relations hold equally well for all the types of moduli space description; Hilbert series, character HWG and mHL HWG.
\begin{equation} 
\label{eq:BCDseries10a}
\begin{aligned}
g_{({2^2})}^{{D_2}} & = g_{(2)}^{{A_1}} \otimes g_{({1^2})}^{{A_1}} + g_{({1^2})}^{{A_1}} \otimes g_{(2)}^{{A_1}} - g_{({1^4})}^{{D_2}}\\
 \end{aligned}
\end{equation}

\begin{equation} 
\label{eq:BCDseries10b}
\begin{aligned}
g_{({2^4})}^{{D_4}} & = {\left. {g_{(3,{1^5})}^{{D_4}}} \right|_{\scriptstyle{m_1} \Leftrightarrow {m_3}\hfill \atop
\scriptstyle{h_1} \Leftrightarrow {h_3}\hfill}} + {\left. {g_{(3,{1^5})}^{{D_4}}} \right|_{\scriptstyle{m_1} \Leftrightarrow {m_4}\hfill \atop
\scriptstyle{h_1} \Leftrightarrow {h_4}\hfill}} - g_{({2^2},{1^4})}^{{D_4}}\\
 \end{aligned}
\end{equation}

\begin{equation} 
\label{eq:BCDseries10c}
\begin{aligned}
g_{({4^2})}^{{D_4}} & = {\left. {g_{(5,{1^3})}^{{D_4}}} \right|_{\scriptstyle{m_1} \Leftrightarrow {m_3}\hfill\atop
\scriptstyle{h_1} \Leftrightarrow {h_3}\hfill}} + {\left. {g_{(5,{1^3})}^{{D_4}}} \right|_{\scriptstyle{m_1} \Leftrightarrow {m_4}\hfill \atop
\scriptstyle{h_1} \Leftrightarrow {h_4}\hfill}} - g_{({3^2},{1^2})}^{{D_4}}
 \end{aligned}
\end{equation}
In all these cases, the non-palindromic moduli space is the union of two palindromic moduli spaces (i.e. their sum less their intersection, given by the palindromic nilpotent orbit of lower dimension). We anticipate this analysis generalises to $D_{2n}$.

There are three remaining non-palindromic moduli spaces of $BCD$ groups up to rank 4, generated by the quivers $B_3-C_2-B_0$, $B_4-C_3-B_1-C_1-B_0$ and $C_4-D_2-C_1$. We can identify relationships between these non-palindromic quivers and the non-palindromic {\it spinor pair} quivers of $D_{2n}$ discussed above. Specifically, 
\begin{enumerate}
\item the quivers $B_3-C_2-B_0$ and $B_4-C_3-B_1-C_1-B_0$ are related to the non-palindromic $D_4-C_2$ and $D_6-C_3$, under character maps between vector representations $D_4 \to B_3 \otimes B_0$ and $D_6 \to B_4 \otimes B_1$, and

\item$C_4-D_2-C_1$ contains the non-palindromic $D_2-C_1$ as a subchain.
\end{enumerate}
These non-palindromic nilpotent orbits of classical groups up to rank 4 are precisely those tabulated as being unions of orbits in \cite{Kraft:1982fk} based on a geometric analysis. We anticipate that this structure extends to higher rank groups.

\begin{sidewaystable}[htp]
\caption{Generalised $B$ Series Nilpotent Orbit Moduli Spaces}
\begin{center}
\begin{tabular}{|c|c|c|c|c|c|}
\hline
${\text{Orbit}}$&${\text{Dimension}}$&$\text{Quiver}$&$\text{Hilbert Series}$&$\text{PL[Character HWG]}$&$\text{mHL HWG}$\\
\hline
$\text{Trivial} $&$0$&$B_n $&$1$&$1$&$ \ldots $ \\
\hline
$\text{Minimal}$&${4n-4}$&$B_n, C_1$&$ \ldots $&$m_2 t^2 $&$ \ldots $\\
\hline
$\text{
Supra~Minimal}$&$4n-2$&$B_n, C_1, B_0$&$ \ldots $&$ m_2 t^2 + m_1^2 t^4 $&$ \ldots $\\
\hline
\hline
$\begin{array}{c} \text{2-Node Quiver}\\n > 2k\end{array}$
&
$4k(n-k)$
&
$B_n-C_k $
&
$ \ldots $
&
$
{\sum\limits_{i = 1}^k {{m_{2i}}} {t^{2i}}}
$
&
 $ \ldots $\\
\hline
$\begin{array}{c} \text{2-Node Quiver}\\n = 2k \end{array}$
&
$4k(n-k)$
&
$B_n-C_k $
&
$ \ldots $
&
${\sum\limits_{i = 1}^{k-1} {{m_{2i}}} {t^{2i}} + m_n^2{t^n}}$
 &
 $ \ldots $\\
\hline
\hline
$\text{Sub~Regular}$&$2n^2-2$&$B_n,C_{n-1},\ldots, B_0$&$ \ldots $&$ \ldots $&${1 - {h_1}{t^{2n}}}$\\
\hline
$\text{Maximal}$&$2n^2$&$B_n,C_n,\ldots, B_0$&$
\frac{{\prod\limits_{i = 1}^n {\left( {1 - {t^{4i}}} \right)} }}{{{{\left( {1 - {t^2}} \right)}^{n\left( {2n + 1} \right)}}}}
$&$ \ldots $&$1$\\
\hline
\end{tabular}
\end{center}
\label{tab:Bseries3}
\text {B/D gauge nodes in a quiver indicate the corresponding O gauge groups.}\\
\text{The maximal and sub-regular quivers contain maximal chains of O/USp gauge groups,}\\
\text{in which substitutions between $B_{k-1}$ and $D_{k}$ gauge groups are permitted.}
\text{Assumes rank $> 2$}
\end{sidewaystable}%

\begin{sidewaystable}[htp]
\caption{Generalised $C$ Series Nilpotent Orbit Moduli Spaces}
\begin{center}
\begin{tabular}{|c|c|c|c|c|c|}
\hline
$\text{Orbit}$&$\text{Dimension}$&$\text{Quiver}$&$\text{Hilbert Series}$&$\text{PL[Character HWG]}$&$\text{mHL HWG}$\\
\hline
$\text{Trivial} $&$0$&$ C_n $&$1$&$1$&$ \ldots $ \\
\hline
$\text{Minimal}$&$ 2n$&$ C_n-B_0 $&$ \ldots $&$m_1^2 t^2 $&$ \ldots $\\
\hline
$
\begin{array}{c}
\text{Supra}\\
\text{Minimal}\\
n \ge 2
\end{array}
$&$
4n-2
$&$ C_n-D_1 $&$ \ldots $&$ m_1^2 t^2 + m_2^2 t^4 $&$ \ldots $\\
\hline
$
\begin{array}{c}
\text{Supra}\\
\text{Supra}\\
\text{Minimal}\\
n \ge 3
\end{array}
$&$ 6n-6 $&$ C_n-B_1 $&$ \ldots $&$ m_1^2 t^2 + m_2^2 t^4+ m_3^2 t^6 $&$ \ldots $\\
\hline
\hline
$\begin{array}{c} \text{2-Node Quiver}\\n \ge k\end{array}
$&
$ k(2n-k+1) $&$C_n-O_k $&$ \ldots $&
$\sum\limits_{i = 1}^{k} {m_{i}^2}{t^{2i}}$
&$ \ldots $\\
\hline
\hline
$\text{Sub-Regular}$&$ 2n^2-2
$&$
\begin{array}{c}
C_n,D_{n-1},C_{n-2}\\
\ldots D_2,C_1,D_1 
\end{array}
$&$ \ldots $&$ \ldots $&${1 - {h_2}{t^{4n-4}}}$\\
\hline
$\text{Maximal}$&$ 2n^2  $&$
\begin{array}{c}
C_n,B_{n-1},C_{n-1}\\
\ldots B_1,C_1,B_0
\end{array}
$&$
\frac{{\prod\limits_{i = 1}^n {\left( {1 - {t^{4i}}} \right)} }}{{{{\left( {1 - {t^2}} \right)}^{n\left( {2n + 1} \right)}}}}
$&$ \ldots $&$1$\\
\hline
\end{tabular}
\end{center}
\label{tab:Cseries3}
\text {B/D gauge nodes in a quiver indicate the corresponding O gauge groups.}\\
\text{The maximal and sub-regular quivers contain maximal chains of O/USp gauge groups,}\\
\text{in which substitutions between $B_{k-1}$ and $D_{k}$ gauge groups are permitted.}
\end{sidewaystable}%

\begin{sidewaystable}[htp]
\caption{Generalised $D$ Series Nilpotent Orbit Moduli Spaces}
\begin{center}
\begin{tabular}{|c|c|c|c|c|c|}
\hline
$\text{Orbit}$&$\text{Dimension}$&$\text{Quiver}$&$\text{Hilbert Series}$&$\text{PL[Character HWG]}$&$\text{mHL HWG}$\\
\hline
$\text{Trivial} $&$0$&$D_n $&$1$&$1$&$ \ldots $ \\
\hline
$\text{Minimal}$&$4n-6$&$D_n, C_1$&$ \ldots $&$m_2 t^2 $&$ \ldots $\\
\hline
$\text{Supra~Minimal}$&$4n-4$&$D_n, C_1, B_0$&$ \ldots $&$ m_2 t^2 + m_1^2 t^4 $&$ \ldots $\\
\hline
\hline
$\begin{array}{c} \text{2-Node Quiver}\\n \ge 2k+2 \end{array}$
&$2k(2n-2k-1)$&$D_n-C_k $&$ \ldots $&${\sum\limits_{i = 1}^k {{m_{2i}}} {t^{2i}}}$&$ \ldots $\\
\hline
$\begin{array}{c} \text{2-Node Quiver}\\n = 2k+1 \end{array}$
&
$2k(2n-2k-1)$
&
$D_n-C_k $
&
$ \ldots $
&
$
{\sum\limits_{i = 1}^{k-1} {{m_{2i}}} {t^{2i}} + {\rm{ }}{m_{n - 1}}{m_n}{t^{n-1}}}
 $
 &
 $ \ldots $\\
\hline
$\begin{array}{c} \text{2-Node Quiver}\\n = 2k \end{array}$
&
$2k(2n-2k-1)$
&
$D_n-C_k $
&
$ \ldots $
&
$\begin{array}{c}
\sum\limits_{i = 1}^{k-1} {{m_{2i}}} {t^{2i}} 
+ m_{n - 1}^2{t^n} \\+ m_n^2{t^n} - m_{n - 1}^2m_n^2{t^{2n}}
\end{array}
$
 &
 $ \ldots $\\
\hline
\hline
$\text{Sub~Regular}$&$2n(n-1)-2$&$D_n,C_{n-2},\ldots, D_1$&$ \ldots $&$ \ldots $&${1 - {h_1}{t^{2n}}}$\\
\hline
$\text{Maximal}$&$2n(n-1)$&$D_n,C_{n-1},\ldots, D_1$&$
 \frac{{\left( {1 - {t^{2n}}} \right)\prod\limits_{i = 1}^{n - 1} {\left( {1 - {t^{4i}}} \right)} }}{{{{\left( {1 - {t^2}} \right)}^{n\left( {2n - 1} \right)}}}}
$&$ \ldots $&$1$\\
\hline
\end{tabular}
\end{center}
\text {B/D gauge nodes in a quiver indicate the corresponding O gauge groups.}\\
\text{The maximal and sub-regular quivers contain maximal chains of O/USp gauge groups,}\\
\text{in which substitutions between $B_{k-1}$ and $D_{k}$ gauge groups are permitted.}
\text{Assumes rank  $\ge 3$}
\label{tab:Dseries3}
\end{sidewaystable}%

\FloatBarrier

Based on the analysis, we can generalise the structure and representation content of the Hilbert series for certain characteristic nilpotent orbits to higher rank groups, as set out in tables \ref{tab:Bseries3}, \ref{tab:Cseries3} and \ref{tab:Dseries3}. The minimal nilpotent orbit is the RSIMS, as discussed earlier. For B/D groups, the supra-minimal nilpotent orbit has dimension two more than the minimal and its characters are generated by the adjoint representation and the graviton representation. The maximal orbit is a complete intersection \cite{Hanany:2011db} and the sub-regular nilpotent orbit differs from the maximal nilpotent orbit by $mHL^{B/D}_{[1,0,\ldots]} t^{2n}$ or $mHL^{C}_{[0,1,0,\ldots]} t^{4n-4}$. For the $C$ series, we can generalise the structure of nilpotent orbits further inside the body of the Hasse diagram.

Interestingly, we can also generalise, to any rank, the character HWGs for all $O/USp$ quivers with only two nodes. The patterns of HWG generators for 2-node quivers with $SO$ flavour groups follow from the antisymmetric invariants of even degree of $USp$ fundamentals; the patterns for 2-node quivers with $USp$ flavour groups follow from the invariants of mixed symmetry of $O$ vectors \cite{Hanany:2014dia, Benvenuti:2010pq, Hanany:2008kn}. While there are several similarities between the forms of these HWGs for $B_n$ and $D_n$ flavour groups, there are differences in relation to the appearance of spinors, as can be seen from tables \ref{tab:Bseries3} and \ref{tab:Dseries3}.
\subsection{Coulomb Branch and Mirror Symmetry: $BCD$ Series}
\label{subsec:$BCD$ Mirrors}
Quivers whose Coulomb branches yield the moduli spaces of minimal nilpotent orbits of $BCD$ groups are known, being given by extended or untwisted affine Dynkin diagrams, as discussed in \cite{Cremonesi:2014xha, Intriligator:1996ex}. By principles of $3d$ mirror symmetry, these Coulomb branch quivers may be mirror dual to the Higgs branch quivers for minimal nilpotent orbits of $BCD$ groups analysed above.

The structure of Coulomb branch quivers for general $BCD$ nilpotent orbits is, however, problematic for a number of reasons. Firstly, the proposals for mirror symmetric duals of $BCD$ Higgs branch linear quivers via brane manipulations \cite{Cremonesi:2013lqa, Cremonesi:2014uva, Feng:2000eq} can lead to Coulomb branch quivers with non-unitary gauge nodes that are not equal in number to the simple roots of the $BCD$ group, and which cannot therefore be calculated using the monopole formula with unitary gauge nodes. Secondly, while versions of the monopole formula with non-unitary gauge nodes have been proposed \cite{Cremonesi:2014kwa}, these have not been successful at generating moduli spaces whose refined Hilbert series match those of the purported mirrors. Indeed, one method currently used for working with the moduli spaces of Coulomb branch $BCD$ quivers for maximal nilpotent orbits ($T(G)$ theories) \cite{Cremonesi:2014kwa, Cremonesi:2014uva} is simply to bypass the problem, by conjecturing the equivalence of the unknown quivers to $BCD$ modified Hall Littlewood polynomials.

This contrasts with the situation for the $A$ series nilpotent orbits, where all the Higgs branch quivers have Coulomb branch mirrors \cite{Hanany:1996ie}, with gauge nodes equal in number to the simple roots, which can be evaluated using the unitary monopole formula to obtain identical Hilbert series. Accordingly, we aim to find Coulomb branch constructions for general $BCD$ series nilpotent orbits in a manner consistent with the $A$ series.

Affine Dynkin diagrams play a pivotal role in the calculation of relationships between moduli spaces. They are instrumental in the Coulomb branch construction of RSIMS \cite{Cremonesi:2014xha}. They encode subgroup branching relationships \cite{Dynkin:1957um}. They encode the logic of gluing constructions, whereby Coulomb branch quivers can be obtained by combining modified Hall Littlewood polynomials  \cite{Gadde:2011uv, Hanany:2015hxa}. In this section we shall show how {\it twisted} affine Dynkin diagrams permit the construction of the moduli spaces of some $BCD$ nilpotent orbits above the minimal nilpotent orbit. We start with a brief recapitulation of twisted affine Dynkin diagrams and of the monopole formula.

\subsubsection{Twisted Affine Dynkin Diagrams}
Affine Dynkin diagrams encode a particular class of degenerate extensions of the Cartan matrix of a Lie algebra. They correspond to those Dynkin diagrams that can be obtained by attaching a single extra node to the regular Dynkin diagram of a group, subject to the constraints (i) that the links are of a type permitted in a regular Dynkin diagram and (ii) that the resulting Cartan matrix, which acquires an extra row and column, is positive semi-definite, having one zero eigenvalue. 

\begin{figure}[htbp]
\begin{center}
\includegraphics[scale=0.40]{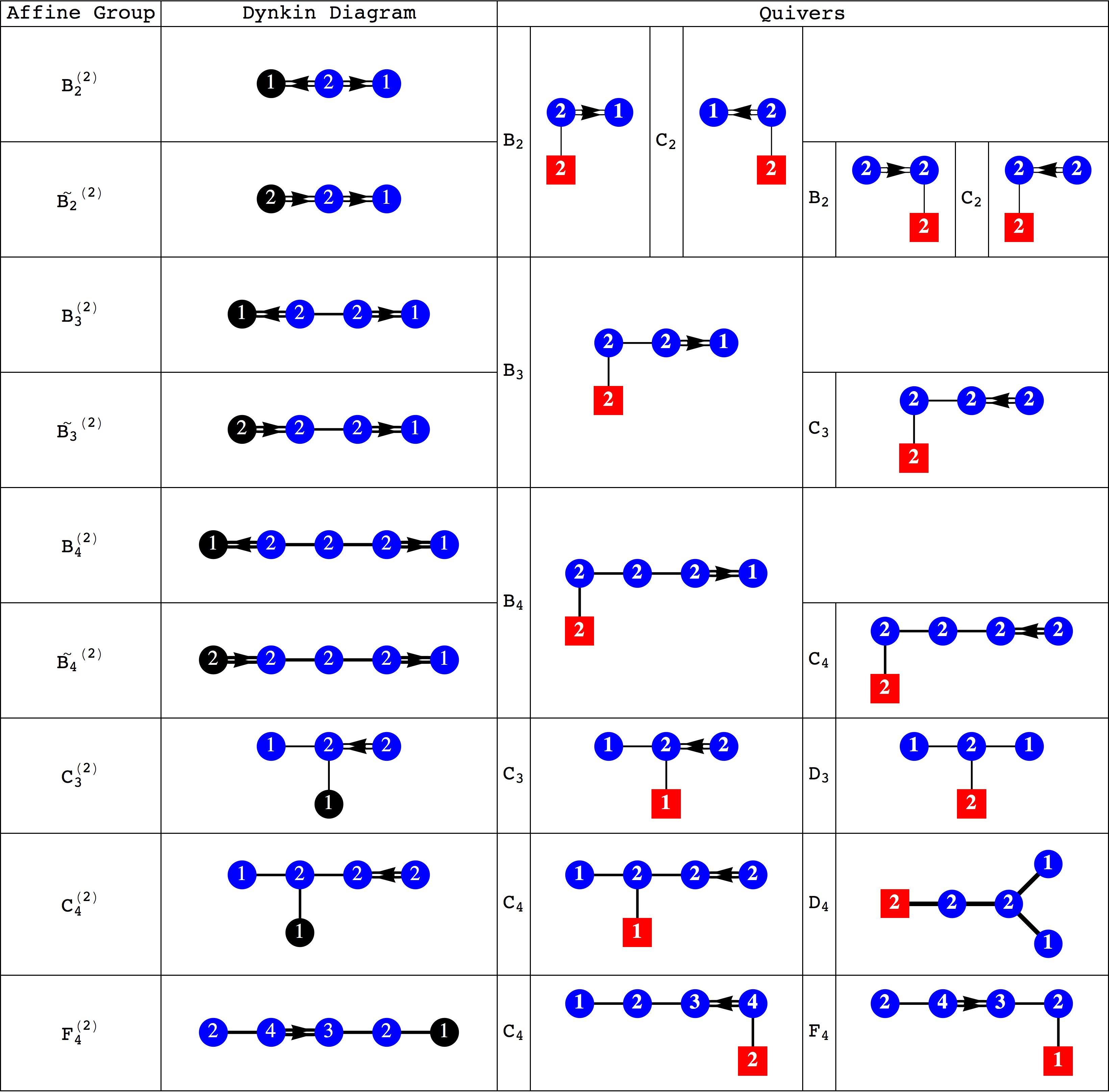}\\
\caption[Quivers from $BCF$ Series Twisted Affine Dynkin Diagrams]{Quivers from $BCF$ Series Twisted Affine Dynkin Diagrams. The affine groups are labelled using the notation of \cite{Fuchs:1997bb}. Round (blue) nodes denote gauge nodes in the regular Dynkin diagram. The twisted affine diagram is obtained by adding a gauge node (black). The dual Coxeter labels of each gauge node are shown. Square (red) nodes in a quiver denote flavour nodes. When a short root attached to a long root in the affine diagram is taken as the flavour node in a quiver, its rank is doubled.}
\label{fig:twistedDynkins}
\end{center}
\end{figure}

In a normal affine or extended Dynkin diagram, the extra node is attached to the adjoint node of the regular Dynkin diagram and the dual Coxter labels of existing nodes are unchanged, with the new node acquiring a dual Coxeter label of 1. This follows from the dual Coxeter labels of the affine Dynkin diagram being the column eigenvector of the affine Cartan matrix with zero eigenvalue (or kernel). In a {\it twisted} affine Dynkin diagram, however, the extra node is attached to some other node of the regular Dynkin diagram \cite{Fuchs:1997bb} and the dual Coxeter labels follow as the kernel of the twisted affine Cartan matrix. A twisted affine Cartan matrix takes the form:
\begin{equation} 
\label{eq:BCDseries11}
\begin{aligned}
{A^{ij}_{Affine}} & = \left( {\begin{array}{*{20}{c}}{{A^{ij}}}&{[ {col} ]}\\{ - [ {irrep} ]}&2\end{array}} \right),
 \end{aligned}
\end{equation}
where the column vector $[col]$ is obtained by transposing the Dynkin labels of $[irrep]$ and replacing any non-zero entries with one of $\{-1,-2,-3,-4\}$, such that ${A^{ij}_{Affine}}$ becomes positive semi-defininite. There are six types of twisted affine Dynkin diagram, with three of these, $B^{(2)}_n, \tilde B^{(2)}_n$ and $C^{(2)}_n$, forming infinite families, plus three unique cases, $A^{(2)}_1, F^{(2)}_4$ and $G^{(3)}_2$. Figure \ref{fig:twistedDynkins} shows the $BCF$ twisted affine Dynkin diagrams, relevant to our study, using the naming convention of \cite{Fuchs:1997bb}.

The degeneracy of an affine Dynkin diagram permits us to make a gauge choice and to eliminate one of the nodes. The other nodes then become the nodes of a new Dynkin diagram. By judicious elimination, we can obtain a simple algebra of the same rank as that of the starting algebra. The node that is eliminated is treated as a flavour node (with zero background charge) in the new quiver diagram. Figure \ref{fig:twistedDynkins} shows the branching options for some twisted affine Dynkin diagrams,\footnote{The corresponding analysis for normal affine or extended Dynkin diagrams was set out in \cite{Hanany:2015hxa}} expressed in terms of the Coulomb branch quivers to which they give rise. The most interesting quiver diagrams for our purposes are those for the three infinite families and $F^{(2)}_4$. These lead, via the monopole formula, to the moduli spaces of certain nilpotent orbits of $BCD$ groups.

It is significant that all the Dynkin diagrams and also the gauge nodes of quiver diagrams each have zero balance, $\forall i: Balance_G(i)=0$, providing the concept of balance, introduced in section \ref{sec:Aseries} for simply laced groups, is adapted to reflect the different root lengths encoded in the off-diagonal terms of the affine Cartan matrix of $G$:
\begin{equation} 
\label{eq:BCDseries12}
\begin{aligned}
Balance_G(i) \equiv  - \sum\limits_j {{A^{ij}_{Affine~G}}} {N_j}.
 \end{aligned}
\end{equation}
As before, $N_j$ is the one dimensional kernel of $ {{A^{ij}_{Affine~G}}}$.
\FloatBarrier

\subsubsection{Monopole Formula}

The unitary monopole formula, in the absence of external charges, can be summarised as:
\begin{equation} 
\label{eq:BCDseries13}
\begin{aligned}
g^G_{Coulomb}( {{{\cal X}}(z),t} ) = \sum\limits_q^{} {{P_{U( N )}}( {q,t^2} ){z^q}{t^{2 \Delta ( q )}}},
 \end{aligned}
\end{equation}
where $q \equiv (q_{1,N_1},\ldots, q_{r, N_r})$ is a set of $U(N)$ monopole charges attaching to the simple roots with fugacities $z \equiv (z_1, \ldots,z_r)$, $P_{U(N)}(q,t^2)$ is the $U(N)$ symmetry factor following from the symmetries of each set of monopole charges $q$ and $\Delta(q)$ is their conformal dimension. The reader is referred to \cite{Hanany:2015hxa} for more detail.\footnote{Note that, {\it in this paper}, we are using $t^2$ rather than $t$ as the fugacity within the RHS of the monopole formula, to give consistency between Higgs branch and Coulomb branch constructions.} We refer to this version of the monopole formula as the {\it unitary} monopole formula, as distinct from versions that have been proposed using other gauge groups \cite{Cremonesi:2013lqa}.

As an example, we give the calculation for the $B_2$ {\it twisted} affine Dynkin diagram ${(1)\Leftarrow(2)\Rightarrow(1)}$, which is mapped to the quiver ${[2] - (2)\Rightarrow (1)}$ by taking the twisted affine node as the zero node. The monopole formula yields:
\begin{equation} 
\label{eq:BCDseries14}
\begin{aligned}
g^{B_2}_{Coulomb} ({\cal X}\left( z \right),t ) = \sum\limits_{{q_{1,1}} = - \infty }^\infty {\sum\limits_{{q_{1,2}} = - \infty }^{q_{1,1}} {\sum\limits_{{q_2} = - \infty }^\infty {P\left( {q,t^2} \right)}{{{z}}_{{1}}}^{{q_{1,1}} + {q_{1,2}}}} } {z_{2}}^{q_2}{t^{2 \Delta (q)}},
 \end{aligned}
\end{equation}
where
\begin{equation} 
\label{eq:BCDseries15}
\begin{aligned}
\Delta ( q ) = \frac{1}{2}( \left| 2{q_{1,1}} \right| + \left| 2{q_{1,2}} \right| + \left| 2{q_{1,1}} - {q_2} \right| + \left| {2{q_{1,2}} - {q_2}} \right| ) - \left| {{q_{1,1}} - {q_{1,2}}} \right|
 \end{aligned}
\end{equation}
and
\begin{equation} 
\label{eq:BCDseries16}
\begin{aligned}
{P_{U( N )}}( {q,t} ) = \left\{ \begin{array}{l}
{q_{1,1}} = {q_{1,2}}:1/\left( {{{\left( {1 - t} \right)}^2}\left( {1 - {t^2}} \right)} \right)\\
{q_{1,1}} \ne {q_{1,2}}:1/{\left( {1 - t} \right)^3}
\end{array} \right..\\
 \end{aligned}
\end{equation}
It is important to note that, under the monopole formula, the quivers ${[1] \Leftarrow (2) \Rightarrow (1)}$ and ${[2] - (2) \Rightarrow (1)}$ are equivalent for an uncharged flavour node. Evaluating the sums analytically and replacing the simple root fugacities of $B_2$ by weight space coordinates $\{z_1,z_2 \} \to \{x^2/y^2,y^2/x \}$, we obtain:
\begin{equation} 
\label{eq:BCDseries17}
\begin{aligned}
g^{B_2}_{Coulomb} (x,y,t ) =
\frac{x^3 y^4  \left(t^2+1\right) \left(t^8 x y^2+t^6 x y^2-t^4 x^2 y^2-t^4 x^2-t^4 y^4-t^4 y^2+t^2 x y^2+x y^2\right)} {\left(t^2-x\right) \left(t^2 x-1\right) \left(t^2-y^2\right) \left(t^2 y^2-1\right) \left(t^2 x-y^2\right) \left(t^2 x^2-y^2\right) \left(t^2 y^2-x\right) \left(t^2 y^2-x^2\right)}
 \end{aligned}
\end{equation}
As before, we can restate this in terms of an unrefined Hilbert series and in terms of a character HWG:
\begin{equation} 
\label{eq:BCDseries18}
\begin{aligned}
g^{B_2}_{Coulomb} (1,t) =\frac{(1+t^2) (1+3t^2+t^4)}{(1-t^2)^6},
 \end{aligned}
\end{equation}
\begin{equation} 
\label{eq:BCDseries19}
\begin{aligned}
g^{B_2}_{Coulomb} (m_1,m_2,t ) =\frac{1}{ (1-{m_2}^2 t^2) (1-{m_1}^{2} t^{4})},
 \end{aligned}
\end{equation}
Comparison with table \ref{tab:Bseries1} shows that we have obtained the moduli space for the 6 dimensional sub-regular nilpotent orbit of $B_2$.

\begin{figure}[htbp]
\begin{center}
\includegraphics[scale=0.40]{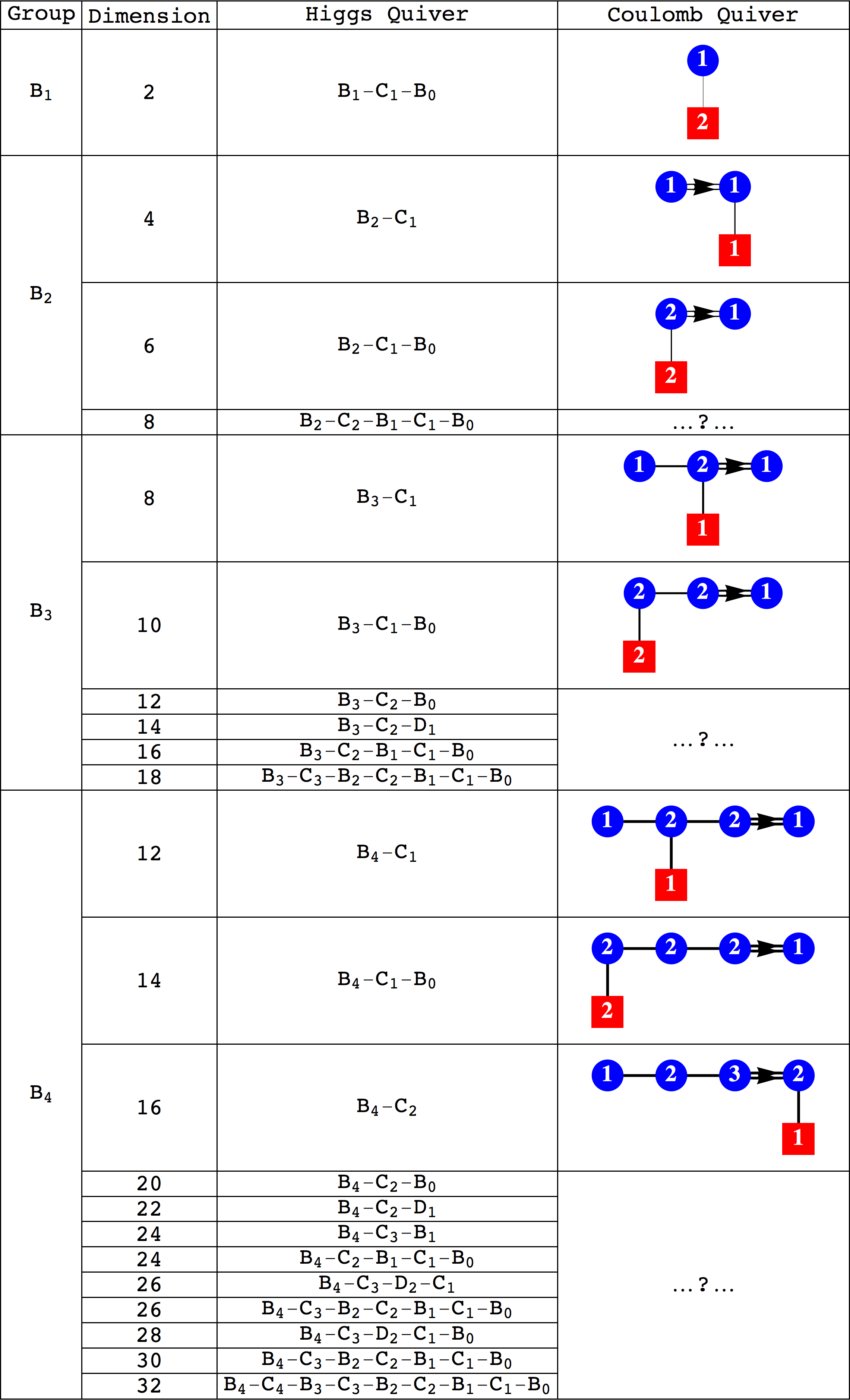}\\
\caption[Higgs/Coulomb Quivers for $B$ Series Nilpotent Orbits]{Higgs/Coulomb Quivers for $B$ Series Nilpotent Orbits up to rank 4. $B/D$ gauge nodes in a Higgs quiver indicate the corresponding O group. Round (blue) nodes denote U(N) gauge nodes. Square (red) nodes denote flavour nodes. The flavour nodes in these Coulomb branch quivers do not carry external charges. The moduli spaces defined by the Nilpotent Orbits can be calculated from either the Higgs or Coulomb branches of the dual quivers using the Higgs branch or monopole formulae, respectively.}
\label{fig:Bmirrors}
\end{center}
\end{figure}

\begin{figure}[htbp]
\begin{center}
\includegraphics[scale=0.35]{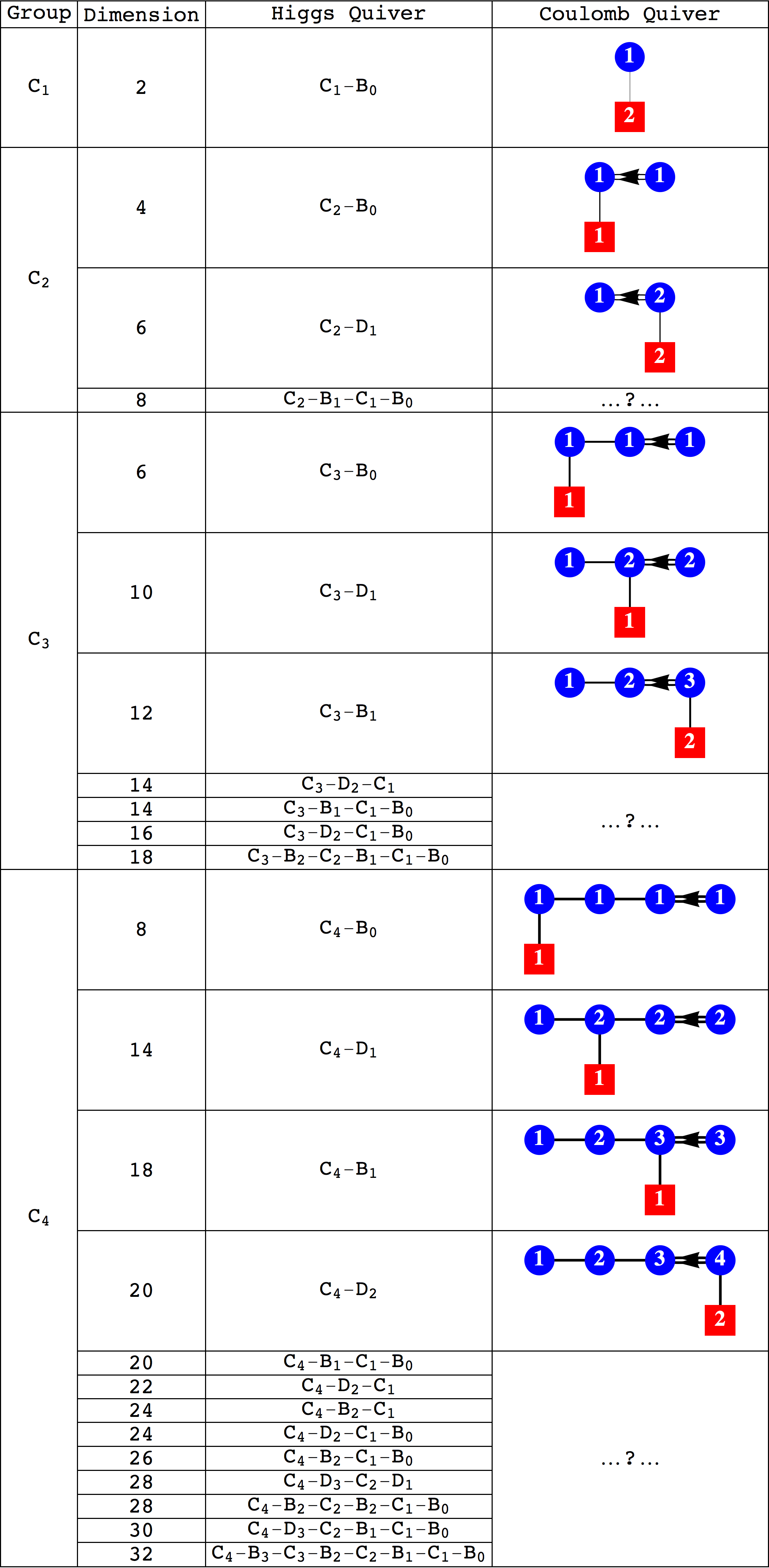}\\
\caption[Higgs/Coulomb Quivers for $C$ Series Nilpotent Orbits]{Higgs/Coulomb Quivers for $C$ Series Nilpotent Orbits up to rank 4. $B/D$ gauge nodes in a Higgs quiver indicate the corresponding O group. Round (blue) nodes denote U(N) gauge nodes. Square (red) nodes denote flavour nodes. The flavour nodes in these Coulomb branch quivers do not carry external charges. The moduli spaces defined by the Nilpotent Orbits can be calculated from either the Higgs or Coulomb branches of the dual quivers using the Higgs branch or monopole formulae, respectively.}
\label{fig:Cmirrors}
\end{center}
\end{figure}

\begin{figure}[htbp]
\begin{center}
\includegraphics[scale=0.40]{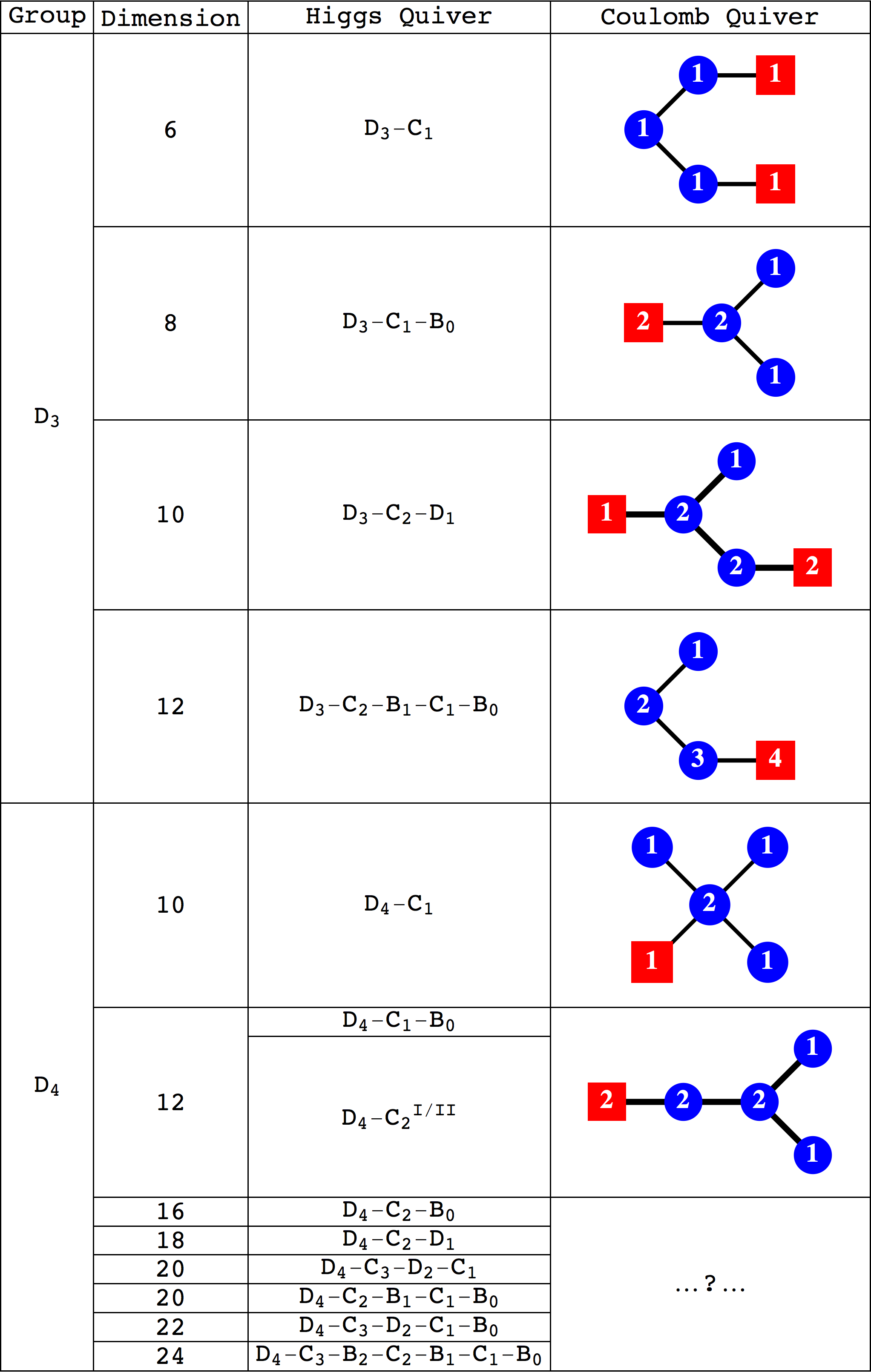}\\
\caption[Higgs/Coulomb Quivers for $D$ Series Nilpotent Orbits]{Higgs/Coulomb Quivers for $D$ Series Nilpotent Orbits up to rank 4. $B/D$ gauge nodes in a Higgs quiver indicate the corresponding O group. Round (blue) nodes denote U(N) gauge nodes. Square (red) nodes denote flavour nodes. The flavour nodes in these Coulomb branch quivers do not carry external charges. The moduli spaces defined by the Nilpotent Orbits can be calculated from either the Higgs or Coulomb branches of the dual quivers using the Higgs branch or monopole formulae, respectively. The three 12 dimensional nilpotent orbits of $D_4$ are related by triality.}
\label{fig:Dmirrors}
\end{center}
\end{figure}

We can repeat this process for the quivers identified in figure \ref{fig:twistedDynkins}. We find a match between the moduli spaces on the Coulomb branches of these quivers and those on the Higgs branches of $BCD$ linear quivers for supra-minimal nilpotent orbits. We summarise this in figures \ref{fig:Bmirrors}, \ref{fig:Cmirrors} and \ref{fig:Dmirrors}, giving the dimensions of the nilpotent orbits, their Higgs branch quivers and their equivalent Coulomb branch quivers. We also present a construction for the 20 dimensional nilpotent orbit of $C_4$, based on a rearrangement of the $F_4$ twisted affine Dynkin diagram.\footnote{This also leads to the 22 dimensional nilpotent orbit of $F_4$, which is a new construction.} For reference, we also show Coulomb branch quivers for nilpotent orbits based on untwisted affine Dynkin diagrams \cite{Hanany:2015hxa}, including the 16 dimensional nilpotent orbit of $B_4$, which is based on a rearrangement of the $F_4$ untwisted affine Dynkin diagram.

Turning to the $D_n$ nilpotent orbits associated with pairs of spinor partitions, these Coulomb branch quivers can generate palindromic moduli spaces centred on the spinor representations. Thus, in the case of $D_4$, the Coulomb branch quiver for the 12 dimensional $D_4-C_1-B_0$ nilpotent orbit is related by triality to two further 12 dimensional moduli spaces, the union of which becomes the $D_4-C_2$ spinor pair of nilpotent orbits in \ref{eq:BCDseries10b}.
\FloatBarrier
We can observe a remarkable correspondence between the numbers of the flavour nodes and gauge nodes of lower dimensional nilpotent orbits, shown in figures \ref{fig:Amirrors}, \ref{fig:Bmirrors}, \ref{fig:Cmirrors} and \ref{fig:Dmirrors}, and the respective root and weight maps, presented in appendix \ref{apxHom}. This correspondence applies for those Coulomb branch quivers whose moduli spaces have HWGs of a freely generated type, without numerator terms. Since the Coulomb branch (unitary) monopole formula leads to a moduli space whose complex dimension is exactly twice that of the sum of the unitary gauge nodes in a quiver \cite{Hanany:2015hxa}, this correspondence only appears for nilpotent orbits whose complex dimension is exactly twice that of the sum of the Dynkin labels in the nilpotent orbit weight map. 

We include in figure \ref{fig:Cmirrors} the Coulomb branch quivers for the 12 and 18 dimensional nilpotent orbits of $C_3$ and $C_4$, respectively, which can be found from Appendix  \ref{apxHom} by this rule. In the case of higher dimensioned nilpotent orbits, the moduli spaces are complicated by relations between generators, so the Coulomb branch quivers (where they are known) have gauge nodes that no longer correspond exactly to the Dynkin labels of nilpotent orbit weight maps. As a corollary, not all the quivers from twisted affine Dynkin diagrams lead to nilpotent orbits. For example, the quivers for $B$ and $C$ groups in figure \ref{fig:twistedDynkins}, in which all the gauge nodes carry U(2) monopole charges, do not match up with any nilpotent orbits.

All the quivers in figures \ref{fig:Bmirrors}, \ref{fig:Cmirrors} and \ref{fig:Dmirrors} are balanced and their moduli spaces match those of the Higgs branch constructions. We anticipate that these relationships between the Coulomb branches of quivers drawn from affine Dynkin diagrams (or the weight and root maps of SU(2) homomorphisms) and the moduli spaces of minimal and near-minimal nilpotent orbits extend systematically to higher rank $BCD$ groups \footnote{These relationships also extend to some near-minimal nilpotent orbits of Exceptional Groups, although these are not the focus of this study}.


\section{Discussion and Conclusions}
\label{sec:conclusions}

The methods set out above for constructing $BCD$ nilpotent orbits resolve a number of difficulties with previously proposed constructions. When working on the Higgs branch, we take $G$ as the flavour group, and when working on the Coulomb branch, we apply the monopole formula to the simple roots of $G$, treating them as unitary gauge nodes. This provides an unambiguous link from the nilpotent orbits of $G$ to their moduli spaces, which contain representations of $G$. 

In particular, we have been able to avoid working with dual groups of $G$, which can lead to difficulties in matching the results obtained to the canonical dimensions of nilpotent orbits of $G$ \footnote{Some of these moduli spaces, described by their unrefined Hilbert series, have been calculated in \cite{Cremonesi:2014kwa, Cremonesi:2014uva}. However, their description and labelling therein is different to the canonical scheme from the mathematical literature used herein.}.

Our approach does not depend on the Spaltenstein map \cite{Chacaltana:2012zy, Chacaltana:2013oka, Cremonesi:2014kwa, Cremonesi:2014uva}, which is many to one, and has the problematic feature of conflating, through $B/C/D$ collapses, nilpotent orbits with different dimensions.

Also, our approach uses quivers which combine $BCD$ gauge groups, rather than shifting the dimensions of gauge nodes to achieve DC series only $({T_{\rho  + }})$ or BC series only $({T_{\rho -}})$ quivers, as discussed in \cite{Benini:2010uu}. Consideration of the dimension formulae \ref{eq:orbits11} and \ref{eq:orbits12} entails that, except for certain shifts, such as those within maximal sub-chains, shifting gauge nodes $B_{n}$ to $D_{n+1}$ for $({T_{\rho  + }})$ or $D_{n}$ to $B_{n}$ for $({T_{\rho -}})$ will displace the dimensions of a nilpotent orbit, as discussed in section \ref{subsec:quivers}. For example, the nilpotent orbits $D_4-C_2-B_0$ and $D_4-C_2-D_1$ can be related by such node shifting, but are not the same, as can be seen from table \ref{tab:Dseries2}.


Our Higgs branch moduli spaces cover the full set of nilpotent orbits for Classical groups and yield palindromic HyperK\"ahler cones in almost all instances. In the few non-palindromic cases, we have been able to identify how the nilpotent orbits are formed as unions of such HyperK\"ahler cones, or, how they are related to such unions. Importantly, the partial ordering of these Higgs branch quivers, using inclusion relations either between the group structures of quiver chains, or between their Hilbert series or character HWGs, matches the canonical ordering of nilpotent orbits into Hasse diagrams by traditional methods \cite{Collingwood:1993fk, Kraft:1982fk}. By way of further confirmation of our constructions, the dualities and relationships between nilpotent orbits, calculated from these Higgs branch moduli spaces, are consistent with relationships identified through geometric reasoning \cite{kobak1996classical}, as elaborated below.

It is clear that the map from Higgs branch quivers to nilpotent orbits is many to one, in that multiple quivers can lead to the same nilpotent orbit and Hilbert series. Indeed, there are many dualities and other relationships between the nilpotent orbits of different groups that can be identified from our analysis of these moduli spaces. (We refer to two quivers as dual if they have isomorphic Higgs branch moduli spaces.) These relationships can be classified into different categories including:

\begin{enumerate}
\item Dualities between $A$ series quivers described by non-canonical partition orderings (see section \ref{sec:AseriesGeneral}),

\item Dualities between quivers containing maximal $D-C$ or $B-C$ subchains,

\item Dualities between quivers from isomorphic Classical flavour groups,

\item Pairs of quivers related by HyperK\"ahler quotients by some compact group and/or discrete quotients \cite{kobak1996classical}. Within these, sub-categories can be identified, as discussed below.
\end{enumerate}

We set out in table \ref{tab:disc1} the main dualities between pairs of quivers for nilpotent orbits $\{{\cal O}_1, {\cal O}_2\}$ of low rank groups that involve isomorphisms and/or maximal $D-C$ or $B-C$ subchains.

\begin{table}[htp]
\caption{Quiver Dualities for Nilpotent Orbits of Low Rank Classical Groups}
\begin{center}
\begin{tabular}{|c|c|c|}
\hline
Dimension & Quiver ${\cal O}_1$ & Quiver $ {\cal O}_2$\\
\hline
2 &$\begin{array}{c}B_1-C_1-B_0\\B_1-C_1-D_1\end{array}$ & $C_1-B_0$~ \vline ~$[2]-(1)$\\
\hline
4 &$B_2-C_1$ & $C_2-B_0$ \\
\hline
6 & $\begin{array}{c} B_2-C_1-B_0 \\ B_2-C_1-D_1 \end{array} $& $C_2-D_1$ \\
\hline
8 & $\begin{array}{c} B_2-C_2-B_1-C_1-B_0 \\ B_2-C_2-D_2-C_1-D_1 \end{array} $& $\begin{array}{c} C_2-B_1-C_1-B_0 \\ C_2-D_2-C_1-D_1 \end{array} $ \\
\hline
\hline
4 &$\begin{array}{c}D_2-C_1-B_0\\D_2-C_1-D_1 \end{array} $ & $[2]-(1) \otimes [2]-(1)$\\
\hline
6 &$D_3-C_1$ & $[4]-(1)$ \\
\hline
8 &$\begin{array}{c}D_3-C_1-B_0\\D_3-C_1-D_1 \end{array} $ & $[4]-(2)$ \\
\hline
10 &$\begin{array}{c}D_3-C_2-D_1 \end{array} $ & $[4]-(2)-(1)$ \\
\hline
12 &$\begin{array}{c}D_3-C_2-B_1-C_1-B_0\\D_3-C_2-D_2-C_1-D_1 \end{array} $ & $[4]-(3)-(2)-(1)$ \\
\hline
\end{tabular}
\end{center}
\text {Higgs branch moduli spaces are isomorphic along rows and identical within cells}\\
\text {B/D gauge groups indicate the corresponding O gauge group}\\
\text {[N] indicates SU(N) flavour and (N) indicates U(N) gauge groups}\\
\label{tab:disc1}
\end{table}
\FloatBarrier

Table \ref{tab:disc2} sets out a selection of pairs of nilpotent orbits that are related by HyperK\"ahler and/or discrete quotients, largely drawn from \cite{kobak1996classical}. These have been rearranged using the dualities in table \ref{tab:disc1}. The relationship between each pair $\{{\cal O}_G \equiv g_{Higgs}^G, {\cal O}_K \equiv g_{Higgs}^{K=G/H}\}$ can be described by a character map from the group $G$ of the parent nilpotent orbit to a product of its subgroups ${\cal X}_G \to {\cal X}_{K \otimes H_1 \otimes \ldots H_m}$, followed by a HKQ by the subgroup $H \equiv H_1 \otimes \ldots H_m$ and/or the action of a finite $\mathbb{Z}_n$ factor:
\begin{equation} 
\label{eq:disc1}
\begin{aligned}
g_{Higgs}^K\left( {{{{\cal X}}_K},t} \right) = \frac{1}{{\left| {{\mathbb{Z}_n}} \right|}}\sum\limits_{{\mathbb{Z}_n}} {\oint\limits_{{H_1} \otimes  \ldots {H_m}} {d\mu} } ~ \frac{{g_{Higgs}^G\left( {{{{\cal X}}_{K \otimes {H_1} \otimes  \ldots {H_m}}},t} \right)}}{{\prod\limits_{i = 1}^m {PE\left[ {{{\left[ {adj} \right]}_{{H_i}}},{t^2}} \right]} }}
 \end{aligned}
\end{equation}
The precise implementation of the $\mathbb{Z}_n$ group average differs from case to case, but can be carried out after calculating the HWG for $g_{Higgs}^K\left( {{{{\cal X}}_K},t} \right)$.
\begin{table}[htp]
\caption{HyperK\"ahler Quotients between Nilpotent Orbits of Low Rank Classical Groups}
{\scriptsize 
\begin{center}
\begin{tabular}{|c|c|c|c|c|c|}
\hline
${\cal O}_G$ &Dim.&  ${\cal X}_G \to {\cal X}_{K \otimes H}$& Quotient & ${\cal O}_K$&Dim.\\
\hline
$ B_2-C_1 $&$ 4 $&$ [1,0]_B \to [1,1]_D \oplus 1$&$\mathbb{Z}_2 $&$ D_2-C_1-B_0 $&$4$\\
\hline
$ D_3-C_1$&$6$&$ [1,0,0]_D \to [1,0]_B \oplus 1$&$\mathbb{Z}_2$&$B_2-C_1-B_0$&$6$\\
\hline
$[4]-(2)$&$8$&$ [1,0,0] \to [1,0]q \oplus \frac{1}{q^3} $&$U(1)$&$[3]-(2)-(1)$&$6$\\
\hline
$ D_4-C_1$&$10$&$ [1,0,0,0]_D \to [1,0]q \oplus [0,1] \frac{1}{q}+(\frac {q}{q_1}+\frac {q_1}{q}) $&$U(1) \otimes U(1)$&$[3]-(2)-(1)$&$6$\\
\hline
$[8]-(1)$&$14$&$ [1,0,0,0,0,0,0] \to [1,0][1]q \oplus[1] \frac{1}{q^3}$&$SU(2) \otimes U(1)$&$[3]-(2)-(1)$&$6$\\
\hline
$ B_3-C_1 $&$8$&$ [1,0,0]_B \to [1,0,0]_D \oplus 1 $&$\mathbb{Z}_2 $&$ D_3-C_1-B_0 $&$8$\\
\hline
$ D_4-C_1$&$10$&$ [1,0,0,0]_D \to [1,0,0]_D  \oplus(q+ \frac{1}{q})$&$O(2)$&$D_3-C_1-D_1$&$8$\\
\hline
$[8]-(1)$&$14$&$ [1,0,0,0,0,0,0] \to [1,0,0][1]$&$SU(2)$&$[4]-(2)$&$8$\\
\hline
$[6]-(1)$&$10$&$ [1,0,0,0,0] \to [1,0,0]_C$&$\mathbb{Z}_2$&$C_3-D_1$&$10$\\
\hline
$ D_4-C_1$&$10$&$ [1,0,0,0]_D \to [1,0,0]_B \oplus 1$&$\mathbb{Z}_2$&$B_3-C_1-B_0$&$10$\\
\hline
$ D_4-C_1-B_0$&$12$&$ [1,0,0,0]_D \to [1,0,0] q \oplus [0,0,1] \frac{1}{q}$&$U(1)$&$[4]-(2)-(1)$&$10$\\
\hline
$[5]-(2)$&$12$&$ [1,0,0,0] \to [1,0,0] q \oplus \frac{1}{q^4} $&$U(1)$&$[4]-(2)-(1)$&$10$\\
\hline
$ D_5-C_1$&$14$&$ [1,0,0,0,0]_D \to [1,0,0] q \oplus [0,0,1] \frac{1}{q}+(\frac {q^4}{q_1^4}+\frac {q_1^4}{q^4})$&$U(1) \otimes U(1)$&$[4]-(2)-(1)$&$10$\\
\hline
$[10]-(1)$&$18$&$ [1,0,0,0,0,0,0,0,0] \to [1,0,0][1]q+[1] \frac{1}{q^4}$&$SU(2) \otimes U(1)$&$[4]-(2)-(1)$&$10$\\
\hline
$ B_4-C_2 $&$16$&$ [1,0,0,0]_B \to [1,0,0]_B \oplus (q+\frac{1}{q}) $&$O(2)$&$ B_3-C_2-D_1 $&$14$\\
\hline
$ C_4-D_2$&$20$&$ [1,0,0,0]_C \to [1,0,0]_C \oplus [1]_C$&$C_1$&$C_3-D_2-C_1$&$14$\\
\hline
\end{tabular}
\end{center}
\text {$B/D$ gauge groups indicate the corresponding $O$ gauge group},\\
\text {$[N]$ indicates $SU(N)$ flavour and $(N)$ indicates $U(N)$ gauge groups},\\
\text{Dynkin labels are $A$ series unless otherwise indicated},\\
\text{$U(1)$ or $O(2)$ fugacities in the character map are denoted $q_i$.}
}
\label{tab:disc2}
\end{table}
\FloatBarrier
The above constitute only a sample of the possible HyperK\"ahler quotients between nilpotent orbit moduli spaces, but serve to exemplify some particular types of relationship. These include:
\begin{enumerate}
\item 2-node quivers with flavour group symmetry breaking (9 examples). The fundamental of the flavour group is broken to a sum of fundamentals of groups of the same type ($O/Sp/U$). The HKQ is taken over the lower rank group, with the quotient for $B_0 \cong O(1)$ given by a $\mathbb{Z}_2$ factor. There are conditions that follow from the requirement that the new quiver should be based on a well ordered partition. Possibilities for Classical flavour groups are shown in table \ref{tab:disc3}. In all cases the reduction in complex dimension of the nilpotent orbit is equal to twice the dimension of the HKQ gauge group.
\begin{table}[htp]
\caption{Some Generalised HyperK\"ahler Quotients between Nilpotent Orbits}
{\scriptsize 
\begin{center}
\begin{tabular}{|c|c|c|c|c|c|}
\hline
${\cal O}_G$ &Dim. ${\cal O}_G$ &  HKQ & ${\cal O}_K$&Dim. ${\cal O}_K$& Conditions\\
\hline
\hline
$[n+1]-(k)$&$ 2k(n+1-k) $&$U(N) $&$[n+1-N]-(k)-(N)$&$2k(n+1-k)- 2{N}^2$&$n+1 \ge 2 k \ge 4 N$\\
\hline
$SO(N)-C_k$&$ 2k(N-2k-1) $&$O(N') $&$ SO(N-N')-C_k-O(N') $&$2k(N-2k-1)-N'(N'-1)$&$N \ge 4k \ge 4 N'$\\
\hline
$C_k-O(N)$&$ N(2k-N+1) $&$C_{k'} $&$C_{k-k'}-O(N)-C_{k'} $&$N(2k-N+1)-2k'(2k'+1)$&$k \ge N \ge 4 k'$\\
\hline
\hline
$[2k]-(1)$&$ 2(2k-1) $&$\mathbb{Z}_2$&$C_k-D_1$&$2(2k-1)$&$k>1$\\
\hline
\end{tabular}
\end{center}
\text {$B/D$ gauge groups indicate the corresponding $O$ gauge group},\\
\text {$[N]$ indicates $SU(N)$ flavour and $(N)$ indicates $U(N)$ gauge groups}.\\
}
\label{tab:disc3}
\end{table}
\item $SU(2k)$ RSIMS folding to the supra minimal nilpotent orbit of $C_k$ (1 example). Consider the RSIMS quiver $SU(2k) - U(1)$. The complex character of the flavour group fundamental representation can be mapped to the pseudo real $C_k$ fundamental. The gauge group maps from $U(1)$ to $O(2)$. The HKQ is a $\mathbb{Z}_2$ factor, as shown in table \ref{tab:disc3}.
\item Flavour group branching to product group. The HKQ is taken over all the members of the product group other than the new flavour group. Considering that the product group need not be semi-simple, there are many possibilities for branching a group into its subgroups \cite{Dynkin:1957um}. The possibilities are compounded by the alternative choices of HKQ and only some of these combinations lead to nilpotent orbits of the new flavour group (rather than more general moduli spaces).
\end{enumerate}
The generalisations in table \ref{tab:disc3} extend the results of \cite{kobak1996classical} to a wide class of relationships involving nilpotent orbits based on the Higgs branches of 2-node quivers.

The Coulomb branch quivers for $A$ series nilpotent orbits herein follow the established principles of $3d$ mirror symmetry and/or affine Dynkin diagrams. These Coulomb branch constructions generalise to cover all the nilpotent orbits of the $A$ series, and the $3d$ mirror symmetry that relates these Coulomb and Higgs branch quivers is well established \cite{Hanany:2011db}. 

In the case of the $BCD$ series, we have been obliged to combine a number of methods. The minimal nilpotent orbits of the $BCD$ series are given by Coulomb branches of quivers based on affine Dynkin diagrams \cite{Cremonesi:2013lqa}, and we have found that supra minimal and other near-minimal nilpotent orbits can be found from twisted affine Dynkin diagrams. Coulomb branch quivers for near-minimal nilpotent orbits can also be identified directly from the Dynkin labels of the root and weight maps associated with nilpotent orbit partitions. Taken together, these Coulomb branch constructions cover the minimal, supra-minimal and other near-minimal nilpotent orbits of the $BCD$ series.

We have shown how the moduli spaces of these nilpotent orbits and their relationships can be analysed in terms of (unrefined) Hilbert series, and {\it highest weight generating} functions giving their decompositions into irreps and/or modified Hall Littlewood polynomials, where it is noteworthy that all the Classical nilpotent orbits studied can be expressed as finite expansions in $mHL^G_{[n]}$. In the course of this we have been able to formulate general conjectures for the moduli spaces of several types of nilpotent orbit, as summarised in tables \ref{tab:Aseries3} and \ref{tab:Bseries3} - \ref{tab:Dseries3}. 
\FloatBarrier

\paragraph{Further Work} 

While we have been able to show how to construct the Higgs branch quivers for any Classical group nilpotent orbit, and also the Coulomb branch quivers for $A$ series and near-minimal $BCD$ series nilpotent orbits, it appears that the higher dimensioned nilpotent orbits of the $BCD$ series contain relations, which obstruct their construction from the unitary monopole formula, and we are not, at this time, able to identify their Coulomb branch quivers. Such Coulomb branch quivers for higher dimensioned nilpotent orbits of the $BCD$ series would complete this study and perhaps illuminate a route to Coulomb branch quivers for $BCD$ $mHL$ polynomials generally.

At this juncture, we are also unable to encapsulate the transformations between Higgs and Coulomb branch quivers for $BCD$ series nilpotent orbits in a set of rules. For example, the $3d$ mirror transformations in \cite{Cremonesi:2014uva} lead, when applied to $BCD$ Higgs branch quivers, to non-unitary gauge nodes that do not correspond to simple roots. Furthermore, we are unable to examine the mirror symmetry of the dual Higgs/Coulomb quivers for $BCD$ series nilpotent orbits, since we cannot apply the unitary monopole formula to the $BCD$ series Higgs branch quivers, which contain non-unitary gauge groups, and we do not know how to adapt the HKQ formula \ref{eq:orbits5a} to $BC$ series Coulomb branch quivers, which contain non-simply laced links.

While we have given examples, we have not attempted a complete enumeration of the possible types of HKQ relationship between the moduli spaces of Classical group nilpotent orbits. This would appear to present a large field for further study.

Finally, since the Higgs branch constructions for minimal nilpotent orbits or RSIMS of Exceptional groups are not known, the identification of quivers for Exceptional group nilpotent orbits is clearly a non-trivial problem. It would be interesting to see how far this problem can be addressed, by drawing upon the Coulomb branch and other methods discussed herein.

\paragraph{Acknowldgements}
Rudolph Kalveks is grateful to Marcus Sperling, Institut f\"ur Theoretische Physik, Leibniz Universit\"at, Hannover and to Santiago Cabrera Marquez, Imperial College, London for valuable discussions.
%

\appendix
\section{Hall Littlewood Polynomials}
\label{apxHLP}

The families of orthogonal Hall-Littlewood polynomials $H{L^G}$ and modified Hall-Littlewood polynomials $mH{L^G}$ of a group $G$, having rank $r$, root space $\Phi$, weight space coordinates $x \equiv \left( {{x_1}, \ldots ,{x_r}} \right)$, positive roots $\{x^{\alpha}: {\alpha \in \Phi + }\}$ and Dynkin labels $[n] \equiv {[{n_1}, \ldots ,{n_r}]}$, can be defined as:

\begin{equation}
\label{eq:apx1}
\begin{aligned}
H{L}_{[n]}^G \left(x, t \right) &= \sum\limits_{w \in Weyl\left[ G \right]} {w\left( {x_1^{{n_1}} \ldots x_r^{{n_r}}\prod\limits_{\alpha \in \Phi + } {\frac{{1 - t{x^{ - \alpha }}}}{{1 - {x^{ - \alpha }}}}} } \right)},
\end{aligned}
\end{equation}
and 
\begin{equation}
\label{eq:apx2}
\begin{aligned}
mH{L}_{[n]}^G \left(x, t \right) &=\left( \prod\limits_{\alpha \in \Phi } {\frac{1}{{1 - t{x^{ \alpha }}}}}\right) \sum\limits_{w \in Weyl\left[ G \right]} {w\left( {x_1^{{n_1}} \ldots x_r^{{n_r}}\prod\limits_{\alpha \in \Phi + } {\frac{{1 - t{x^{ - \alpha }}}}{{1 - {x^{ - \alpha }}}}} } \right)},
\end{aligned}
\end{equation}
 where the sums are taken over the action of the Weyl group of $G$ and we use the fugacity $t$. The orthogonality between the ${(m)HL_\lambda }$ and their complex conjugates, under an inner product incorporating the (modified) Hall-Littlewood measure $d{\mu ^G}_{(m)HL}$, is given by \footnote{In \cite{Macdonald:1995fk} the ${(m)HL_{[n]} }$ are normalised by dividing by ${v_{[n]}}(t)$ and in \cite{Gadde:2011uv} they are normalised by dividing by ${\sqrt {{v_{[n]} }(t)}}$. We do not use either of these schemes, consistent with the approach in \cite{Cremonesi:2014kwa}}:
\begin{equation}
\label{eq:apx3}
\oint\limits_G {d{\mu_{HL} ^G}}~H{L_{[n]} ^G}\left( x*,t \right)~H{L_{[m]}^G }\left(x,t \right) = {\delta _{[n] [m] }}~{v}_{[n]}^G \left( t \right),
\end{equation}
and
\begin{equation}
\label{eq:apx4}
\oint\limits_G {d{\mu_{mHL} ^G}}~mH{L_{[n]}^G }\left( x*,t \right)~mH{L_{[m]}^G }\left(x,t \right) = {\delta _{[n] [m] }}~{v}_{[n]}^G \left( t \right),
\end{equation}
where we are using notation ${d{\mu_{(m)HL} ^G}}$ for the (modified) Hall-Littlewood measure:
\begin{equation}
\label{eq:apx5}
d\mu _{HL}^G \equiv \frac{1}{{\left| {Weyl\left[ G \right]} \right|}}\left( {\prod\limits_{i = 1}^r {\frac{{d{x_i}}}{{{x_i}}}} } \right)\left( {\prod\limits_{\alpha \in \Phi } {\left( {1 - {x^\alpha }} \right)} } \right)\left( {\prod\limits_{\alpha \in \Phi } {\frac{1}{{1 - t{x^\alpha }}}} } \right)
\end{equation}
and
\begin{equation}
\label{eq:apx6}
d\mu _{mHL}^G \equiv \frac{1}{{\left| {Weyl\left[ G \right]} \right|}}\left( {\prod\limits_{i = 1}^r {\frac{{d{x_i}}}{{{x_i}}}} } \right)\left( {\prod\limits_{\alpha \in \Phi } {\left( {1 - {x^\alpha }} \right)} } \right)\left( {\prod\limits_{\alpha \in \Phi } {\left( {1 - t{x^\alpha }} \right)} } \right).
\end{equation}
The factors ${v}_{[n]}^G (t)$ relate to the symmetric Casimirs of $G$ or its subgroups, and depend on any zero Dynkin labels in the representation $[n]$, being given by:
\begin{equation}
\label{eq:apx7}
\begin{aligned}
v_{[n]}^G\left( t \right) = \prod\limits_{C \in Casimirs\left( {G/[n]} \right)} {\left( {\frac{{1 - {t^{degree \left( C \right)}}}}{{1 - t}}} \right)}.
\end{aligned}
\end{equation}

The subgroup $G/[n]$ is defined by the Dynkin diagram that remains after eliminating from the Dynkin diagram of $G$ any nodes which correspond to non-zero Dynkin labels of $[n]$. Thus, the ${v}_{[n]}^G (t)$ incorporate all the Casimirs of $G$ if the Dynkin labels of $[n]$ are all zero and reduce to unity if the Dynkin labels are all non-zero. For example, the representation $[0,0,0,0]$ of $D_4$ has the ${v}_{[n]}^G (t)$ factor $\frac{\left(1-t^2\right) \left(1-t^4\right)^2 \left(1-t^6\right)}{(1-t)^4}$, while $[0,1,0,0]$ of $D_4$ has the factor $\frac{\left(1-t^2\right)^3}{(1-t)^3}$ and $[1,1,1,1]$ has the factor $1$.

In the limit where $t \to 0$, the (modified) Hall-Littlewood polynomials reduce to the characters of $G$, the (modified) Hall-Littlewood measure reduces to the Haar measure for $G$, and the factors ${v_{[n]}}^{G}(0)$ reduce to unity.

We now introduce the fugacities $\left\{ {{h_1}, \ldots ,{h_r}} \right\}$ for the highest weight Dynkin labels of the (modified) Hall-Littlewood polynomials and define and construct their generating functions:

\begin{equation}
\label{eq:apx8}
\begin{aligned}
g_{HL}^G\left( {x,t,h} \right) &\equiv \sum\limits_{[n]= [0]}^{\left[ \infty \right]} {HL_{\left[ n \right]}^G} \left( {x,t} \right)~h^n\\
& = \sum\limits_{w \in Weyl\left[ G \right]} {w\left( {\left( {\prod\limits_{i = 1}^r {\frac{1}{{1 - {x_i}{h_i}}}} } \right)\prod\limits_{\alpha \in \Phi + } {\frac{{1 - t{x^{ - \alpha }}}}{{1 - {x^{ - \alpha }}}}} } \right)}
\end{aligned}
\end{equation}
and 
\begin{equation}
\label{eq:apx9}
\begin{aligned}
g_{mHL}^G\left( {x,t,h} \right) &\equiv \sum\limits_{[n] = [0]}^{\left[ \infty \right]} {mHL_{\left[ n \right]}^G} \left( {x,t} \right)~h^n\\
 &= \left( {\prod\limits_{\alpha \in \Phi } {\frac{1}{{1 - t{x^\alpha }}}} }\right) g_{HL}^G\left( {x,t,h} \right),
\end{aligned}
\end{equation}
where we have defined ${h^n} \equiv \prod\limits_{i = 1}^r {h_i^{{n_i}}}$.

From \ref{eq:apx3} and \ref{eq:apx4}, it follows that the generating functions $g_{(m)HL}^G({x},t,{h})$ have the orthogonality property with the $(m)H{L_{[n]}^G}$:
\begin{equation}
\label{eq:apx10}
\begin{aligned}
\oint\limits_G {d{\mu _{(m)HL}}}~g_{(m)HL}^G\left( {{x^*},t,h} \right)~(m)H{L_{[n]}^G}\left( {x,t} \right) = {v_{[n]}^G}\left( t \right){h^n}.
\end{aligned}
\end{equation}

We can obtain more useful contragredient generating functions $\overline {g_{(m)HL}^G}\left(x,t,h \right)$, which generate polynomials that are orthonormal (rather than just orthogonal) to the ${(m)HL_{[n]}^G}$, by gluing together the $g_{(m)HL}^G({x*},t,{h})$ with generating functions for the $1/{{v_\lambda}^{G}(t)}$, as described in \cite{Hanany:2015hxa}. These have the orthonormality:
\begin{equation}
\label{eq:apx11}
\begin{aligned}
\oint\limits_G {d{\mu _{(m)HL}}}~\overline {g_{(m)HL}^G} \left(x,t,h \right)~(m)H{L_{[n]}^G}\left( {x,t} \right) ={h^n},
\end{aligned}
\end{equation}

Since the (modified) Hall-Littlewood polynomials provide a complete basis for class functions that combine the characters of a group $G$ with coefficients given by polynomials in the fugacity $t$, we can use these generating functions and orthonormality relationships to decompose any such class function $F^G \left( {{x},t} \right)$ into (modified) Hall-Littlewood polynomials. We first define the decomposition coefficients ${C_{[n]}(t)}$ from:
\begin{equation}
\label{eq:apx12}
\begin{aligned}
F^G \left( {x,t} \right) \equiv \sum\limits_{[n]}^{} {C_{[n]}\left(t\right)(m)H{L_{[n]}^G}\left( {x,t} \right)}.
\end{aligned}
\end{equation}
We can then find a HWG ${C(t,h)}$ for the ${C_{[n]}}(t)$, using the contragredient generating functions and their orthonormality: 
\begin{equation}
\label{eq:apx13}
\begin{aligned}
C(t,h) & \equiv \sum\limits_{[n]}^{} {{C_{[n]}}\left( t \right){h^n}}\\
&= \oint\limits_G {d\mu _{(m)HL}^G} ~\overline {g_{(m)HL}^G} \left(x,t,h \right)~F\left({x,t} \right).
\end{aligned}
\end{equation} 
Individual ${C_{[n]} }(t)$ can be extracted from ${C(t,h)}$ by Taylor expansion, followed by matching the coefficients of the monomials ${h^n}$. The reader is referred to \cite{Hanany:2015hxa} for additional explanation on our use of highest weight generating functions or to \cite{Macdonald:1995fk} for mathematical background.

In this study, we work with the modified Hall Littlewood polynomials since these typically provide more concise HWGs ${C(t,h)}$ for the decomposition coefficients of nilpotent orbits. Many residue calculations are typically required by the contour integrations involved and we use customised {\it Mathematica} routines to assist in this.

\clearpage

\section{Nilpotent Orbits and $SU(2)$ Homomorphisms}
 \label{apxHom}

\subsection{A Series }
\label{apxA}

\begin{center}
~\\
\includegraphics[scale=1]{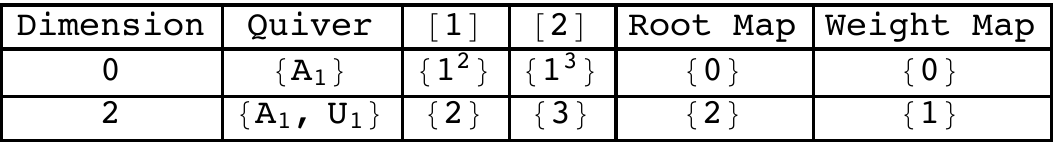}\\
~\\
\includegraphics[scale=1]{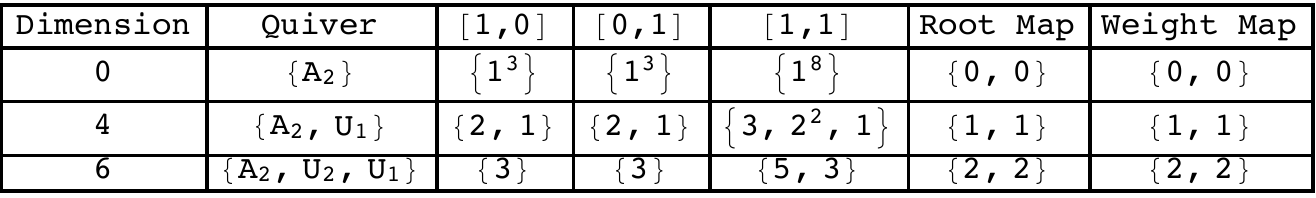}\\
~\\
\includegraphics[scale=1]{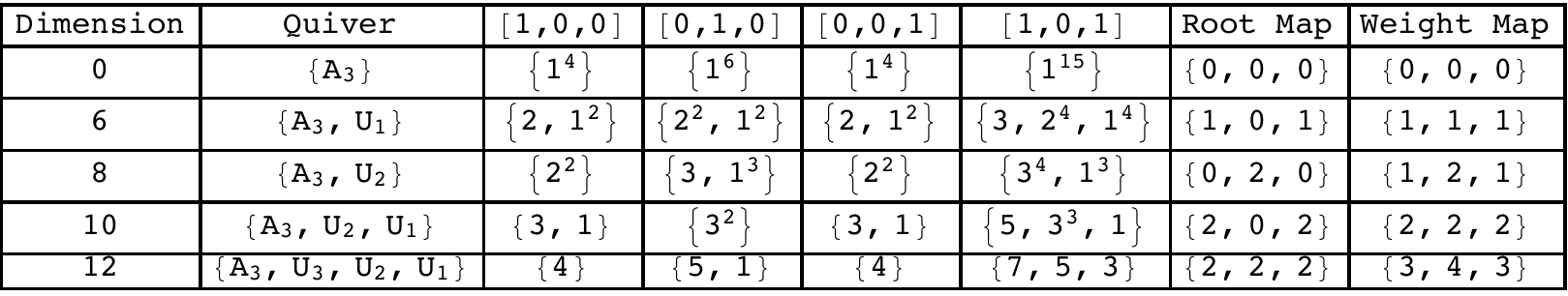}\\
~\\
\includegraphics[scale=.875]{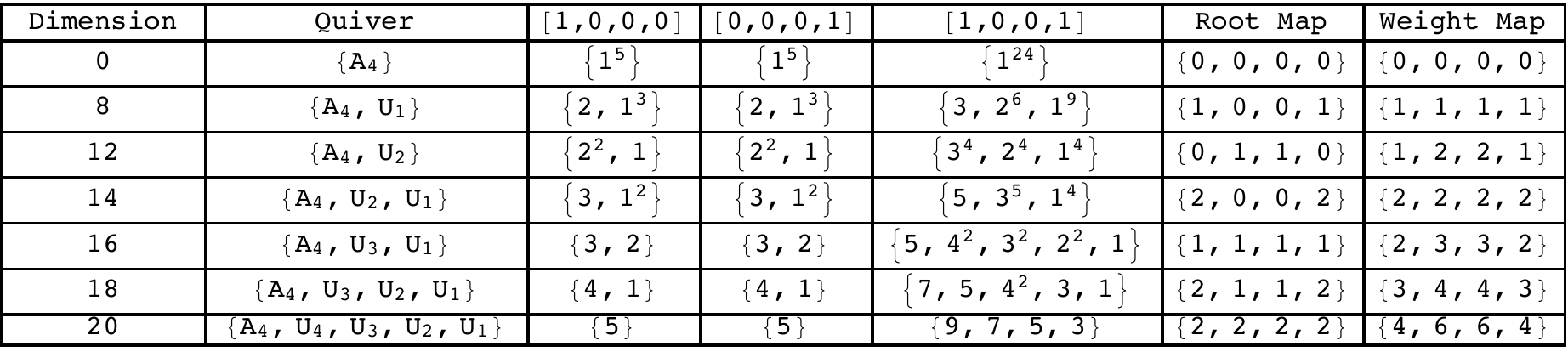}\\
~\\
\includegraphics[scale=.725]{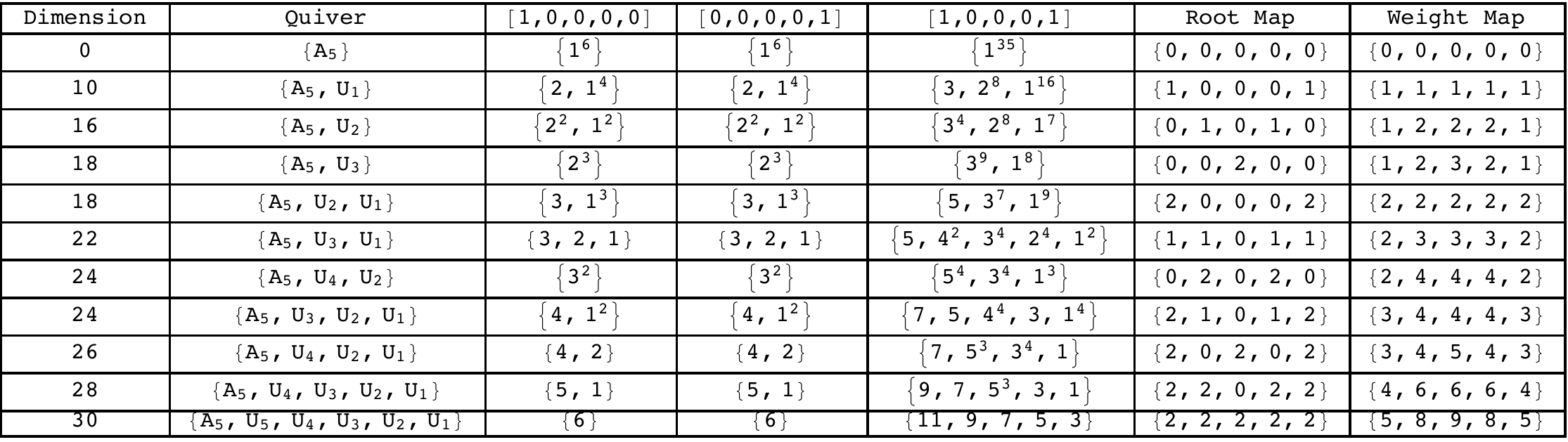}\\
\end{center}
Partitions are shown under each homomorphism for the fundamental, anti-fundamental and adjoint representations. For $A_3$, the vector representation partitions are also shown.


\subsection{B Series}
\label{apxB}

\begin{center}
~\\
\includegraphics[scale=1]{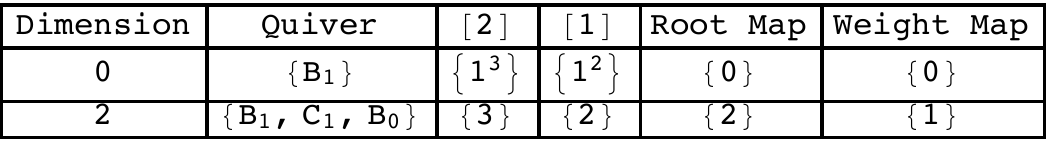}\\
~\\
\includegraphics[scale=1]{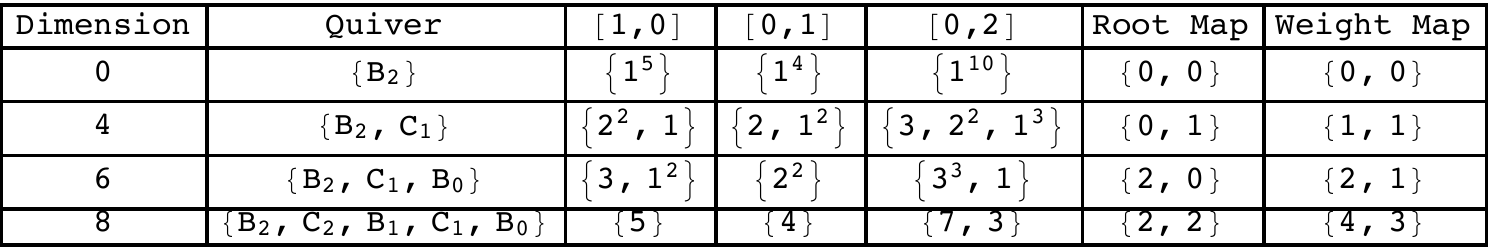}\\
~\\
\includegraphics[scale=.8]{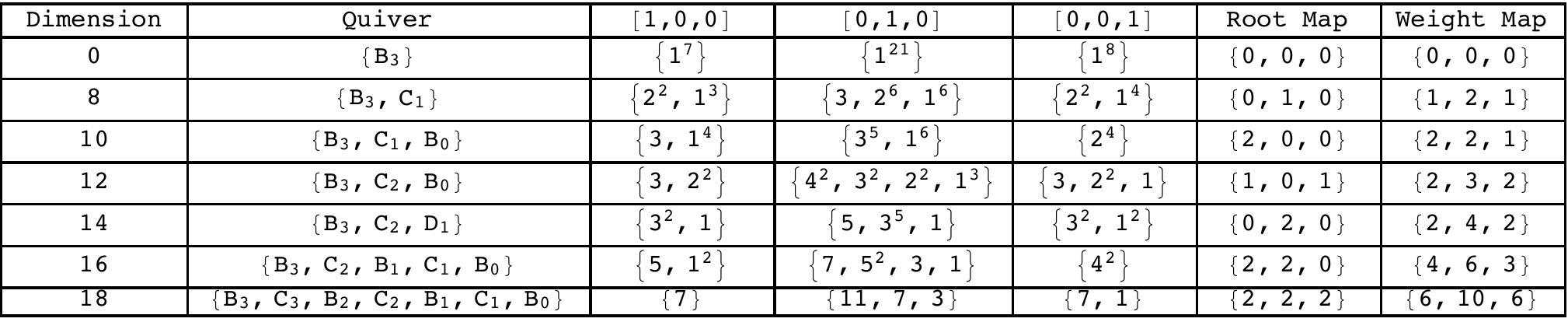}\\
~\\
\includegraphics[scale=.71]{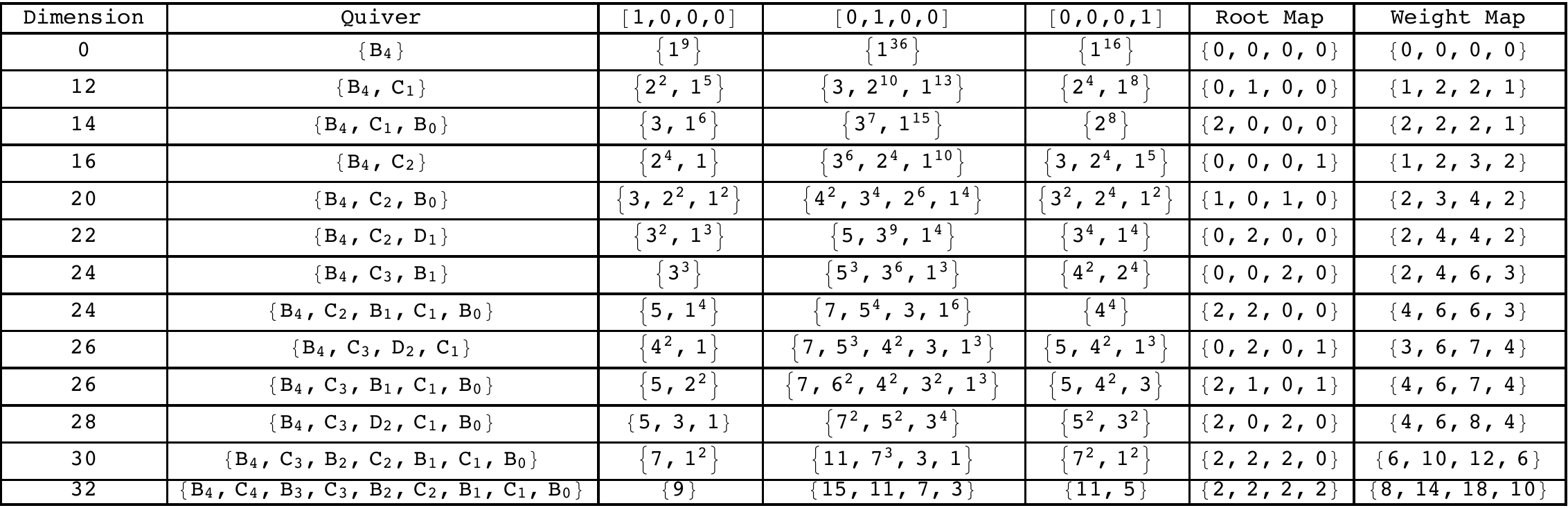}\\
~\\
\includegraphics[scale=.55]{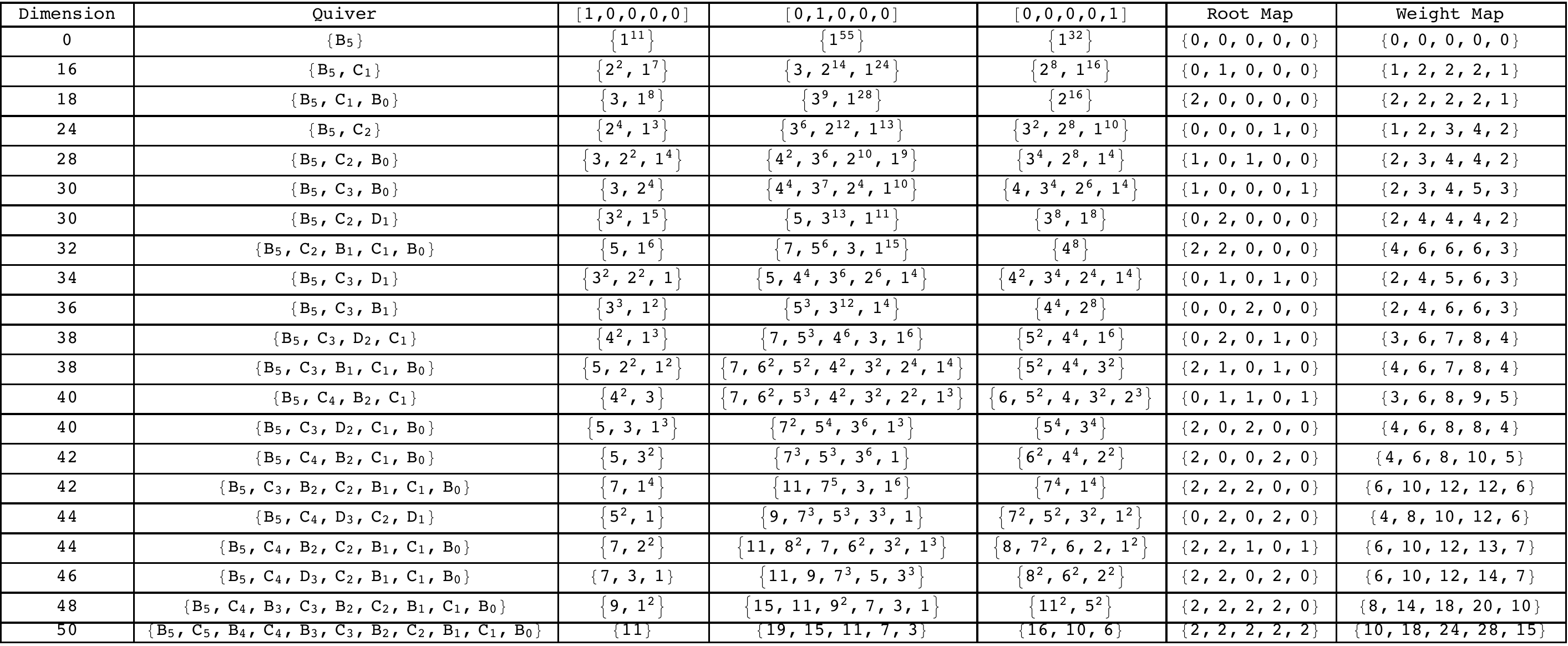}\\
\end{center}
Partitions are shown under each homomorphism for the vector, adjoint and spinor representations.
~\\
\subsection{C Series}
\label{apxC}

\begin{center}
~\\
\includegraphics[scale=1]{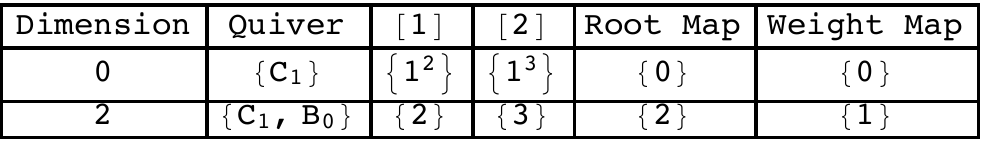}\\
~\\
\includegraphics[scale=1]{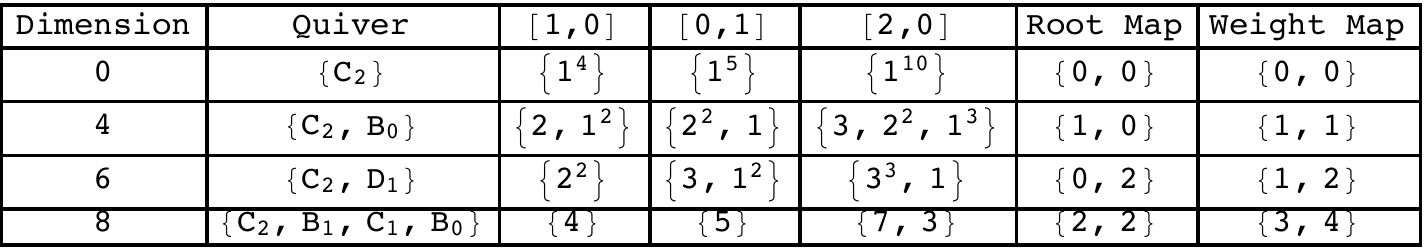}\\
~\\
\includegraphics[scale=1]{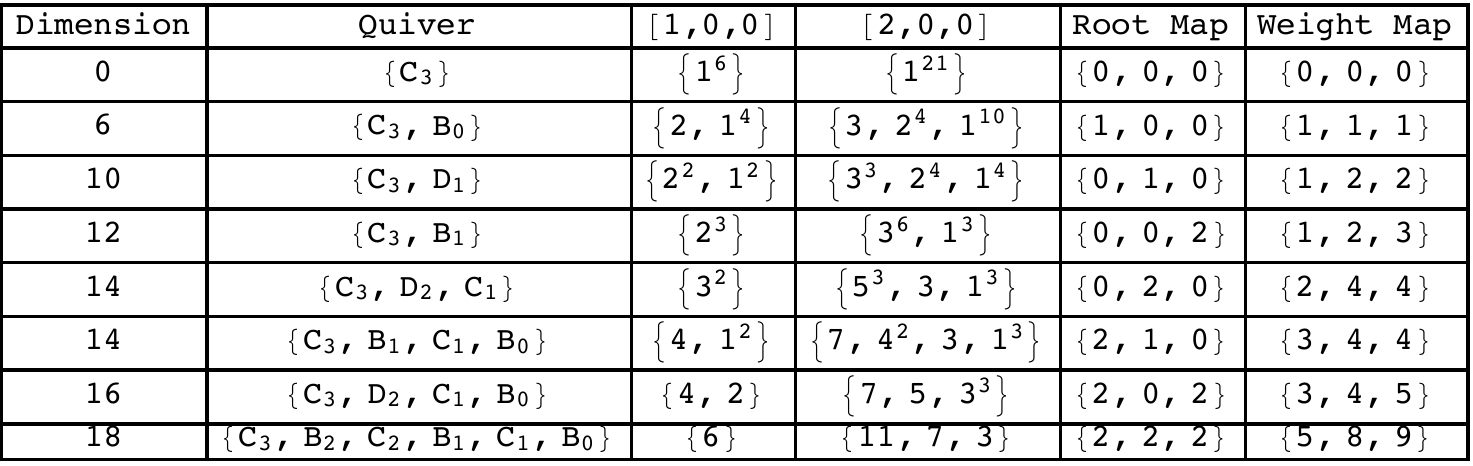}\\
~\\
\includegraphics[scale=.79]{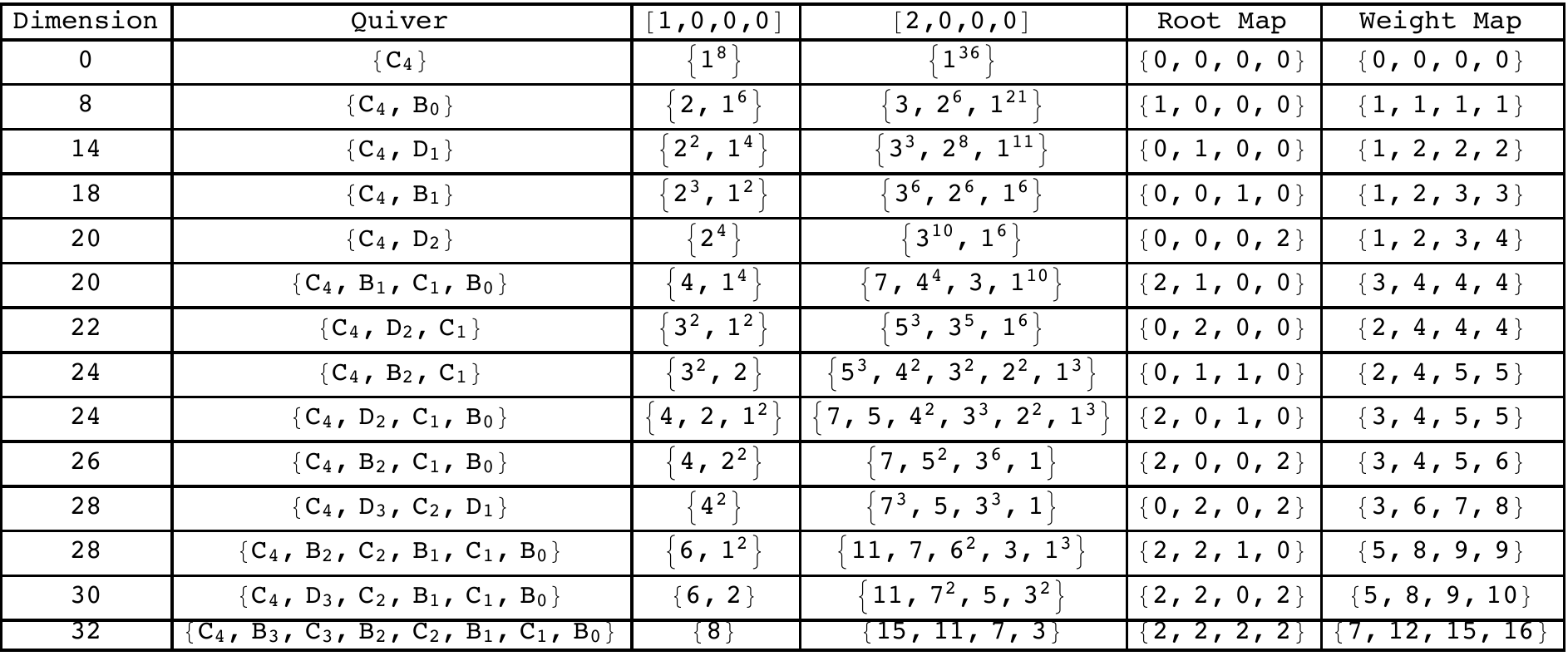}\\
~\\
\includegraphics[scale=.66]{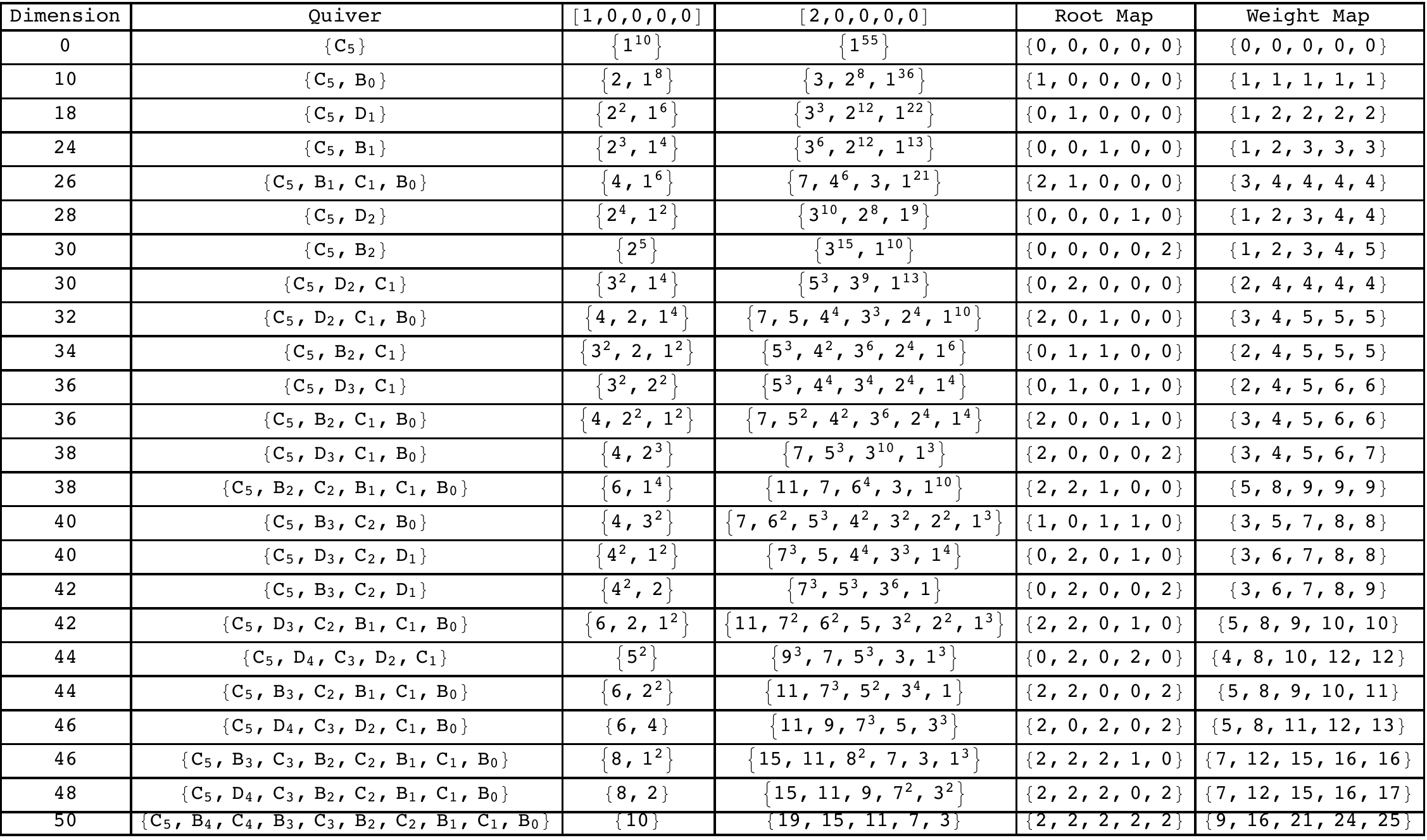}\\
\end{center}
Partitions are shown under each homomorphism for the symplectic vector and adjoint representations. For $C_2$, the partition of the $[0,1]$ representation is also shown.
\clearpage


\subsection{D Series}
\label{apxD}

\begin{center}
~\\
\includegraphics[scale=1]{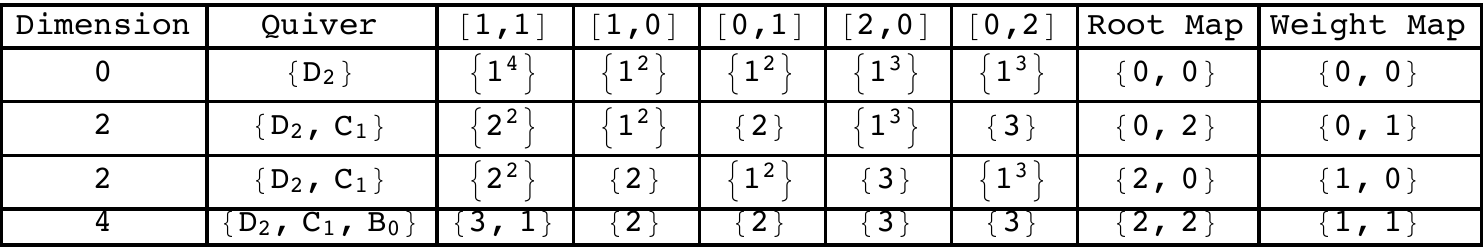}\\
~\\
\includegraphics[scale=.89]{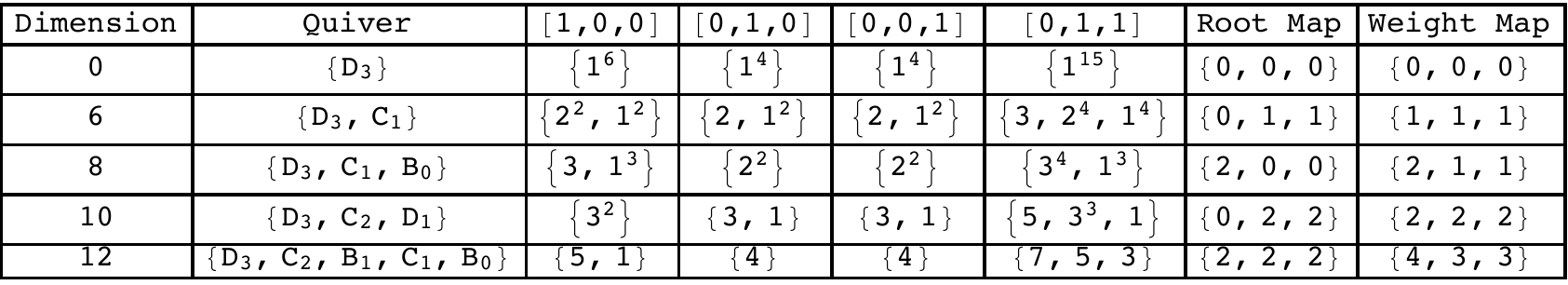}\\
~\\
\includegraphics[scale=.73]{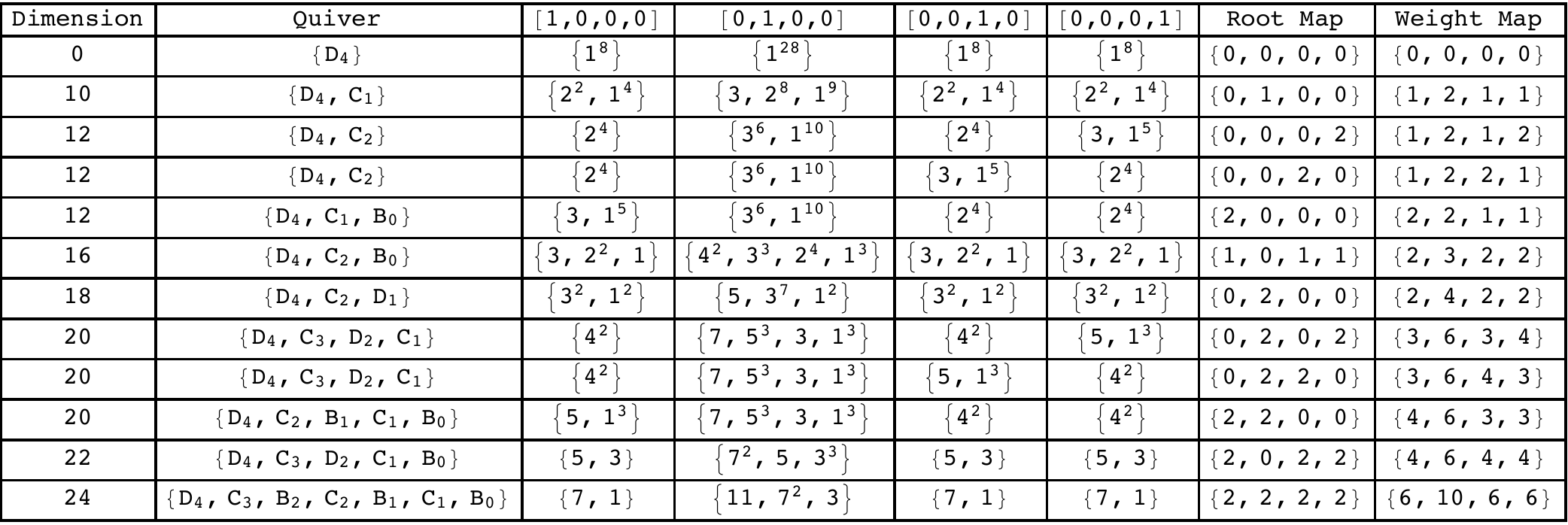}\\
~\\
\includegraphics[scale=.56]{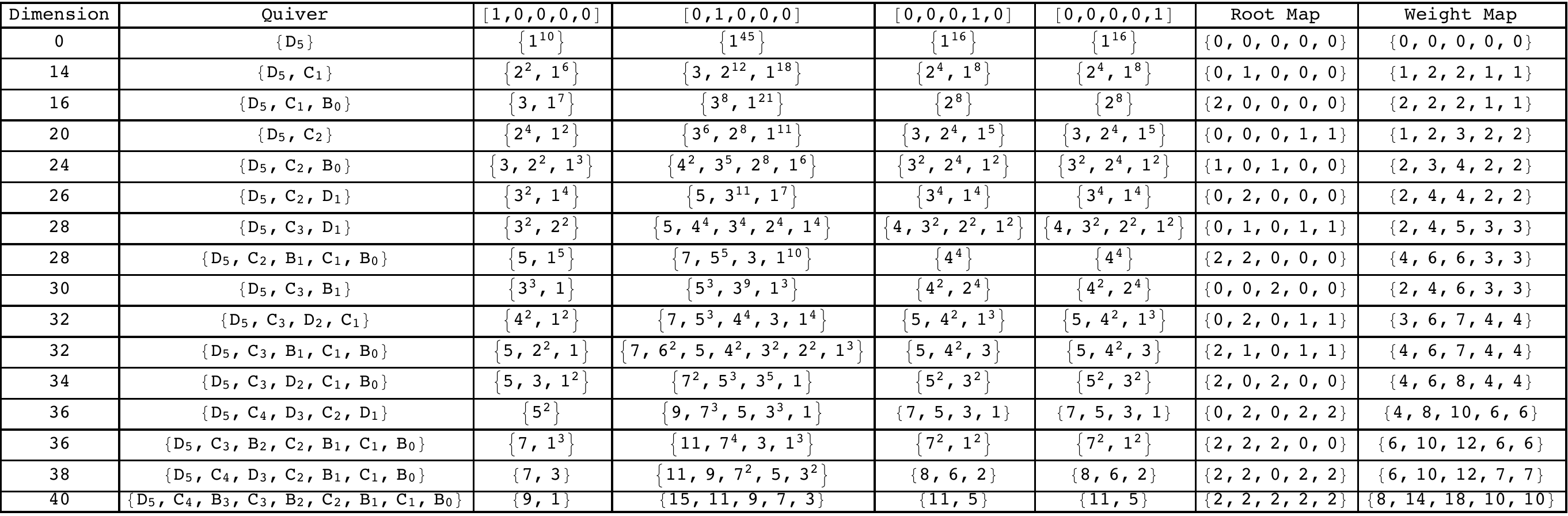}\\
\end{center}
Partitions are shown under each homomorphism for the vector, spinor and adjoint representations.

\clearpage


\bibliographystyle{JHEP}
\bibliography{RJKBibLib}


\end{document}